\documentclass[a4paper,11pt]{article}
\pdfoutput=1 

\usepackage{jheppub} 

\usepackage[T1]{fontenc} 
\usepackage[mathletters]{ucs}
\usepackage[utf8x]{inputenc}
\usepackage{multirow,booktabs,array}
\usepackage{lmodern}
\usepackage{braket}
\usepackage{verbatim}
\usepackage{multirow}
\usepackage{graphics,graphicx}
\usepackage{psfrag}
\usepackage{subcaption}
\usepackage{latexsym,slashed}
\usepackage{bbold}

\usepackage{mathtools}
\newcommand{\norm}[1]{\left\lVert#1\right\rVert}

\newcommand\al{{\alpha}}
\newcommand\ep{\epsilon}
\newcommand\si{\sigma}
\newcommand\Si{\Sigma}

\newcommand\de{{\ensuremath{{\delta}}}}

\newcommand\ka{\kappa}
\newcommand\be{\beta}

\newcommand{\bee}{\begin{equation}}
\newcommand{\ee}{\end{equation}}
\def\ba{\begin{array}}
\def\ea{\end{array}}

\def\bo1{ \left | B^0 (p^+) \right \rangle}

\newcommand{\bea}{\begin{eqnarray}}
\newcommand{\eea}{\end{eqnarray}}

\def\<{ \langle }
\def\>{ \rangle }

\usepackage{amsmath,latexsym,amssymb,slashed}

\title{One point functions for black hole microstates}

\author{Joan Garcia i Tormo}
\author{and Marika Taylor}
\affiliation{Mathematical Sciences and STAG Research Centre, University of Southampton, \\
Highfield, Southampton, SO17 1BJ, UK.}

\emailAdd{jgt1e15@soton.ac.uk}
\emailAdd{m.m.taylor@soton.ac.uk}

\abstract{We compute one point functions of chiral primary operators in the D1-D5 orbifold CFT, in classes of states corresponding to microstates of two and three charge black holes. Black hole microstates describable by supergravity solutions correspond to coherent superpositions of states in the orbifold theory and we develop methods for approximating one point functions in such superpositions in the large $N$ limit. We show that microstates built from long strings (large twist operators) have one point functions that are suppressed by powers of $N$. Accordingly, even when these microstates admit supergravity descriptions, the characteristic scales in these solutions are comparable to higher derivative corrections to supergravity. 
}

\begin{document} 
\maketitle
\flushbottom

\section{Introduction}\label{section:introduction}

The microscopic origin of black hole entropy has been at the forefront of research ever since the discovery of Hawking radiation \cite{Hawking:1974rv} and the formulation of the information loss paradox \cite{Hawking:1976ra}. Twenty years ago, Strominger and Vafa showed that the entropy of a class of supersymmetric black holes in string theory could be understood microscopically by counting states in a dual conformal field theory \cite{Strominger:1996sh}. These results, and their generalisations to other near supersymmetric black holes in string theory, were later understood to be part of the AdS/CFT correspondence discovered by Maldacena \cite{Maldacena:1997re}. These black holes have anti-de Sitter regions in their interiors that are dual to conformal field theories. 

AdS/CFT provides a microscopic explanation of the origin of black hole microstates in terms of states in the dual CFT. Holography also settles the longstanding information loss question:  since the dual quantum field theory is unitary, the evolution of black holes must be unitary. However, neither of these answers is entirely satisfactory from the gravity perspective. In the field theory one can describe both individual black hole microstates and the thermal ensemble, and one can find computables that distinguish between individual states. The recovery of information in the quantum field theory is associated with the unitary evolution of pure states: the radiation emitted is not exactly thermal, but carries information about the specific state.

On the gravity side information loss is inextricably related with the causal structure of the black hole. This led Lunin, Mathur and collaborators to postulate the fuzzball proposal \cite{Lunin:2001fv, Lunin:2001jy,Lunin:2002qf,Mathur:2002ie,Lunin:2002bj,Mathur:2005zp,Mathur:2008nj}: each individual black microstate should be described by a horizonless non-singular solution that differs from the black hole only at (sub)horizon scales. It is important to note that fuzzballs are not generically solutions of supergravity but rather solutions of the full quantum string theory. The fuzzball proposal directly addresses the issues of both black hole entropy and information loss: the entropy relates to the number of fuzzballs for a given black hole, while information is manifestly not lost as fuzzballs have no horizons and the radiation emitted depends on the specific microstate represented by the fuzzball. 

Note that the fuzzball proposal is not the only proposal for black hole physics that postulates qualitative changes in the spacetime at or behind the black hole horizon. For example, firewalls at the horizon \cite{Almheiri:2012rt,Almheiri:2013hfa} were proposed to resolve the black hole information loss paradox, while in the SYK duality it has been suggested that individual microstates are associated with shock waves behind the horizon \cite{Kourkoulou:2017zaj}. 

The fuzzball proposal has been extensively explored in the context of near supersymmetric black holes, particularly in the D1-D5 system originally studied by \cite{Strominger:1996sh}. There has been considerable work on constructing black hole microstate solutions of supergravity for this system. For 1/4 BPS black holes, namely the D1-D5 system with zero momentum, there are sufficient supergravity solutions to 
span all the microstates of the system \cite{Lunin:2001fv,Lunin:2002iz,Taylor:2005db,Kanitscheider:2007wq}, although generically the solutions are actually only extrapolations to supergravity, as higher derivative corrections are non-negligible. The latter relates to the fact that the 1/4 BPS black holes do not have macroscopic event horizons; the event horizon is only manifest after taking into account higher derivative corrections to supergravity.  

Nevertheless the 1/4 BPS black holes are an important arena for exploring the fuzzball proposal. The black hole microstates can be geometrically quantised and counted \cite{Rychkov:2005ji}, giving rise to an entropy that matches the result from the corresponding D1-D5 CFT. The microstates have interior AdS$_3$ regions and one can thus use the AdS$_3$/CFT$_2$ duality to explore their properties. In particular, one can use the precise holographic dictionary to relate data in the asymptotically AdS$_3$ region of the geometry to one point functions of chiral primary operators in the CFT \cite{Kanitscheider:2006zf,Kanitscheider:2007wq,Skenderis:2008qn}. This technique provides detailed matching between geometries and CFT states that goes beyond conserved charges, to complete Kaluza-Klein towers of operators. 

Let $ | \Psi \rangle $ represent the microstate of interest and ${\cal O}_s$ represent a specific single particle chiral primary operator of dimension $\Delta$, dual to a supergravity mode. In general the microstate $| \Psi \rangle$ is completely determined by the expectation values of all local operators in that state. The number of chiral primary operators (single plus multiple particle) grows exponentially as $\sqrt{N}$; the spectral flow of these operators gives the 1/4 BPS black hole microstates, which are the Ramond ground states of the dual CFT. In a 1/4 BPS microstate only chiral primaries can acquire expectation values; operators preserving less supersymmetry or no supersymmetry cannot acquire expectation values. Thus knowledge of the expectation values of chiral primaries is enough to determine the state. 

The expectation value of a scalar single particle chiral primary operator in a 1/4 BPS microstate can be expressed as:
\begin{equation}
\langle {\cal O}_s \rangle_{\Psi} = {\cal N}_{\Psi}  l_s^{\Delta} C_{\Psi}, \label{Cpsi}
\end{equation}
where ${\cal N}_{\Psi}$ is the operator normalisation and $l_s$ is the scale\footnote{The explicit factor of ${\cal N}_{\Psi}$ is included as the normalisation of the operators in supergravity differs by a factor of $N$ from the standard CFT normalisation.}. Here $C_{\Psi}$ is dimensionless, and is the fusion coefficient of the associated three point function in the conformal vacuum. 

One finds that  \cite{Kanitscheider:2006zf,Kanitscheider:2007wq,Skenderis:2008qn}
\begin{equation}
C_{\Psi} = \frac{c_{\Psi}}{N}, \label{cpsi}
\end{equation}
where $c_{\Psi}$ is order one for long string microstates while for short string microstates $c_{\Psi}$ can be large, of order $N$. Here $N = N_1 N_5$ is the rank of the dual CFT, with $N_1$ and $N_5$ being the numbers of D1-branes and D5-branes, respectively. The interpretation of these results from the gravity side is the following. Fuzzball solutions representing long string microstates can be described only as extrapolations of supergravity solutions: the characteristic scales in the supergravity solutions are of order $1/N$, and are hence comparable to higher derivative corrections to supergravity. Short string microstates are atypical but are well captured by supergravity solutions, as their characteristic scales are large compared to higher derivative corrections. 

\bigskip

Now let us turn to the case of 1/8 BPS black holes, the D1-D5 system with momentum $P$ studied in \cite{Strominger:1996sh}. There has been a long history of constructing supergravity solutions corresponding to D1-D5-P microstates, see
\cite{Mathur:2003hj,Giusto:2004ip,Bena:2004tk,Lunin:2004uu,Balasubramanian:2005qu,Berglund:2005vb,Jejjala:2005yu,Srivastava:2006xn,Bena:2007kg,deBoer:2009un,Giusto:2009qq,Bena:2011uw,Giusto:2011fy,Giusto:2012gt,Giusto:2012yz,Giusto:2012jx,Giusto:2013bda,Bena:2013dka,Bossard:2014ola,Bena:2015bea,Bena:2016agb,Bena:2016ypk,Bena:2017geu,Martinec:2017ztd,Bena:2017upb,Bena:2017fvm,Bena:2017xbt,Hampton:2018ygz,Tyukov2018,Ceplak:2018pws,Heidmann:2019zws}. The precision holography dictionary has also been extended to 1/8 BPS black hole microstates in \cite{Giusto:2015dfa}. 
In contrast to the 1/4 BPS case, it has not been possible to carry out geometric quantisation and count explicitly the number of black hole microstates visible in supergravity. 

However, one would not expect the supergravity solutions to account for a representative fraction of the black hole entropy. Just as in the case of a 1/4 BPS microstate, a 1/8 BPS microstate is characterised by the expectation values of all operators in that state. Both chiral primaries, single and multiple particle, and 1/8 BPS operators can acquire expectation values in 1/8 BPS microstates. Supergravity solutions can encode only the expectation values of the chiral primaries, not those of 1/8 BPS operators. The number of 1/8 BPS microstates with momentum $P$ grows exponentially as $\sqrt{N P}$ and one would not expect that information carried in the expectation values of chiral primaries (the number of which grows exponentially as $\sqrt{N}$) can suffice to capture all 1/8 BPS states. 

Microstates of 1/8 BPS black holes that can be described in supergravity are atypical but are nonetheless useful prototypes for exploring features of the fuzzball proposal such as information recovery. As described above, for supergravity solutions to provide a reliable description of a given black hole microstate, at least some of the generic single particle chiral primaries \eqref{Cpsi} must have expectation values such that 
\begin{equation}
C_{\Psi} \gg \frac{1}{N}, \label{crit}
\end{equation}
i.e. they must be large relative to the scale of higher derivative corrections to supergravity. (Here by generic we mean chiral primaries that do not have expectation values in the black hole i.e. we exclude operators associated with the conserved charges of the black hole.)

\bigskip

The objective of this paper is to explore one point functions of chiral primaries in 1/4 BPS and 1/8 BPS states in the D1-D5 CFT. We work in the orbifold limit of the CFT. Since the three point functions of (single particle) chiral primaries are protected in this theory (see \cite{Pakman:2007hn,Dabholkar:2007ey,Gaberdiel:2007vu,Taylor:2007hs}), the one point functions in 1/4 BPS states are not renormalised away from the orbifold point. While there are no proofs of non-renormalisation for three point functions involving two 1/8 BPS and one 1/4 BPS operators, the results of  \cite{Giusto:2015dfa} suggest that there may also be non-renormalisation in this sector and our results will allow this issue to be explored further. 

As explained above, our primary motivation for exploring one point functions in such states is black hole physics. In cases where supergravity representations of the black hole microstates exist, our results will allow detailed tests of their identification in terms of states in the D1-D5 CFT, using the methods of \cite{Kanitscheider:2006zf,Kanitscheider:2007wq,Skenderis:2008qn}. Perhaps more importantly our results lead to a better understanding of which classes of 1/8 BPS microstates are accessible within supergravity: microstates which do not satisfy \eqref{crit} cannot be captured in supergravity. 

Black hole microstates represented by supergravity geometries are dual to certain coherent superpositions of states in the D1-D5 CFT. In the case of 1/4 BPS black holes, there is a direct relationship between the curves defining the supergravity geometries and the corresponding superpositions of Ramond ground states in the dual CFT, as discussed and tested in \cite{Kanitscheider:2006zf,Kanitscheider:2007wq}. For 1/8 BPS microstates, analogous maps between supergravity geometries and superpositions of CFT states were explored in \cite{Giusto:2015dfa}. 

Here we calculate one point functions of chiral primaries in superpositions of 1/4 BPS microstates, and corresponding 1/8 BPS microstates obtained by adding momentum excitations. The latter states have a similar structure to those explored in \cite{Giusto:2015dfa}. 
Our results build on our recent work \cite{Tormo:2018fnt}, in which general correlation functions in the orbifold CFT were derived for processes involving $n$ strands being joined by a twist $n$ operator.

The D1-D5 CFT in the orbifold limit is a free theory, and therefore computation of the one point functions is essentially a problem in combinatorics. However, the combinatorics problem is complicated: the microstates under consideration are viewed as superpositions of a large number of orbifold states:
\begin{equation}
| \Psi \rangle = \sum_a {\cal A}_a | \Psi_a \rangle
\end{equation}
(with appropriate coefficients ${\cal A}_a$, see section \ref{joiningsec:setup}) and correspondingly the one point functions \eqref{Cpsi} involve summations over many contributions. In the case of operator expectation values in 1/4 BPS states studied in  \cite{Kanitscheider:2006zf,Kanitscheider:2007wq,Skenderis:2008qn} one could obtain exact results; here we are also able to obtain exact results for one point functions in 1/4 BPS states for a wide class of chiral primary operators. 

For 1/8 BPS states we do not work out exact results for the required combinatoric problems but instead develop approximation methods that make use of the large $N$ limit, together with estimates of the dominant contributions to the correlation functions. These approximation methods would be applicable to other calculations of correlation functions in orbifold CFTs, and are hence of interest beyond the black hole microstate programme.

One of our main results is the suppression of one point functions in microstates built on long strings, relative to those built on short strings. 
Just as in 1/4 BPS black holes, the 1/8 BPS microstates associated with long strings have one point functions that are parametrically smaller than those in short string microstates; they are suppressed by factors of $N$. As we discuss in the conclusions, these results imply that such long string microstates have at best an extrapolation to supergravity: the characteristic scales in the supergravity solutions are comparable to higher derivative corrections to supergravity.

\bigskip

The plan of this paper is as follows. In section \ref{section:setup} we summarise relevant features of the D1-D5 CFT. 
In section \ref{1pfsection:1pf} we calculate one point functions for representative single particle chiral primary operators in the short strand limit,
namely in the limit where the strand lengths are of order one and the number of strands is of order $N$.
We first give an overview of the procedure and then we calculate some examples explicitly for untwisted and twisted operators.
Section \ref{1pfsection:long1pf} contains one point functions for the same operators as the previous section but in the long strand case (strand length of order $N$).
We separate this section into two-charge states and three-charge states, as for the two-charge case we can give exact results whereas in the three-charge case we need to make approximations.
In section \ref{1pfsection:discussion} we briefly discuss the results obtained for the one point functions.
After that, we move to the calculation of multi particle one point functions, by considering products of twists operators.
In section \ref{joiningsec:setup} we present the state with which we calculate all results and we then derive the $n$-point function corresponding to the creation of a strand by joining strands two by two. At the end of this section, in subsection \ref{1pfsubsec:allwaystojoin}, we comment on all other possible ways of joining $n$ strands.
We conclude with a summary of all results in section \ref{section:resultreview} and a discussion of the implications of the results in section \ref{section:conc}.

\section{D1-D5 orbifold CFT}\label{section:setup}
Consider type IIB string theory compactified on $X \times S^1$, with $X$ being $\mathbb T^4$ or $K3$.
Let $N_5$ D5-branes wrap the five compact dimensions and $N_1$ D1-branes wrap the $S^1$.
$X$ is taken to be string scale and the scale of the $S^1$ is assumed to be much larger (so that the circle can effectively be treated as non-compact).
D1-D5 black hole solutions in the supergravity limit are asymptotic to $M^{4,1}\times S^1 \times X$.
The geometry of the decoupled near horizon limit is AdS$_3\times S^3 \times X$, and there is supersymmetry enhancement (see \cite{Boonstra:1997dy} and references therein).

The CFT dual to the decoupling region geometry is a two-dimensional superconformal field theory (SCFT).
In what follows, the focus is put on the theory for $X = \mathbb T^4$, although much of the later analysis of this paper also holds for $K3$, i.e. it does not rely on features specific to $\mathbb T^4$.
For toroidal compactifications, the SCFT is an $\mathcal N = (4,4)$ superconformal sigma model with central charges $c = \tilde c = 6 N_1 N_5$; this theory can be viewed as a deformation of a free orbifold CFT with target space $(\mathbb T^4)^{N_1 N_5}/S( N_1 N_5)$, where $S(n)$ is the symmetric group.
Three point functions of (single trace) chiral primaries are protected in this theory (see \cite{Pakman:2007hn,Dabholkar:2007ey,Gaberdiel:2007vu,Taylor:2007hs}), and thus agree with the corresponding three point functions calculated in supergravity.

In sections \ref{1pfsection:1pf} and \ref{1pfsection:long1pf} we calculate one point functions involving chiral primary operators in the field theory.
The states involved in such computations have a chiral primary operator $O$ within them.
Thus, in general the one point functions can be written as
\begin{equation}
\bra{O_1} O_2(y) \ket{O_3}.
\end{equation}
Therefore, using the relation between states and operators the one point functions calculated in this paper can easily be related to three point functions of the form
\begin{equation}
\braket{O_1(x) O_2(y) O_3(z)}.
\end{equation}
For some of them the dual supergravity one point functions are known, and for the others the matching is yet to be done.
We work in Euclidean signature on a cylinder which is parametrised as
\bee
w = \tau + i \sigma
\ee
where $0 \le \sigma < 2 \pi$ and $ -\infty < \tau < \infty$.

For the calculations presented in this paper it is crucial to understand the orbifold description of the theory, so let us review it.
The Hilbert space of the orbifold theory decomposes into twisted sectors, which come from the action of the symmetry group $S(N_1 N_5)$.
They are thus labelled by the conjugacy classes of the group, which consist of cyclic subgroups of various lengths.
See the discussion in \cite{Jevicki:1998bm} for further details.
Let $N := N_1 N_5$ be the total number of copies of the CFT, $m_i$ the lengths of the different cycles and $n_i$ their multiplicity.
Then, in order to have full physical states in the theory the conjugacy classes must satisfy the constraint
\begin{equation}
\sum_i n_i m_i = N,
\label{constraint}
\end{equation}
where the sum is over all the cycles.
There is a direct correspondence between the conjugacy classes and the long/short string picture of the D1-D5 system \cite{Maldacena:1996ds}.
The symmetry group of the SCFT is $SU(1,1|2) \times SU(1,1|2)$, which breaks down into the following parts.
The $SO(4)_E$ isometry of the $S^3$ in the gravity side is identified with the $SO(4)$ R-symmetry in the $\mathcal N = (4,4)$ superconformal algebra.
The other $SO(4)$ symmetry of the field theory is identified with the $SO(4)_I$ of the torus.
In subsection \ref{subsec:ffd} an explicit index description with free fields is given.

At the orbifold point of the theory chiral primaries can be precisely described, as they are associated with the cohomology of $X$.
Hence, chiral primaries in the NS sector are labelled as $\mathcal O_m^{(p,q)}$, where $m$ is the twist of that chiral primary and $(p, q)$ refers to its associated cohomology class.
The conformal weights $(h, \tilde h)$ and the R charges $(j_3, \tilde j_3)$ of these chiral primaries are given by
\begin{equation}
h^{\rm NS} = j_3^{\rm NS} = \frac{1}{2} (p + m -1), \qquad \tilde{h}^{\rm NS} = \tilde{j}_3^{\rm NS} = \frac{1}{2} (q + m -1). \label{NS-charge}
\end{equation}
Recalling the constraint for the cycles mentioned above, the complete set of chiral primaries is built from products adding up to the total twist,
\begin{equation}
\prod_l ({\cal O}^{(p_l,q_l)}_{m_l})^{n_l}, \qquad \sum_l n_l m_l = N,
\label{d1d5eq:stateconstr}
\end{equation}
with symmetrisation over $N$ copies of the CFT implicit.

Chiral primaries in the NS sector are mapped to ground states in the R sector via spectral flow.
Spectral flow is a deformation of the algebra, under which the quantum numbers of the R ground states transform as
\begin{equation}
h^{\rm R} = h^{\rm NS} - j_3^{\rm NS} + \frac{c}{24}, \qquad j_3^{\rm R} = j_3^{\rm NS} - \frac{c}{12},
\end{equation}
where $c$ is the central charge of the CFT.
For chiral primaries of associated twist $m$ the central charge is $c = 6m$; the central charge of the full theory is $c = 6N_1 N_5$.
Analogous expressions hold for the right moving sector.
As we just said, NS chiral primaries are mapped by spectral flow to R ground state operators,
\begin{equation}
\prod_l ({\cal O}^{(p_l,q_l)}_{m_l})^{n_l} \rightarrow \prod_l ({\cal O}^{{\rm R}(p_l,q_l)}_{m_l})^{n_l}
\end{equation}
with R charges
\begin{equation}
j_3^{\rm R} = \frac{1}{2} \sum_l (p_l - 1) n_l, \qquad \tilde{j}_3^{\rm R} = \frac{1}{2} \sum_l (q_l - 1) n_l.
\end{equation}
Note that the Ramond operators obtained from primaries associated with the $(1,1)$ cohomology have zero R charge.

The microstates of the 2-charge D1-D5 black hole are Ramond ground states.
The entropy associated to the microstates is
\begin{equation}
S = 2 \pi \sqrt{\frac{C(X) N}{6}}
\end{equation}
where $C(X)$ is determined by the cohomology.
$C = 12$ for $K3$ and $C = 24$ for $\mathbb{T}^4$.
However, the corresponding black holes do not have macroscopic horizons.
The famous 3-charge black holes with macroscopic horizons discussed in \cite{Strominger:1996sh} are obtained by exciting the left moving sector with momentum $P$, and they do have macroscopic horizons.
The entropy is then
\begin{equation}
S =  2 \pi \sqrt{\frac{C(X) N P}{6}},
\end{equation}
where implicitly it is assumed that $P \gg N$.
The generic structure of the 3-charge microstates is thus
\begin{equation}
{\cal O}_{P} \prod_l ({\cal O}^{{\rm R}(p_l,q_l)}_{m_l})^{n_l},
\label{d1d5eq:gen3chargemicro}
\end{equation}
where ${\cal O}_P$ describes the excitation of momentum $P$.
As discussed in early works such as \cite{Maldacena:1996ds}, most of the 3-charge microstates are associated with excitations over maximal and near maximal twist ground states (``long strings'') as there are more ways to fractionate the momentum over such states.
This is our motivation to calculate one point functions of chiral primaries for short and long strings and compare their results.

Throughout this paper we will be using nomenclature and results of the theory of integer partitions.
We introduce all necessary concepts as they are needed.
However, we include a short and introductory section on the topic, with the basic definition and references for further reading.
The reader familiar with the concept can skip the next section.

\subsection{Integer partitions}\label{1pfsubsec:integerpartitions}
Let $n$ be a positive integer.
A partition of $n$ is a finite non-increasing sequence of positive integers $m_1, ..., m_r$ such that they add up to $n$,
\begin{equation}
\sum_{i = 1}^r m_i = n.
\end{equation}
The $m_i$ are called the parts of the partition.
The total number of partitions of $n$ is denoted by $p(n)$ and is called the partition function.
Let us write the partitions for the first six natural numbers as an example.
\begin{align}
p(0) = 1 & \qquad \text{(the empty sequence)} \nonumber \\
p(1) = 1 &; \;\;\; 1 \nonumber \\
p(2) = 2 &; \;\;\; 2, \;\; 1+1; \nonumber \\
p(3) = 3 &; \;\;\; 3, \;\; 2+1, \;\; 1+1+1; \nonumber \\
p(4) = 5 &; \;\;\; 4, \;\; 3+1, \;\; 2+2, \;\; 2+1+1, \;\; 1+1+1+1 \nonumber \\
p(5) = 7 &; \;\;\; 5, \;\; 4+1, \;\; 3+2, \;\; 3+1+1, \;\; 2+2+1, \;\; 2+1+1+1, \;\; 1+1+1+1+1.
\label{1pfeq:intpartex}
\end{align}
For an exhaustive explanation and results on integer partitions see, for instance, \cite{andrews1976}.
As we said, we introduce more definitions and theorems that derive from this definition in each section, when they are needed.
Let us now go back to reviewing the orbifold CFT, by introducing the free field description.

\subsection{Free field description}\label{subsec:ffd}
In this paper we do not use explicitly the free field description of the theory for most calculations, but it is necessary to introduce part of it to define the chiral primary operators with which we work.
Details of this description and also a complete classification of all chiral primaries for this theory can be found in \cite{David:2002wn}.
Our notation follows closely that of \cite{Giusto:2015dfa} and \cite{Bena:2016agb}.

Let us first define our notation and conventions.
Recalling that $SO(4) \simeq SU(2)\times SU(2)$ we write the $SO(4)$ symmetry associated with the torus as $SU(2)_{\mathcal C} \times SU(2)_{\mathcal A}$.
We label the $SU(2)_{\mathcal C}$ group with $\dot A = {\dot 1, \dot 2}$, and the $SU(2)_{\mathcal A}$ with $A = {1, 2}$.
The R-symmetry $SO(4)$ group also splits into two $SU(2)$ subgroups, corresponding to the left and right R-symmetry.
We use the labels $\al, \dot \al$ to identify them, with $\al = \{+, -\}$ and $\dot \al = \{\dot +, \dot -\}$.
To refer to the copies of the torus we use a subindex $(r)$, which runs from 1 to $N_1 N_5$.
Fields and operators corresponding to the right moving sector are denoted with a tilde.

At the orbifold point, the CFT has free fields
\begin{equation}
\left(X^{\dot A A}_{(r)}(w, \bar w ), \psi^{\al\dot A}_{(r)}(w), \tilde{\psi}^{\dot \al  \dot A}_{(r)}(\bar w )\right),
\end{equation}
that is, four bosons and four doublets of fermions.
The mode expansion of the fermions in the Ramond sector is
\begin{equation}
\psi^{\al \dot A}_{(r)}(w) = \sum_{n \in \mathbb Z} \psi^{\al \dot A}_{n (r)} e^{-nw}, \qquad \tilde \psi^{\dot \al \dot A}_{(r)}(\bar w) = \sum_{n \in \mathbb Z} \tilde \psi^{\dot \al \dot A}_{n (r)}e^{-n \bar w},
\end{equation}
and they satisfy the following Hermitian properties,
\begin{equation}
\psi^{+ \dot 1 \dagger}_{n (r)} = -\psi^{- \dot 2}_{-n (r)}, \qquad \psi^{+ \dot 2 \dagger}_{n (r)} = \psi^{- \dot 1}_{-n (r)}.
\end{equation}
The right-moving sector is completely analogous.
The Ramond vacuum state, which we denote as $\ket{++}_{(r)}$, is defined by
\begin{equation}
\psi^{+ \dot 1}_{0 (r)}\ket{++}_{(r)} = \psi^{+ \dot 2}_{0 (r)}\ket{++}_{(r)} = 0, \qquad \tilde \psi^{\dot + \dot 1}_{0 (r)}\ket{++}_{(r)} = \tilde \psi^{\dot + \dot 2}_{0 (r)}\ket{++}_{(r)} = 0.
\label{corfun++def}
\end{equation}
The R-symmetry currents in terms of the free fermions read
\begin{equation}
J^{\al \be}_{(r)} (w) = \frac 1 2 \psi^{\al \dot A}_{(r)} (w) \epsilon_{\dot A \dot B} \psi^{\be \dot B}_{(r)} (w), \qquad \tilde J^{\dot \al \dot \be}_{(r)}(\bar w) = \frac 1 2 \tilde \psi^{\dot \al \dot A}_{(r)} (\bar w)\epsilon_{\dot A \dot B}\tilde \psi^{\dot \be \dot B}_{(r)}(\bar w),
\end{equation}
where the operators are normal-ordered with respect to the $\ket{++}_{(r)}$ ground state.
Another operator we are interested in is
\begin{equation}
\mathcal O^{\al \dot \al}_{(r)} := - \frac{i}{\sqrt 2} \psi^{\al \dot A}_{(r)} \ep_{\dot A \dot B} \tilde \psi^{\dot \al \dot \be}_{(r)}.
\end{equation}
Notice that all these operators have been defined to be unitary.
Using its zero mode and the zero modes of the standard $SU(2)$ generators of the left R-symmetry current, which are defined as
\begin{equation}
J^3_{(r)} := -J^{+-}_{(r)} + \frac 1 2, \qquad J^+_{(r)} := J^{++}_{(r)} \qquad \mathrm{and} \qquad J^-_{(r)} := -J^{--}_{(r)}
\end{equation}
the other R ground states can be written as
\begin{equation}
\ket{-+}_{(r)} := J^-_{0(r)}\ket{++}_{(r)}, \qquad \ket{+-}_{(r)} := \tilde{J}^-_{0(r)}\ket{++}_{(r)}
\label{corfun--def}
\end{equation}
and
\begin{equation}
\ket{00}_{(r)} := \lim_{z\to 0}\mathcal O^{-\dot -}_{00(r)} \ket{++}_{(r)} = \frac{1}{\sqrt 2}\psi^{-\dot A}_{0(r)}\epsilon_{\dot A\dot B}\tilde{\psi}^{\dot -\dot B}_{0(r)}\ket{++}_{(r)}.
\label{corfun00def}
\end{equation}
The operator modes change when moving from the NS to the R sector.
This change is determined by the spectral flow operation mentioned above. 

Let us turn our attention now to the twisted sector of the theory.
The twist (or gluing) operator, which is denoted by $\Si^{\al \dot \al}_\ka$, is an operator which induces a cyclic permutation of $\ka\geq 2$ copies of elementary fields.
This operator is also a chiral primary, and it generates the twisted states.
That is, it generates the cycles of length $\ka$.
In other words, the twist operator joins $\ka$ strings of winding one into a single string of winding $\ka$.
The strands of length $\ka$ are defined as
\begin{equation}
\ket{++}_\ka := \lim_{z\to 0}|z|^{\ka - 1}\Si^{- \dot -}_\ka (z, \bar z)\prod_{r = 1}^\ka \ket{++}_{(r)},
\label{corfunstrandk}
\end{equation}
where $\Si^{- \dot -}_\ka$ is the lowest weight state in the $\Si_\ka$ multiplet.
This operator has conformal dimensions ($\frac{\ka - 1}{2}$, $\frac{\ka - 1}{2}$) and, as can be read from the definition above, the state $\ket{++}_\ka$ in the Ramond sector has spin ($\frac 1 2$, $\frac 1 2$) and winding $\ka$.
To write the expression of the twist operator $\Si_{\ka}$ in terms of free fields it is necessary to bosonise the fermions.
Details on bosonisation and the expression of the twist operator in terms of free fields are not explicitly necessary for the calculations in this paper, and so we do not include them.
We refer to \cite{Burrington:2012yq} for the exact relations.
The normalisation of the twist operator is dealt with in section \ref{1pfsubsec:excitedstates}.

Now that the gluing operator has been introduced we can study the twisted sector of the theory.
First of all, in order to facilitate the calculations, the fermion basis needs to be changed, in order to obtain independent fields for this sector.
To do so we consider the following combinations, which diagonalise the boundary conditions:
\begin{equation}
\psi^{\al\dot A}_{\rho}(z) = \frac{1}{\sqrt{\ka}}\sum_{r = 1}^{\ka} e^{-2\pi i\frac{r\rho}{\ka}} \psi^{\al\dot A}_{(r)}(z), \qquad \mathrm{with} \qquad \rho = 0, 1, ..., \ka - 1.
\label{corfundiagonalfermions}
\end{equation}
To obtain the other R ground states in the twisted sector the zero modes of the fermions are used, just as in the untwisted case,
\begin{equation}
\ket{-+}_\ka = J^-_{0 \rho = 0}\ket{++}_\ka, \qquad \ket{+-}_\ka = \tilde{J}^-_{0 \rho = 0}\ket{++}_\ka.
\end{equation}
Similarly, for the spin zero R ground states we act with the zero mode of $\sum_{r = 1}^ \ka \mathcal O^{- \dot -}_{(r)}$,
\begin{equation}
\ket{00}_\ka = -\frac{i}{\sqrt 2} \psi^{- \dot A}_{0 \rho=0} \epsilon_{\dot A\dot B}\tilde{\psi}^{\dot -\dot B}_{0 \rho=0} \ket{++}_\ka.
\end{equation}
The relation between ground states and their associated cohomologies is
\begin{equation}
{\mathcal O}^{R (2,2)}_{\kappa} \leftrightarrow \ket{ + +}_\ka, \qquad {\mathcal O}^{R (1,1)}_{\kappa} \leftrightarrow \ket{ 0 0 }_\ka, \qquad {\mathcal O}^{R (0,0)}_{\kappa} \leftrightarrow \ket{ - - }_\ka,
\label{eq:cohomologyrelations1}
\end{equation}
and
\begin{equation}
{\mathcal O}^{R (2,0)}_{\kappa} \leftrightarrow \ket{ + -}_\ka, \qquad {\mathcal O}^{R (0,2)}_{\kappa} \leftrightarrow \ket{ - +}_\ka.
\label{eq:cohomologyrelations2}
\end{equation}
In this paper we only consider ground states associated with even cohomology classes.
Henceforth, in order to see the effects of the operators more explicitly we drop the cohomology notation.

Now that we have twisted sectors we should also give the definitions of the Ramond sector and the Neveu-Schwarz sector in the twisted case.
Analogous to the usual definition for the untwisted case, which is
\begin{equation}
\psi^{\mu} (w + 2\pi) = e^{2\pi i \nu} \psi^{\mu} (w), \qquad \tilde{\psi}^{\mu} (\bar w + 2\pi) = e^{-2\pi i \tilde{\    nu}} \tilde{\psi}^{\mu}(\bar w),
\label{introeq:rnssectors}
\end{equation}
with $\nu = \tilde{\nu} = 0$ being the R sector and $\nu = \tilde{\nu} = 1/2$ being the NS sector, we now define
\begin{align}
\text{R sector:} &\qquad \psi_{(r)} \to \psi_{(r + 1)}, \quad ...\; , \quad \psi_{(n)} \to \psi_{(1)} \nonumber \\
\text{NS sector:} &\qquad \psi_{(r)} \to \psi_{(r + 1)}, \quad ...\; , \quad \psi_{(n)} \to - \psi_{(1)}.
\end{align}

So far we have defined the ground states for a single copy, namely, in the untwisted sector, and we have also seen how to construct the twisted sector ones.
We have also defined all the operators for which we will compute one point functions explicitly, but only in the untwisted sector.
Most of the explicit calculations in this paper will be using these untwisted operators, but we are also interested in one point functions for twisted operators.
Therefore, we define them in the next section.

\subsection{Operators in the twisted sector}\label{1pfsubsec:twistedoperators}
Let us now focus on the chiral primary operators defined in the twisted sector.
All chiral primaries from single particle states of the SCFT are listed in \cite{David:2002wn}, and we refer there for the exhaustive list.
We calculate one point functions for a subset of them in this paper, but the methods can be easily extended to the rest.

Keeping in mind that the twist operator $\Si_\ka$ acts on a product of $\ka$ copies, the first twisted operator that would come to mind to extend $\mathcal O^{\al \dot \al}_{(r)}$ to $\ka$ copies would be to join $\ka$ $\mathcal O^{\al \dot \al}_{(r)}$ operators using a twist of length $\ka$.
This process creates chiral primary operators, but not coming from single particle states.
We can see this explicitly by looking at the quantum numbers.
Consider for instance the operator
\begin{equation}
\mathcal O_{\ka}^{+ \dot +,*} := \Si_\ka \bigotimes_{r = 1}^{\ka} \mathcal O^{+ \dot +}_{(r)}.
\label{1pfeq:twistedheavyO}
\end{equation}
It is a chiral primary, as it has
\begin{align}
(h,j) & = \left(\frac{\ka - 1}{2}, \frac{\ka - 1}{2}\right) + \ka \cdot \left(\frac 1 2, \frac 1 2\right) = \left(\ka -\frac 1 2, \ka -\frac 1 2\right) \nonumber \\
(\tilde h,\tilde j) & = \left(\frac{\ka - 1}{2}, \frac{\ka - 1}{2}\right) + \ka \cdot \left(\frac 1 2, \frac 1 2\right) = \left(\ka -\frac 1 2, \ka -\frac 1 2\right).
\end{align}
It is not however a ``single particle'' operator, as we have the product of many $\mathcal O^{\al \dot \al}_{(r)}$ operators.
Analogous definitions can be done for the rest of the $\mathcal O^{\al \dot \al}$ operators.
These operators are similar, but not exactly equal to the heavy states considered in \cite{Galliani:2016cai}.
The operators considered there are the same without the twist operator.
That also generates heavy chiral primaries, but untwisted ones.
We leave for the end of this section the discussion on the heavy and light operator nomenclature.

We can also create analogous chiral primaries with the $J$ currents.
For instance, we can consider the $\ka$-twist operator
\begin{equation}
J^{+,*}_{\ka} := \Si_\ka \bigotimes_{r = 1}^{\ka} J^+_{(r)},
\end{equation}
which has $(h,j) = ((3\ka-1)/2, (3\ka -1)/2)$ and $(\tilde h, \tilde j) = ((\ka - 1)/2, (\ka - 1)/2)$ and so is also a chiral primary.
We can analogously construct the rest of the $J$ operators this way, and also without the twist operator.
The methods for computing one point functions showed in this paper extending the work of \cite{Giusto:2015dfa} can be easily applied to these cases.
We will comment on this further in subsection \ref{1pfsubsec:multishort}.
Hence, it would be interesting to have a better understanding of the gravity duals of all these one point functions.

However, in this paper, in sections \ref{1pfsection:1pf} and \ref{1pfsection:long1pf} we are concerned with chiral primaries from single particle states.
Before we go into details, let us define some conventions that will simplify the expressions and ease the notation.
In what follows we will define operators like
\begin{equation}
\Si_\ka \sum_{r = 1}^{\ka} \mathbb{1}_{(1)} \otimes ... \otimes \mathbb 1_{(r - 1)} \otimes \mathcal O^{\al \dot \al}_{(r)} \otimes \mathbb 1_{(r + 1)} \otimes ... \otimes \mathbb 1_{(\ka)}.
\end{equation}
That means, we act trivially on all copies except one, on which we act with the $\mathcal O^{\al \dot \al}_{(r)}$ operator.
In order to avoid long expressions, we will leave all the identity operators implicit, and instead write
\begin{equation}
\Si_\ka \sum_{r = 1}^{\ka} \mathcal O^{\al \dot \al}_{(r)}.
\label{defOnsingle}
\end{equation}
From now on, and throughout this paper, every time we write an expression like \eqref{defOnsingle} we are leaving all the identity operators implicit.
Let us now write the chiral primaries from single particle states.
As we said, they are all listed in \cite{David:2002wn}, and we give only the subset with which we work.

\paragraph{Chiral primaries from single particle states with $h - \tilde h = 0$:}
The four chiral primaries with $h - \tilde h = 0$ corresponding to the (1,1) cohomology are
\begin{equation}
\Si^{\frac{\ka - 1}{2}}_\ka \sum_{r = 1}^\ka \psi^{+ \dot 1}_{(r)} \tilde{\psi}^{\dot + \dot 1}_{(r)}, \qquad \Si^{\frac{\ka - 1}{2}}_\ka \sum_{r = 1}^{\ka} \psi^{+ \dot 1}_{(r)} \tilde{\psi}^{\dot + \dot 2}_{(r)}, \qquad \Si^{\frac{\ka - 1}{2}}_\ka \sum_{r = 1}^\ka \psi^{+ \dot 2}_{(r)} \tilde{\psi}^{\dot + \dot 1}_{(r)}, \qquad \Si^{\frac{\ka - 1}{2}}_\ka \sum_{r = 1}^\ka \psi^{+ \dot 2}_{(r)} \tilde{\psi}^{\dot + \dot 2}_{(r)},
\label{1pfeq:twistedOdef}
\end{equation}
where the superindex in the twist operator corresponds to its conformal dimension.
These chiral primaries have conformal dimension $(\ka/2,\ka/2)$.
They have one fermion of the left sector and one of the right sector, and so they correspond to the $\mathcal O^{\al \dot \al}_n$ operators.
Therefore, taking combinations\footnote{Note that the operators given in \eqref{1pfeq:twistedOdef} correspond only to the case $\al = +$, $\dot \al = \dot +$ of the $\mathcal O^{\al \dot \al}_{\ka}$ operator.
That is because, as is usual in the literature, in \cite{David:2002wn} only the bottom component of the multiplet is given explicitly.
We obtain the other $\mathcal O^{\al \dot \al}_{\ka}$ operators that we have written above by taking other components of that same multiplet.}
we see that the twisted sector generalisation of the chiral primary $\sum_{r = 1}^{\ka} \mathcal O^{\al \dot \al}_{(r)}$ is
\begin{equation}
\mathcal O^{\al \dot \al}_{\ka} := \Si^{\frac{\ka - 1}{2}}_\ka \sum_{r = 1}^{\ka} \mathcal O^{\al \dot \al}_{(r)}.
\label{1pfeq:twistedOdefi}
\end{equation}
The other operators with $h - \tilde h = 0$ are obtained with other components of the short multiplet of the twist operator.
There is another chiral primary associated to the (0,0) cohomology, which is
\begin{equation}
\Si^{\frac{\ka - 2}{2}} \sum_{r = 1}^{\ka} \psi^{+ \dot 1}_{(r)} \psi^{+ \dot 2}_{(r)} \tilde \psi^{\dot + \dot 1}_{(r)} \tilde \psi^{\dot + \dot 2}_{(r)}.
\end{equation}
We do not use this operator in this paper because we only want to raise the spin on the left, as we mentioned before.
The sixth and last chiral primary with $h = \tilde h$ is the twist operator $\Si^{\frac{\ka}{2}}_{\ka}$, which also has conformal dimension $(\ka/2, \ka/2)$.

\paragraph{Chiral primaries from single particle states with $h - \tilde h = 1$:}
In this case we have
\begin{equation}
\Si^{\frac{\ka}{2}}_{\ka} \sum_{r = 1}^{\ka} \psi^{+ \dot 1}_{(r)} \psi^{+ \dot 2}_{(r)},
\end{equation}
from which we construct the $J_n$ currents.
Thus, we define
\begin{equation}
J^i_{\ka} := \Si^{\frac{\ka}{2}}_{\ka} \sum_{r = 1}^{\ka} J^i_{(r)}.
\end{equation}
Notice that $J^i_{\ka}$ and $J^{i,*}_{\ka}$ have the same transformation under spectral flow, just like $\mathcal O^{\al \dot \al}_{\ka}$ and $\mathcal O^{\al \dot \al, *}_{\ka}$ do.
To finish this section let us connect these operators to other notation found in the literature (see, for instance, \cite{Galliani:2016cai}).

\paragraph{Heavy and light operators}
In the context of one point functions in the D1-D5 system, and more generally in holographic CFT calculations, it is common in some literature to define heavy and light operators \cite{Galliani:2016cai,Galliani:2017jlg,Bombini:2017sge}.
These are all chiral primaries, with different conformal weights.
Light operators are operators with low conformal dimension relative to the central charge $c$, and heavy operators have large conformal dimension (of order $c$).
With the definitions that we have given above, if we consider the operators alone it is natural to say that heavy operators are the ones we have constructed in the twisted case, using the gluing operator $\Si$ and combinations of fermions.
Light operators would then correspond to single trace operators in the untwisted sector.

When looking at the one point functions that we compute in section \ref{1pfsection:1pf} we need to take into account that, in the definition of our states (which we will see in subsection \ref{1pfsubsection:statesreview}), we have some operators inside the definition of the strands.
As we will see in the following sections, the calculations that we make can be used to calculate both heavy and light one point functions.
Now that we have defined all the strands and operators we will briefly review how to construct 1/8-BPS strands, that is, strands where we raise the left R symmetry charge.
Further details can be found in \cite{Bena:2015bea, Chakrabarty:2015foa,Giusto:2015dfa, Bena:2016agb,Bena:2016ypk}.

\subsection{Creating 1/8-BPS strands}\label{1pfsubsec:1/8bpsstatedef}
So far we have only defined the two-charge strands in section \ref{subsec:ffd}.
In this section we define three-charge strands.
To construct the most general 1/8-BPS strands considered so far in the literature first we need to introduce fractional modes.
In a sector of twist $\ka$ of our theory one can define \cite{Lunin:2001pw}
\begin{equation}
J^+_{- \frac{n}{\ka}} := \oint \frac{\mathrm d z}{2 \pi i} \sum_{r = 1}^{\ka} J^+_{(r)} (z) e^{-2 \pi i \frac{n}{\ka} (r - 1)} z^{- \frac{n}{\ka}},
\label{1pfeq:fractionalmode}
\end{equation}
where $n$ is an integer. 
These modes allow us to increase the R charge of a state by one unit while only raising the conformal dimension by $n/\ka$.
Then, given the R ground states that we defined in section \ref{subsec:ffd}, one can add momentum excitations by acting with these fractional modes,
\begin{equation}
\left( J^+_{-\frac{n_{\kappa}}{\kappa}} \right)^{m_{\kappa}} \ket{00}_{\ka}. \label{3cm}
\end{equation}
We will be writing strands generically like this in section \ref{1pfsection:1pf}, but it is important to keep in mind that the total momentum added to the state has to be integral \cite{Maldacena:1996ds}.
In this paper the attention is focused on the case when $n = \ka$, even though the calculation can be easily extended to generic fractional modes.
This state also shows us why we expect most of the 3-charge microstates to be associated with long strings: the greater $\ka$ is, the more possibilities we have to distribute the momentum excitations within the strand.
We will see this very clearly when we consider the norm of the states in the next section.
This long and short string difference is also our motivation to calculate one point functions in the long strand case.

We now have all the definitions for the strands and operators that we need to calculate all the one point functions that we consider explicitly in this paper.
However, so far we have only considered building blocks of our states.
Recalling equation \eqref{constraint}, we see that any state of the full theory must have strands adding up to $N$.
We call states which satisfy this condition physical states, and in the next section we construct all the ones that we use in this paper, and give their norm.

\subsection{Physical states}\label{1pfsubsection:statesreview}
This section is very closely related to section 3 of \cite{Giusto:2015dfa}, but we include it here for completeness.
Before we start constructing states let us introduce some notation.
We denote by $N$ the total winding number, and $\ket{gs}_{(r)}$ denotes any of the ground states, i.e., $\ket{\pm, \pm}_{(r)}$ or $\ket{00}_{(r)}$, on the copy $r$ of the CFT.
When instead of writing a number in parenthesis in the subindex we write a number $\ka$ then it denotes a strand of length $\ka$.
We consider several copies of each strand, in order to satisfy the condition \eqref{constraint} and be able to have generic strand lengths.
We denote the number of copies of each strand by $N_\ka^{(gs)}$, where $\ka$ is the length of the strand $\ket{gs}_\ka$.
To get the $\frac 1 8$-BPS states we use the R-symmetry current, as explained in section \ref{1pfsubsec:1/8bpsstatedef}.
We denote by $N_{\ka}^{m_{\ka}(00)}$ the number of copies of the three-charge strand, where the new index stands for the number of insertions of the $J^+$ mode.

Now that we have this notation, let us write a full 1/4-BPS state.
Keeping in mind that the total winding number is $N$ we define
\begin{equation}
\psi(\{N_\ka^{(gs)}\}) := \prod_{gs, \ka} (\ket{gs}_\ka)^{N_\ka^{(gs)}}, \qquad \mathrm{with} \qquad \sum_{gs, \ka}\ka N_\ka^{(gs)} = N,
\label{gmr1/4def}
\end{equation}
where $\{N_\ka^{(gs)}\}$ denotes the partition that satisfies the condition of the total winding being $N$.
We denote by $\mathcal N(\{N_\ka^{(gs)}\})$ the norm of this state.
This norm is taken to be the number of combinations in which one can produce $N_\ka^{(gs)}$ strands $\ket{gs}_\ka$ starting from the state
\begin{equation}
\bigotimes_{r = 1}^N \ket{++}_{(r)}.
\label{1pfeq:initialstate}
\end{equation}
Recall that the Ramond ground states in a single copy have unit norm.
Also, once we have created the different strands, there is a unique way to transform them to the desired ground state.
We do this by acting with the $\mathcal O$ and $J$ operators defined in \ref{subsec:ffd}, which are already normalised.
Therefore we only need to consider the creation of the twisted sectors.
Starting from the state \eqref{1pfeq:initialstate} there are $\frac{N!}{(N - \ka)! \ka}$ possible ways in which we can choose $\ka$ of these copies up to cyclic permutations.
Taking this number into account every time we construct another twisted sector, we produce the following number of terms
\begin{equation}
\frac{N!}{(N - \ka_1)! \ka_1} \frac{(N - \ka_1)!}{(N - \ka_1 - \ka_2)! \ka_2} \cdot ... \cdot \frac{(N - \ka_1 - \ka_2 - ... - \ka_{g - 1})!}{0! \ka_g} = \frac{N!}{\prod_{\ka, S} \ka^{N_\ka^{(S)}}},
\end{equation}
where $g$ simply denotes the last term.
The normalisation of the twist operator is calculated in this way as well, as we show explicitly in the next section.
If we have several strands of the same type it does not matter in what order we got them, and thus we have to divide by an extra $N_\ka^{(S)}!$.
So, the norm of the physical state \eqref{gmr1/4def} is
\begin{equation}
\mathcal N(\{N_\ka^{(gs)}\}) = \frac{N!}{\prod_{gs, \ka} N_\ka^{(gs)}! \ka^{N_\ka^{(gs)}}}.
\label{gmr14norm}
\end{equation}
Notice that the states are orthogonal,
\begin{equation}
\left(\psi_{\{N_\ka^{(gs)}\}}, \psi_{\{N_\ka^{'(gs)}\}}\right) = \de_{\{N_\ka^{(gs)}\}, \{N_\ka^{'(gs)}\}} \mathcal N(\{N_\ka^{(gs)}\}).
\end{equation}
Last, we write a dimensionless coefficient $A_{\ka}^{(gs)}$ in front of each strand, so that our state describes the CFT dual of a black hole microstate geometry.
These coefficients satisfy
\begin{equation}
\sum_{gs, \ka} |A_{\ka}^{(gs)}|^2 = N.
\label{1pfeq:coefconstraint}
\end{equation}
See \cite{Giusto:2015dfa} for the formulas connecting these coefficients to the corresponding supergravity geometry.
Including these $A^{\text(gs)_\ka}$ Fourier parameters the physical states are written as
\begin{equation}
\psi_{\{A_\ka^{(gs)}\}} := \sum_{\{N_\ka^{(gs)}\}} \left(\prod_{gs, \ka} A_\ka^{(gs)} \right)^{N_\ka^{(gs)}} \psi_{\{N_\ka^{(gs)}\}} = \sum_{\{N_\ka^{(gs)}\}} \prod_{gs, \ka} (A_\ka^{(gs)} \ket{gs}_\ka)^{N_\ka^{(gs)}},
\label{gmr14dual}
\end{equation}
where the sum is restricted as in \eqref{gmr1/4def}.

Now that we have described the $\frac 1 4$-BPS state, let us excite it to obtain the $\frac 1 8$-BPS one.
As we mentioned in \eqref{3cm}, to obtain a three-charge solution we raise the momentum of the states using modes of the $J^{\pm}$ operators.
In sections \ref{1pfsection:1pf} and \ref{1pfsection:long1pf} we restrict to the $-1$ mode, as for the one point functions involving only this mode, some of the results have been explicitly matched with its gravity dual \cite{Giusto:2015dfa}.
Thus the three-charge states that we consider are written as
\begin{equation}
\psi_{\{N_{\ka ,m_{\ka}}^{(S)}\}} = \prod_{s = 1}^4 \prod_{\ka} (\ket{s}_{\ka})^{N_{\ka}^{(s)}} \prod_{\ka, m_{\ka}} \left(\frac{1}{m_{\ka}!} (J^+_{-1})^{m_{\ka}} \ket{00}_{\ka}\right)^{N_{\ka,m_{\ka}}^{(00)}},
\label{1pfeq:18state}
\end{equation}
where $m_\ka$ is the number of insertions of $J^+_{-1}$.
As it is important for this paper, let us explain carefully the notation and normalisation of the excited state.
The operator is acting on a strand of length $\ka$ which, with the notation that we introduced above, is equivalent to writing a sum over the copies.
That is, when we write the state above it is shorthand notation for
\begin{equation}
\psi_{\{N_{\ka ,m_{\ka}}^{(S)}\}} = \prod_{s = 1}^4 \prod_{\ka} (\ket{s}_{\ka})^{N_{\ka}^{(s)}} \prod_{\ka, m_{\ka}} \left(\frac{1}{m_{\ka}!} \left(\sum_{r = 1}^{\ka} J^+_{-1 (r)}\right)^{m_{\ka}} \ket{00}_{\ka}\right)^{N_{\ka,m_{\ka}}^{(00)}},
\end{equation}
which, as always, the sum for $J^+$ is over any $\ka$ copies.
Notice that $m_{\ka} \leq \ka$, as otherwise we would have two insertions of the mode on a same copy in every term, and so the resulting term would vanish.
Writing the copies out explicitly we have
\begin{equation}
\left( J^+_{-1(1)} \otimes \mathbb 1_{(2)} \otimes ... \otimes \mathbb 1_{(\ka)} + ... + \mathbb 1_{(1)} \otimes ... \otimes \mathbb 1_{(\ka - 1)} \otimes J^+_{-1 (\ka)} \right)^{m_{\ka}}.
\end{equation}
We will thus have $\binom{\ka}{m_{\ka}}$ terms, up to cyclic permutations of the R-symmetry modes.
We need to divide by $m_{\ka}!$ to get rid of the permutations, as they all correspond to the same state.
Also, let us recall that the sum says is over $\ka$ strands, but it does not give any information on which $\ka$ strands we act on.
This will be taken into account by the normalisation of the state.
Keeping all these comments in mind, the norm of this three charge state is analogous to the previous one, equation \eqref{gmr14norm}, except for an extra factor accounting for the different combinations in which this operator can act on the strands, as we just mentioned.
Therefore, the norm of \eqref{1pfeq:18state} is
\begin{equation}
\mathcal N(\{N_{\ka, m_\ka}^{(S)}\}) = \left(\frac{N!}{\prod_{s, \ka} N_\ka^{(s)}! \ka^{N_\ka^{(s)}}}\right)\left(\frac{1}{\prod_{\ka, m_\ka} N_{\ka, m_\ka}^{(00)}! \ka^{N_{\ka, m_\ka}^{(00)}}}\right) \prod_{\ka, m_\ka} \binom{\ka}{m_\ka}^{N_{\ka, m_\ka}^{(00)}}.
\label{statenormgeneral}
\end{equation}
As a side comment, let us recall that, as we said above equation \eqref{1pfeq:fractionalmode}, in the twisted sector we have fractional modes.
Therefore, when writing the operator in terms of modes of the fermions we have the integer modes, but also all the combinations of the fractional modes that give the desired one.

Let us write now, as in the case of the $\frac 1 4$-BPS state, the state dual to the supergravity geometries.
Analogous to the previous case, we define the supergravity dual as
\begin{align}
\psi(\{A_\ka^{(s)}, B_{\ka, m_\ka}\}) &:= \sum_{\{N_{\ka, m_\ka}^{(S)}\}} \left(\prod_{s, \ka} A_{\ka}^{(s)}\right)^{N_\ka^{(s)}} \left(\prod_{\ka, m_\ka} B_{\ka, m_\ka}\right)^{N_{\ka, m_\ka}^{(00)}} \psi_{\{N_{\ka, m_\ka}^{(S)}\}} = \nonumber \\
& \;= \sum_{\{N_{\ka, m_\ka}^{(S)}\}} \left(\prod_{s, \ka} (A_{\ka}^{(s)} \ket{s}_\ka)^{N_\ka^{(s)}} \prod_{\ka, m_\ka} \left( \frac{B_{\ka, m_\ka}}{m_\ka!} (J^+_{-1})^{m_\ka} \ket{00}_\ka \right)^{N_{\ka, m_\ka}^{(00)}}\right),
\end{align}
where the condition \eqref{1pfeq:coefconstraint} also holds, adding now the new $B_{\ka, m_{\ka}}$ coefficients.
That is, the condition now reads
\begin{equation}
\sum_{gs, \ka} |A^{(gs)}_{\ka}|^2 + \sum_{\ka, m_{\ka}} |B_{\ka, m_{\ka}}|^2 = N.
\label{1pfeq:coefconstraint18}
\end{equation}
The norm of the state is
\begin{equation}
|\psi(\{A_\ka^{(s)}, B_{\ka, m_\ka}\})|^2 = \sum_{\{N_{\ka, m_\ka}^{(S)}\}} \mathcal N(\{N_{\ka, m_\ka}^{(S)}\}) \left(\prod_{s, \ka} |A_\ka^{(s)}|^{2N_\ka^{(s)}}\right) \left(\prod_{\ka, m_\ka} |B_{\ka, m_\ka}|^{2N_{\ka, m_\ka}^{(00)}} \right).
\end{equation}

We have now defined all ground states and three-charge states that we use throughout this paper.
However, in section \ref{1pfsection:1pf} when calculating one point functions of twisted chiral primaries we will end up with other excited states, for which we will need the normalisation.
We dedicate the next section to calculate these normalisations.

\subsection{Normalisations}\label{1pfsubsec:normalisations}
Before we calculate the normalisation for the excited states let us review the normalisations of all the operators and of the one point functions themselves.
As we have seen in the previous sections, the $\mathcal O^{\al \dot \al}$ and the $J^i$ operators are all normalised to one by definition.
Also, in the previous section we have normalised the physical states by counting the number of combinations in which we can create them from the untwisted vacuum.
As usual, we normalise the untwisted vacuum to one, that is,
\begin{equation}
\norm{\ket{++}_{(r)}}^2 = 1.
\end{equation}
Therefore, the norm of the physical states is fully determined by the twist operators $\Si_{\ka}$.
This means that this norm is not trivial, and so we will need it when computing one point functions of twist operators.
Let us calculate its norm.
To calculate the norm of an operator we need to calculate its vacuum expectation value.
To do so we need to calculate a two-point function; otherwise the result will be zero.
First of all let us recall that the vacuum of the Ramond sector is
\begin{equation}
\bigotimes_{r = 1}^N \ket{++}_{(r)}.
\end{equation}
Then, the norm of the twist operator is given by
\begin{equation}
|\Si_{\ka}|^2 = \left( \bigotimes_{r = 1}^N {}_{(r)}\bra{++}\right) \Si^{+ \dot +}_{\ka} \Si^{- \dot -}_{\ka} \left(\bigotimes_{r = 1}^N \ket{++}_{(r)}\right).
\end{equation}
This two-point function is easily calculated with the combinatorics presented in the previous section.
Namely, the number of ways in which the $\Si^{- \dot -}_{\ka}$ operator can act is given by the choices of $\ka$ objects among $N$ up to cyclic permutations.
Formally, the twist operator can act in any of this combinations because when we write the operator $\Si_{\ka}$ this is shorthand notation for
\begin{equation}
\sum_{\{i_1, ..., i_{\ka}\}} \Si_{(i_1 ... i_{\ka})},
\end{equation}
where the sum runs over all possible choices of $\ka$ copies among the $N$ total ones up to cyclic permutations.
The only non-trivial action its complex conjugate can perform is to undo that joining, and so the norm of the twist operator is given by
\begin{equation}
|\Si^{\al \dot \al}_{\ka}|^2 = \frac{N!}{(N - \ka)! \ka}.
\label{1pfeq:sigma2norm}
\end{equation}
This of course coincides with the norm given in equation \eqref{gmr14norm} when we consider the state $\psi = \ket{++}_{\ka} \otimes (\ket{++}_1)^{N - \ka}$.
We can also calculate the norm of the operators in the twisted sector.
Consider for instance the chiral primary
\begin{equation}
\Si_{\ka} \sum_{r = 1}^{\ka} \mathcal O^{\al \dot \al}_{(r)}.
\end{equation}
As we said above, the $\mathcal O^{\al \dot \al}$ operators are normalised to 1, and so only the twist contributes.
Therefore
\begin{equation}
|\Si_{\ka} \sum_{r = 1}^{\ka} \mathcal O^{\al \dot \al}_{(r)}|^2 = \frac{N!}{(N - \ka)! \ka}.
\end{equation}
Let us finish this section with a general comment regarding the normalisation that we use for the one point functions.
We will always normalise them by the norm of the in state\footnote{By in state we mean the state for which we calculate the one point function. We will call out state to the state after we act on it with the operator for which we are calculating the one point function}.
More concretely, if $\mathcal O$ is an operator for which we want to calculate the one point function and $\ket{O}$ is the state we are interested in, the results that we will give will be
\begin{equation}
\frac{\bra{O} \mathcal O(y)\ket{O}}{\braket{O|O}}.
\end{equation}
Now that we have the normalisations of all the operators we can find the norm of other excited states in which we will be interested on.

\subsection{Other excited states}\label{1pfsubsec:excitedstates}
As we have said in section \ref{1pfsubsec:twistedoperators}, in this paper paper we are interested in calculating one point functions for chiral primaries.
The untwisted chiral primaries and twist operators will generate other ground states or three-charge states, and so the same states that we have described before will give a non-zero answer for the one point functions.
However, the state resulting from acting with a twisted operator on a ground state is not any strand we have discussed before, and so we need to introduce these other excited strands.
Consider for instance the operator $\mathcal O^{+ \dot +}_{\ka}$ defined by \eqref{1pfeq:twistedOdefi} acting on a ground state $\ket{++}_{\ka}$.
It will generate an excited state, which we define as
\begin{equation}
\ket{++}^*_{\ka} := \lim_{z \to 0} |z|^{\ka - 1} \Si_{\ka}^{- \dot -} \sum_{r = 1}^{\ka} \mathcal O^{+ \dot +}_{00(r)} \ket{++}_{\ka}.
\end{equation}
In the R sector this is a state with $h = \tilde h = j = \tilde j = 1$.
We do not use its norm in this paper though.
As we will see in section \ref{1pfsubsec:twistedO}, we can show how the calculation would be done for these cases, but we will not give the final result as there is more work to be done in the supergravity side necessary to finish the calculation.
We give more details in that section.

Now that we have given a description of the free field theory we are ready to start calculating the one point functions.
In section \ref{1pfsection:1pf} we study one point functions in the short strand case, and in section \ref{1pfsection:long1pf} we study them in the opposite limit; in the long strand one.

\section{One point functions: short strand case}\label{1pfsection:1pf}
In this section we compute one point functions for all the chiral primaries described in sections \ref{subsec:ffd} and \ref{1pfsubsec:twistedoperators}, for the two and three-charge cases.
Some one point functions have been calculated in \cite{Giusto:2015dfa}, where strands of length one and two were considered, and also strands of arbitrary length for the operator $\Si_2^{- \dot -}$.
This section extends the CFT calculation performed in that paper, by considering arbitrary strand length for all one point functions.
In what follows, first we describe the approximation used in the short strand case, and then we go case by case calculating the one point functions for all different chiral primaries.

\subsection{Approximation used}
In this section we are concerned about the case where the strands are short and we have a large number of them.
That is, we consider the case where $\ka$ is of order one and $N^{(gs)}_{\ka} \lesssim N$.
This is the case considered in \cite{Giusto:2015dfa}.
We take the same approximation, which consists in finding the saddle point on which the sum over the partitions is peaked.
This saddle point is determined by the numerical coefficients $A_{\ka}^{(gs)}$ and $B_{\ka, m_{\ka}}$ accompanying each strand.
In the two-charge case the norm of the state is
\begin{equation}
|\psi(\{A_\ka^{(S)}\})|^2 = \sum_{\{N_\ka^{(S)}\}} \mathcal N(\{N_\ka^{(S)}\}) \prod_{S, \ka} |A_\ka^{(S)}|^{2N_\ka^{(S)}}.
\label{gmr14dualnorm}
\end{equation}
Let $\overline N_\ka^{(S)}$ be the saddle point on which the sum is peaked.
To obtain it we use Stirling's approximation in its weakest form,
\begin{equation}
\log n \approx n \log n - n, \qquad n\in \mathbb N, \; n\gg 1.
\label{stirlingweak}
\end{equation}
Taking the logarithm of each term of \eqref{gmr14dualnorm} and using \eqref{stirlingweak} we obtain
\begin{equation}
N\log N + \sum_{S, \ka} \left(N_\ka^{(S)} \log |A_\ka^{(S)}|^2 - N_\ka^{(S)} \log N_\ka^{(S)} + N_\ka^{(S)} - N_\ka^{(S)}\log \ka \right).
\end{equation}
The stationary point $\overline N_\ka^{(S)}$ is then
\begin{equation}
\overline N_\ka^{(S)} = \frac{|A_\ka^{(S)}|^2}{\ka}.
\end{equation}
For the three-charge case we can use an analogous approximation, which results into
\begin{equation}
\overline N_{\ka}^{(s)} = \frac{|A_\ka^{(s)}|^2}{\ka}, \qquad \overline N_{\ka, m_\ka}^{(00)} = \binom{\ka}{m_\ka} \frac{|B_{\ka, m_\ka}|^2}{\ka}.
\label{gmr18dualrelations}
\end{equation}
We use these relations to approximate the resulting sums after we act on the states with the chiral primaries.

Now that we have all the ingredients needed we can start calculating the one point functions in this limit.
Notice that for each chiral primary we can use a state which only has the strands that will come into play.
With the approximation taken in this section the norm of the state will always cancel, and so having extra strands which do not play a role in the process does not affect the result.
To see this clearly and to introduce the calculation and some simplified notation we start with a review example.
Afterwards we will calculate one point functions in more general cases.

\subsection{Review example: \texorpdfstring{$\Si_2^{+ \dot -}$}{Si2+-} operator}\label{1pfsubsec:shortreview}
In this section we calculate the one point function of $\Si_2^{+ \dot -}$, a chiral primary which joins two strands into a single one and increases the left R symmetry charge by 1/2 and lowers the right one by the same amount.
This example will be used as an explanation of how to calculate the one point functions, following closely \cite{Giusto:2015dfa}.
In subsequent sections we will use the notation and methods introduced here directly.
Let us start calculating the one point function.
In order to have a non-trivial answer for this operator we consider the following state,
\begin{align}
\psi(\{A^{(++)}_{n_l}, B^{k(00)}_i\}) = \sum_{N^{1(00)}_{m_l} = 0}^{N/m_l} \sum_{N^{0(00)}_{p_l} = 0}^{\frac{N - N^{1(00)}_{m_l} m_l}{p_l}} & \left(A^{(++)}_{n_l} \ket{++}_{n_l} \right)^{N^{(++)}_{n_l}} \left( B^{0(00)}_{p_l} \ket{00}_{p_l} \right)^{N^{0(00)}_{p_l}} \otimes \nonumber \\
& \otimes \left( B^{1(00)}_{m_l} J^+_{-1} \ket{00}_{m_l} \right)^{N^{1(00)}_{m_l}},
\end{align}
where we take $m_l = n_l + p_l$.
Using equation \eqref{gmr1/4def} we obtain the constraint
\begin{equation}
N^{(++)}_{n_l} n_l + N^{0(00)}_{p_l} p_l + N^{1(00)}_{m_l} m_l = N.
\end{equation}
One of the $N^{(gs)}$ can be related to the others due to equation \eqref{d1d5eq:stateconstr}, but we will not substitute it during calculations to simplify the notation.
From equation \eqref{gmr18dualrelations} we learn that the sum is peaked at
\begin{equation}
n_l \overline{N^{(++)}_{n_l}} = \left|A^{(++)}_{n_l}\right|^2, \qquad p_l \overline{N^{0(00)}_{p_l}} = \left|B^{0(00)}_{p_l}\right|^2, \qquad \overline{N^{1(00)}_{m_l}} = \left|B^{1(00)}_{m_l}\right|^2.
\end{equation}
Also, using \eqref{statenormgeneral} we find the normalisation factor for the state $\psi$ to be
\begin{equation}
\mathcal N \left( N^{(++)_{n_l}, N^{0(00)}_{p_l}}\right)= \frac{N!}{N^{(++)}_{n_l}! N^{0(00)}_{p_l}! N^{1(00)}_{m_l}! n_l^{N^{(++)}_{n_l}} p_l^{N^{0(00)}_{p_l}}}.
\end{equation}
The action of the $\Si_2^{+ \dot -}$ operator on these strands is
\begin{equation}
\Si_2^{+ \dot -} \left( \ket{++}_{n_l} \otimes \ket{00}_{p_l} \right) \to J^+_{-1} \ket{00}_{m_l = n_l + p_l}.
\label{1pfeq:actionexample}
\end{equation}
To write the state resulting from the action of the operator exactly we need to calculate two coefficients in general.
The first one, which we will always denote by $\al$, is to ensure that we have the same number of terms before and after applying the operator, that is, in the l.h.s. and in the r.h.s. of \eqref{1pfeq:actionexample}.
The second coefficient will only be needed when we calculate one point functions of gluing operators.
We will call this second coefficient $c_n$, where the subindex $n$ stands for the number of strands joined together.
This coefficient was first calculated for $n = 2$ in \cite{Carson:2014ena}, and was recently generalised to arbitrary $n$ in \cite{Tormo:2018fnt}.

Let us start by calculating the $\al$ coefficient for this case.
The twist operator $\Si_2^{+ \dot -}$ can act on any of the $N^{(++)}_{n_l}$ strands $\ket{++}_{n_l}$ and on any of the $N^{0(00)}_{p_l}$ strands $\ket{00}_{p_l}$.
So, we need to multiply the l.h.s. by this two numbers.
Also, once the strands are picked, the operator can act on any of the copies within each strand.
Therefore, we also need to multiply by their lengths, $n_l p_l$.
Last, we also need to divide by two, for the following reason: the twist operator acts up to cyclic permutations, and so we need to take the symmetrisation of the strands over which it acts.
More precisely, the one point function that we consider is
\begin{equation}
_{m_l} \bra{00} J^-_{+1} \Si_2^{+ \dot -} \left( \ket{++}_{n_l} \otimes \ket{00}_{p_l} \right)_{\mathrm{Symm.}},
\end{equation}
where
\begin{align}
\left( \ket{++}_{n_l} \otimes \ket{00}_{p_l} \right)_{\mathrm{Symm.}} := & \ket{++}_{(1)} \otimes ... \otimes \ket{++}_{(n_l)} \otimes \ket{00}_{(n_l + 1)} \otimes ... \otimes \ket{00}_{(m_l)} + ... + \nonumber \\
& + \ket{00}_{(1)} \otimes ... \otimes \ket{00}_{(p_l)} \otimes \ket{++}_{(p_l + 1)} \otimes ... \otimes \ket{++}_{(m_l)}
\end{align}
is a symmetrisation over all the copies.
Notice that if we pick strands of the same kind we do not include this factor, as the symmetrisation in that case is already taken into account by the norm of the whole state.
Hence, the $\al$ combinatorial factor is obtained by solving
\begin{equation}
\frac{N^{(++)}_{n_l}n_l N^{0(00)}_{p_l}p_l}{2} \mathcal N(N^{(++)}_{n_l}, N^{0(00)}_{p_l}, N^{1(00)}_{m_l}) = \al \mathcal N(N^{(++)}_{n_l} - 1, N^{0(00)}_{p_l} - 1, N^{1(00)}_{m_l} + 1),
\end{equation}
which gives
\begin{equation}
\al = \frac{N^{1(00)}_{m_l} + 1}{2}.
\end{equation}
If we had included other ground state strands in our state $\psi$ this coefficient would remain the same, as they would be unaltered after the application of the operator and thus they would cancel.
Now, the $c_2$ coefficient was first computed in \cite{Carson:2014ena} and reads
\begin{equation}
c_{n_l, p_l} = \frac{n_l + p_l}{2n_l\; p_l}.
\end{equation}
Last, the commutator of the R-symmetry current with the twist operator is
\begin{equation}
\left[ \left( J^i_n \right)^{\al \be}, \Si_2^{\be \dot \al}(v, u) \right] = \frac 1 2 e^{i n \frac{\sqrt 2 v}{R}} \left( \si^i \right)^{\al \be} \Si_2^{\be \dot \al} (v, u),
\label{1pfeq:jsigmacommutator}
\end{equation}
and so
\begin{align}
& _{m_l} \bra{00} J^-_{+1} \Si_2^{+ \dot -} \left( \ket{++}_{n_l} \otimes \ket{00}_{p_l} \right)_{\mathrm{Symm.}} = \nonumber \\
& \;\;\;\;\;\;\;\; =  e^{i \frac{\sqrt 2 v}{R}} {}_{m_l} \bra{00} \Si^{- \dot -}_2  \left( \ket{++}_{n_l} \otimes \ket{00}_{p_l} \right)_{\mathrm{Symm.}} = {}_{m_l} \braket{00|00}_{m_l} e^{i \frac{\sqrt 2 v}{R}}.
\end{align}
Notice that here we are also implicitly using that, when acting on a ground state $\ket{++}_n$, the operators $\Si^{- \dot -}_2$ and $\mathcal O^{- \dot -}_{(r)}$ commute.
Plugging all the expressions together we see that the action of the twist operator yields
\begin{align}
&\;\; \Si^{+ \dot -}_2 \left[ \left(\ket{++}_{n_l}\right)^{N^{(++)}_{n_l}} \left(\ket{00}_{n_l}\right)^{N^{0(00)}_{n_l}} \left(J^+_{-1} \ket{00}_{m_l}\right)^{N^{1(00)}_{m_l}} \right] = \nonumber \\
= &\;\; e^{i \frac{\sqrt 2 v}{R}} c_{n_l,p_l}\frac{N^{1(00)}_{m_l} + 1}{2} \left[ \left(\ket{++}_{n_l}\right)^{N^{(++)}_{n_l} - 1} \left(\ket{00}_{n_l}\right)^{N^{0(00)}_{n_l} - 1} \left(J^+_{-1} \ket{00}_{m_l}\right)^{N^{1(00)}_{m_l} + 1} \right].
\end{align}
We are now ready to calculate the one point function.
Writing it out explicitly, we get
\begin{align}
\braket{\Si^{+ \dot -}_2} & = |\psi|^{-2} |\Si_2^{+ \dot -}|^{-1} \bra{\psi} \Si^{+ \dot -}_2 \ket{\psi} = \nonumber \\
& = |\psi|^{-2} |\Si_2^{+ \dot -}|^{-1} \psi^{\dagger} \sum_{N^{1(00)}_{m_l} = 0}^{N/m_l} \sum_{N^{0(00)}_{p_l} = 0}^{\frac{N - N^{1(00)}_{m_l} m_l}{p_l}} c_{n_l,p_l}\frac{N^{1(00)}_{m_l} + 1}{2} \left(A^{(++)}_{n_l} \ket{++}_{n_l} \right)^{N^{(++) - 1}_{n_l}} \otimes \nonumber \\
& \;\;\;\;\; \otimes \left( B^{0(00)}_{p_l} \ket{00}_{p_l} \right)^{N^{0(00)}_{p_l} - 1} \left( B^{1(00)}_{m_l} J^+_{-1} \ket{00}_{m_l} \right)^{N^{1(00)}_{m_l} + 1} = \nonumber \\
& = \frac{e^{i\frac{\sqrt 2 v}{R}}}{4} \frac{(n_l + p_l)}{n_l\; p_l} A^{(++)}_{n_l} B^{0(00)}_{p_l} \overline{B^{1(00)}_{n_l + p_l}} \left( \frac{2}{N (N - 1)} \right)^{\frac 1 2},
\label{simplestSigma}
\end{align}
where in the last equality we have used the approximation \eqref{gmr18dualrelations}, cancelled the norm of the out state with the norm of the in state and used equation \eqref{1pfeq:sigma2norm} for the normalisation of the twist operator.
Notice that our result differs from the one in \cite{Giusto:2015dfa} by a factor proportional to $N^{-1}$, which comes from the normalisation of the twist operator.
That normalisation was not considered there, but for the results in this paper will play a crucial role.
Now that we have reviewed the calculation and some of its key elements we will calculate more general one point functions.
We will simplify the notation used where possible, to make the equations less cumbersome.

\subsection{\texorpdfstring{$\mathcal O^{- \dot -}_{(r)}$}{O--} operator}
In this section we calculate the one point function for $\sum_r \mathcal O^{- \dot -}_{(r)}$, which transforms a single strand by lowering the R-symmetry charge by 1/2 on the left and on the right.
Consider the two-charge state
\begin{equation}
\psi (A, B) = \sum_{p = 0}^{\frac N n} \left( A \ket{++}_n \right)^p \left( B \ket{00}_n \right)^{\frac N n - p},
\end{equation}
which has norm
\begin{equation}
\mathcal N(p) = \frac{N!}{p! \left( \frac N n - p \right)! n^{\frac N n}}.
\end{equation}
We are assuming that $n$ is an integer, multiple of $N$.
With this simplified notation we already took into account the constraint \eqref{constraint} for the lengths and number of strands.
The sum over all the possible strand combinations is peaked in this case at, from equation \eqref{gmr18dualrelations},
\begin{equation}
n \bar p = \left| A \right|^2, \qquad N - n \bar p = \left| B \right|^2.
\end{equation}
As this is the first one point function that we calculate where our operator is defined as a sum over copies let us discuss the combinatorics carefully.
The $\al$ coefficient goes as follows.
We choose one of the $p$ strands to act with the operator.
For each strand, we are acting with $n$ terms, corresponding to the sum over copies.
However, only the action on one copy is non-trivial and so there is no extra combinatorial factor.
Another way to think about this is that once we have picked up the copy, the operator can act on any of the copies within the strand.
Again, though, in this case this action is only non-trivial in one copy and so there is no extra factor.
This will not be the case for more complicated operators.
Therefore, the action of the operator on a ground state is straightforward, as the resulting combination is just the definition of a $\ket{00}_n$ ground state,
\begin{align}
&\sum_{r = 1}^n \mathcal O^{- \dot -}_{(r)} \ket{++}_n = \nonumber \\
=&\sum_{r = 1}^n \left( \ket{++}_{(1)} \otimes ... \otimes \ket{++}_{(r - 1)} \otimes \ket{00}_{(r)} \otimes \ket{++}_{(r + 1)} \otimes ... \ket{++}_{(n)}\right) = \ket{00}_n.
\end{align}
Therefore, the $\al$ factor is obtained by solving
\begin{equation}
p \mathcal N (p) = \al \mathcal N (p - 1),
\end{equation}
which gives
\begin{equation}
\al = \frac N n - p + 1.
\end{equation}
In this case there is no $c$ coefficient, as we are not gluing strands.
The action of the operator on the strands is thus
\begin{align}
&\;\; \mathcal O^{- \dot -} \left[ \left(\ket{++}_{n_l}\right)^p \left(\ket{00}_{n_l}\right)^{\frac N n - p} \right] = \nonumber \\
= &\;\; \left(\frac N n - p + 1\right) \left[ \left(\ket{++}_{n_l}\right)^{p - 1} \left(\ket{00}_{n_l}\right)^{\frac N n - p + 1} \right].
\end{align}
The calculation of the one point function is now completely analogous to the one in section \ref{1pfsubsec:shortreview}.
The result is
\begin{equation}
\left\langle \sum_{r = 1}^n \mathcal O^{- \dot -}_{(r)} \right\rangle = \frac{A \bar B}{n}.
\label{1pointfunO--}
\end{equation}

\subsection{\texorpdfstring{$\mathcal O^{+ \dot -}_{(r)}$}{O+-} operator}
Consider now the operator $\sum_r \mathcal O^{+ \dot -}_{(r)}$, which takes a strand and raises its left R charge by 1/2 and it lowers it by the same amount on the right, creating a three-charge strand from a two-charge one.
This one point function is analogous to the previous one but for a 1/8-BPS state.
To calculate its vacuum expectation value, consider the state
\begin{equation}
\psi (A, B^1) = \sum_{p = 0}^{\frac N n} \left( A \ket{++}_n \right)^p \left( J^+_{-1} B^1 \ket{00}_n \right)^{\frac N n - p},
\end{equation}
which has norm
\begin{equation}
\mathcal N(p) = \frac{N!}{p! \left( \frac N n - p \right)! n^p}.
\end{equation}
The process is now
\begin{equation}
\left( \sum_{r = 1}^{n_l} \mathcal O^{+ \dot -}_{(r)} \right) \ket{++}_{n_l} \to \left( J^+_{-1 (r)} \right) \ket{00}_{n_l}.
\end{equation}
To find this one point function we need the commutation relation between the R-symmetry current and the $\mathcal O^{\al \dot \al}$ operator, which is given by
\begin{equation}
\left[ \left( J^i_n \right)^{\al \be}, \mathcal O^{\be \dot \al} (v, u) \right] = \frac 1 2 e^{i n \frac{\sqrt 2 v}{R}} \left( \si^i \right)^{\al \be} \mathcal O^{\be \dot \al} (v, u).
\label{1pfeq:JOcommutator}
\end{equation}
Then,
\begin{align}
_n \bra{00} \left(\sum_{r = 1}^n J^-_{+1 (r)}\right) \left( \sum_{r = 1}^n \mathcal O^{+ \dot -}_{(r)} \right) \ket{++}_n &= e^{i \frac{\sqrt 2 v}{R}} {}_n \bra{00} \left( \sum_{r = 1}^n \mathcal O^{- \dot -}_{(r)} \right) \ket{++}_n = \nonumber \\
& = e^{i \frac{\sqrt 2 v}{R}} {}_n \braket{00|00}_n.
\end{align}
The combinatorial factor $\al$ is analogous to the previous case, as we now have
\begin{equation}
p \mathcal N (p) = \al \mathcal N (p - 1)
\end{equation}
which gives
\begin{equation}
\al = \frac 1 n \left( \frac N n - p + 1 \right).
\end{equation}
Recalling that the sum in the definition of the state is peaked at
\begin{equation}
n \bar p = |A^2|, \qquad \frac N n - \bar p = |B^1|^2
\end{equation}
we find the answer of the one point function to be
\begin{equation}
\left\langle \sum_{r = 1}^n \mathcal O^{+ \dot -}_{(r)} \right\rangle = \frac 1 n A \overline{B^1} e^{i \frac{\sqrt 2 v}{R}}.
\label{eq:oshortgeneral}
\end{equation}
Now that we have computed the one point functions for these untwisted chiral primaries let us go to the one point functions for twist operators.

\subsection{\texorpdfstring{$\Si_n^{- \dot -}$}{Sin--} operator}\label{1pfsubsec:sigman}
\begin{figure}
\centering
\includegraphics[height=3.5cm]{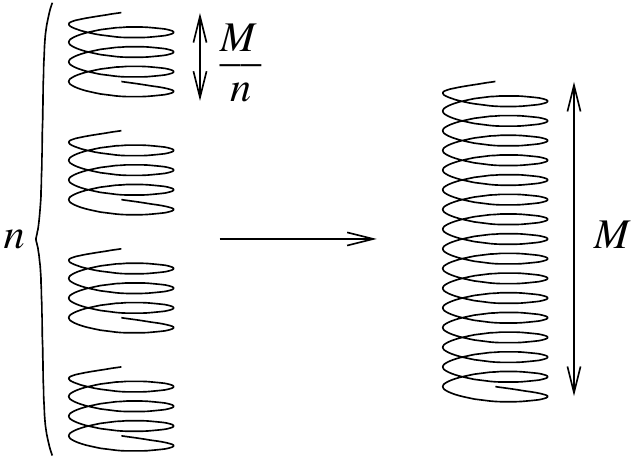}
\caption[Gluing process of the twist one point function.]{Initial and final states of the twist operator action. We start with $n$ strands of length $M/n$ and we join them all together in one step, {\it i.e.} we act on them with a twist $n$ operator to join them in a single strand of length $M$.}
\label{fig:initialstrandsstructure}
\end{figure}

For this first twisted sector one point function consider the operator $\Si_n^{- \dot -}$, which joins $n$ strands into a single one.
We consider the case where we join strands of the same length, as then the formula for the result is much simpler.
The generalisation to different strand lengths does not involve any extra subtleties.
So, consider the state
\begin{equation}
\psi = \sum_{p=0}^{\frac{nN}{M}} \left(A\ket{++}_{\frac M n}\right)^p \left( A_2 \ket{++}_M\right)^{\frac N M -\frac p n},
\end{equation}
where $M$ (and thus also $n$) are small integers.
The norm of this state is
\begin{equation}
\left|\psi\right|^2 = \sum_{p = 0}^{\frac{nN}{M}} |A|^{2p} |A_2|^{\frac N M - \frac p n} \mathcal N(p),
\end{equation}
where
\begin{equation}
\mathcal N(p) = \frac{N!}{p! \left(\frac N M - \frac p n\right)! \left(\frac M n\right)^p M^{\frac N M - \frac p n}}.
\end{equation}
The sum over strands of the norm peaks at
\begin{equation}
\bar p = \frac{n |A|^2}{M} = \frac n M (N - |A_2|^2),
\end{equation}
like in the previous cases.
Now, the action of the operator $\Si_n^{- \dot -}$ on the strands of the state $\psi$ is
\begin{equation}
\Si^{- \dot -}_n \left[\left(\ket{++}_{\frac M n}\right)^p \left(\ket{++}_M\right)^{p_2}\right] = c_n \al \left[\left(\ket{++}_{\frac M n}\right)^{p-n} \left(\ket{++}_M\right)^{p_2 + 1}\right],
\end{equation}
where the $c_n$ coefficient is the generalisation of the $c_2$ used in section \ref{1pfsubsec:shortreview}.
So, we start with $n$ strands of length $M/n$, and at the end we have a single strand of length $M$.
The coefficient $c_n$ was obtained in \cite{Tormo:2018fnt}, and in the case of joining strands of the same length its expression is
\begin{equation}
c_n = M^{1-n} n^{\frac{M}{2} + \frac{n}{2} - \frac{M}{n}} \left (  (n-1)^2 (n \bar{a})^{(n-2)} \Lambda^{-1} \right )^{\frac{1}{2} (n-1) \left (1 - \frac{M}{n} \right )}
\label{1pfeq:cn}
\end{equation}
where
\begin{equation}
\Lambda = \left ( 1 + n (n-1)^{(n-1)} - n^{n-1} \right )
\end{equation}
and
\begin{equation}
\bar{a}^{n-1} = n \left ( 1 - \frac{1}{n} \right )^{n-1} - 1.
\end{equation}
Let us now compute the $\al$ coefficient.
As in the previous cases, it is a combinatorial factor obtained by matching the number of terms (the normalisation of the state) before and after the application of the gluing operator.
Since we join all the strands in a single step, the only freedom we have is where within each strand we insert the operator.
Thus, clearly the gluing operator can act in $\left(\frac M n\right)^n$ ways.
In figure \ref{fig:initialstrandsstructure} we present a picture depicting this process.
Therefore,
\begin{equation}
\binom{p}{n} \left(\frac{M}{n}\right)^n \mathcal N(p) = \al \mathcal N(p - n).
\end{equation}
Solving for $\al$ we obtain
\begin{equation}
\al = \frac{M \left(\frac N M - \frac p n + 1\right)}{n!}.
\end{equation}
Assuming, as we said before, that the sum is peaked at $\bar p$ we obtain
\begin{align}
\langle\Si^{--}_n\rangle & = |\psi|^{-2} |\Si_n^{- \dot -}|^{-1} \bra{\psi}\Si^{--}_n\ket{\psi} = \nonumber \\
& = |\psi|^{-2} |\Si_n^{- \dot -}|^{-1} \psi^{\dagger} \sum_{p = 0}^{\frac{nN}{M}} A^p A_2^{p_2}c_n \frac{M \left(\frac N M - \frac p n + 1\right)}{n!} \left(\ket{++}_{\frac M n}\right)^{p - n}\left(\ket{++}_M\right)^{p_2 + 1} = \nonumber \\
& = |\Si_n^{- \dot -}|^{-1} c_n \frac{N - \frac{\bar p M}{n}}{n!} A^n A_2^{-1} = A^n \bar A_2 \frac{c_n}{n!} \left( \frac{n}{N (N - 1) \cdot ... \cdot (N - n + 1)} \right)^{\frac 1 2},
\label{sigmanresult}
\end{align}
where $c_n$ is given by equation \eqref{1pfeq:cn}.
Let us now go to the analogous 1/8-BPS one point function.

\subsection{\texorpdfstring{$\Si_n^{+ \dot -}$}{Sin+-} operator}\label{1pfsubsec:sinresult}
In this section we present some combinatorics needed for the calculation of the $\Si^{+ \dot -}$ one point function.
We do not give the final result of this one point function as it involves the calculation of a new commutator between operators, which is left as future work.
Consider the state
\begin{equation}
\psi = \sum_{\{\mathcal N\}} \left( A \ket{++}_{\frac M n} \right)^{N^+} \left( B \ket{00}_{\frac M n} \right)^{N^0} \left( C \left( J^+_{-1} \right)^{n - 1} \ket{00}_M \right)^{N^1}
\end{equation}
which has norm
\begin{equation}
\mathcal N (N^+, N^0) = \frac{N!}{N^+! N^0! N^1! \left( \frac M n \right)^{N^+ + N^0} M^{N^1}} \binom{M}{n - 1}^{N^1}.
\end{equation}
Let us recall that if we add other strands to consider a more general state the result is exactly the same using the short strand approximation, and so we only include the ones involved in the correlation at hand to ease the notation.
Thus, in this section we are interested in the one point function of the twist operator $\Si_n^{+ \dot -}$.
Namely, we consider the following process
\begin{equation}
\Si_n^{+ \dot -} \left( \left( \ket{++}_{\frac M n} \right)^{n - 1} \otimes \ket{00}_{\frac M n} \right) \to \left( J^+_{-1} \right)^{n - 1} \ket{00}_M.
\end{equation}
As in the previous section we need the $c_n$ coefficient, which is given in equation \eqref{1pfeq:cn}.
Let us now compute the combinatorial $\al$ factor.
The twist operator acts on $n - 1$ $\ket{++}_{\frac M n}$ states, and on one $\ket{00}_{\frac M n}$ state.
Within each strand, it can act on any of the $M/n$ copies.
However, there is a further subtlety in this case, as we explained in section \ref{1pfsubsec:shortreview}.
Recalling that the twist operator acts up to cyclic permutations, we need to take the symmetrisation of the states over which it acts, as we did in that case.
Notice that in the previous section we did not need to take this into account as all the strands were the same, but now we have two different kinds of strands and need to take that into account.
So, the gluing process is
\begin{equation}
\Si^{+ \dot -}_n \left( \left(\ket{++}_{\frac M n}\right)^{n - 1} \otimes \ket{00}_{\frac M n} \right)_{\mathrm{Symm}} = \left( J^+_{-1} \right)^{n - 1} \ket{00}_M.
\end{equation}
Now, when we take the strands on the left hand side we have already picked the copies on which the gluing operator acts, but they have all possible orderings within all the strands.
Therefore, we need to divide by the right amount to have only the distinct cycles.
The number of ways in which we can order the $M$ copies in the strands is
\begin{equation}
\frac{M!}{\left( \frac M n \right)! \left( (n - 1)\frac M n \right)!},
\end{equation}
which corresponds just to how many ways we can order the two distinct ground states.
Now, we will need to divide by this number, and multiply by the distinct ways in which they can be arranged up to cyclic permutations, since we have the twist $\Si_n$ acting.
This second counting is less direct than the one above, and to obtain the result we need Burnside's lemma or, more generally, the Pólya enumeration theorem.
Using the theorem, we find that there are
\begin{equation}
\frac 1 M \sum_{d|\mathrm{gcd}(\frac M n, (n - 1) \frac M n)} \phi(d) \binom{\frac{M}{n d} + (n - 1)\frac{M}{n d}}{\frac{M}{nd}}
\label{eq:burnside2}
\end{equation}
different combinations up to cyclic permutations.
There are special cases (when the number of strands we join are prime numbers for example) where this formula reduces to a compact expression, however these special cases are not relevant for the calculation at hand.
Therefore, we leave the coefficient as is, defining
\begin{equation}
S :=\frac{
\frac 1 M \sum_{d|\mathrm{gcd}(\frac M n, (n - 1) \frac M n)} \phi(d) \binom{\frac{M}{n d} + (n - 1)\frac{M}{n d}}{\frac{M}{nd}}
}
{
\frac{M!}{\left( \frac M n \right)! \left( (n - 1)\frac M n \right)!},
}
\end{equation}
as a parameter.
Thus, the counting of terms before and after the action of the twist is
\begin{equation}
S\binom{N^+}{n - 1} N^0 \left( \frac M n \right)^n \mathcal N (\psi) = \al \mathcal N (\psi (N^1 + 1)),
\end{equation}
where $\mathcal N (\psi)$ is the combinatorial factor we wrote above, and the one on the right hand side is the one after the action of the twist.
Plugging in the expressions for $\mathcal N$ and simplifying yields
\begin{equation}
\al = S\left( N^1 + 1 \right) \frac{(M - n + 1)!}{M!} M.
\end{equation}
With all this, the only remaining thing is to put everything together to obtain the one point function.
However, to do so we need the commutator between $\Si_n^{+ \dot -}$ and $\otimes_r J_{(r)}$, which in general will have a more complex expression than \eqref{1pfeq:jsigmacommutator}.
This commutator can be obtained by bosonising the fermions for example, but this is beyond the scope of this paper and is left as future work.

\subsection{\texorpdfstring{$(J^+_{-1})^m$}{J+-1m} operator}\label{1pfsubsec:Jshort}
In this section we consider the $m$-point function
\begin{equation}
\left< \left( J^+_{-1} \right)^m \right>. 
\end{equation}
To do so we should a priori consider a state with all possible combinations of powers of the $J$ operator, which is
\begin{equation}
\psi = \sum_{\left\{ \mathcal N \right\} } \left( B \ket{00}_n \right)^p \bigotimes_m \left( B_m \left( J^+_{-1} \right)^m \ket{00}_n \right)^{p_m},
\end{equation}
This is a complicated state, and the equations involved cannot be written easily.
We will consider such a state in section \ref{1pfsubsec:1/8statefull} for the long strand case.
However, as we have seen in the previous sections with the short strand approximation the norm of the state cancels, and thus the result is the same as in an easier case where we only take the strands involved in the process.
Therefore, for this section we consider instead the state
\begin{equation}
\psi = \sum_{p = 0}^{\frac N n} \left( B \ket{00}_n \right)^p \left( B_m \left( J^+_{-1} \right)^m \ket{00}_n \right)^{\frac N n - p},
\end{equation}
which has norm
\begin{equation}
|\psi|^2 = \sum_{p = 0}^{\frac N n} |B|^{2p} |B_m|^{2 \left( \frac N n - p \right)} \mathcal N(p)
\end{equation}
with
\begin{equation}
\mathcal N(p) = \frac{N!}{n^{\frac N n}} \frac{1}{p!} \frac{1}{\left( \frac N n - p \right)!} \binom{n}{m}^{\frac N n - p}.
\end{equation}
This is the basic piece needed to calculate this one point function for more complex states, where we have products of modes.
The $\al$ coefficient for this process is
\begin{equation}
p \binom{n}{m} \mathcal N(p) = \al \mathcal N(p - 1) \qquad \Longrightarrow \qquad \al  = \frac N n - p + 1,
\end{equation}
and the condition for the average number of strands reads
\begin{equation}
\binom{n}{m} |B_m|^2 = n \left(\frac N n - \bar p \right).
\end{equation}
Thus, the result for this $m$-point function in the short strand case is
\begin{equation}
\left< \left( J^+_{-1} \right)^m \right> = B \bar B_m \frac 1 n \binom{n}{m}.
\label{1pfeq:Jshort}
\end{equation}
Let us now comment on some other twisted operators.

\subsection{Twist sector \texorpdfstring{$\mathcal O^{\al \dot \al}_n$}{Onaa} operators}\label{1pfsubsec:twistedO}
As in section \ref{1pfsubsec:sinresult} here we do not give the final result as, in this case, more knowledge of the supergravity side is needed to finish the calculation.
However our objective here is to give some extra tools needed in the CFT calculation of more chiral primary one point functions.
We now consider the state
\begin{equation}
\psi (\{A, C\}) = \sum_{p = 0}^{\frac N n} \left(A \ket{00}_n \right)^{p} \left(C \ket{00}^*_n \right)^{\frac N n - p},
\label{1pfeq:excitedstate}
\end{equation}
where by $\ket{00}^*_n$ we denote a ground state which we have excited by raising its R-symmetry spin $(j,\bar j)$.
In this example we, of course, choose the out state resulting from the application of the operator for which we are calculating the one point function, in order to have a non-trivial result.
We consider the following operator in the twisted sector,
\begin{equation}
\Si_{p_l}^{- \dot -} \left( \sum_{r = 1}^{n_l} \mathcal O^{+\dot +}_{(r)} \right).
\end{equation}
The conformal dimensions of this operator are, as we saw in equation \eqref{1pfeq:twistedOdef}, ($n$/2, $n$/2).
Therefore, we are raising the left and right R-symmetry charges by the same amount in both sides using this operator.
The calculation of this one point function is completely analogous to the previous sections, taking into account that we now have a product of two operators and so in general we need to combine different results from previous sections.
There is a $c_n$ coefficient coming from the action of the twist operator on the $\mathcal O^{\al \dot \al}$ operator, which in this case turns out to be $c_n = 1$ as there is only one choice and ordering of operators.
Thus, taking into account that the norm of the state \eqref{1pfeq:excitedstate} is
\begin{equation}
|\psi|^2 = \sum_{p = 0}^{\frac N n} |A|^{2p} |C|^{2(N - p)} \mathcal N (p),
\end{equation}
where, from equation \eqref{1pfeq:sigma2norm},
\begin{equation}
\mathcal N (p) = \frac{N!}{p! (N - p)! n^{\frac N n}} \left( \frac 1 n N (N - 1) \cdot ... \cdot (N - n + 1) \right)^{\frac N n - p},
\end{equation}
we find that the $\al$ coefficient for this process is
\begin{equation}
p \mathcal N (p) = \al \mathcal N(p - 1) \qquad \Longrightarrow \qquad \al = \frac{N - p + 1}{\frac 1 n N(N - 1) \cdot ... \cdot (N - n + 1)}.
\end{equation}
Now, to give the final answer of the one point function we would need to relate the factors $|A|^2$ and $|C|^2$ to $N$, as we did in the previous sections.
However, to do so we need more insights on the supergravity side.
More precisely, the formula for the relation comes from the mapping of the supergravity coefficients of the corresponding geometry.
In the previous cases we have coherent superpositions of ground states, but in this case we are mixing supergravity operators in the CFT.
The geometries associated to the state we consider have not been investigated, and so the relation between $|C|$ and $N$ needs more work to be obtained.
We thus leave the final answer for this one point function as future work, and move to other cases.

\subsection{Other chiral primaries}\label{1pfsubsec:multishort}
Even if the matching with the supergravity side is not clear, in this section we want to point out that the CFT calculations for other operators are completely analogous.
Consider for instance the chiral primary
\begin{equation}
\bigotimes_{r = 1}^n \mathcal O^{\al \dot \al}_{(r)}.
\end{equation}
Acting on a $\ket{++}_n$ strand it creates again an excited strand, just as in the previous case.
However, we can also consider it acting on $n$ copies of a unit length strand.
That is, we can now easily calculate the process
\begin{equation}
\bigotimes_{r = 1}^n \mathcal O^{\al \dot \al}_{(r)} \left( \ket{++}_1 \right)^n \qquad \to \qquad \left( \ket{00}_1 \right)^n.
\end{equation}
Thus, consider the state
\begin{equation}
\psi = \sum_{p = 0}^N \left( A \ket{++}_1 \right)^p \left( B \ket{00}_1 \right)^{N - p}
\end{equation}
which has norm
\begin{equation}
|\psi|^2 = \sum_{p = 0}^N |A|^{2p} |B|^{2 (N - p)} \frac{N!}{p! (N - p)!}.
\end{equation}
The $\al$ coefficient in this case is
\begin{equation}
\binom{p}{n} \mathcal N(p) = \al \mathcal N (p - n) \qquad \Longrightarrow \qquad \al = \frac{(N - p + n)!}{n! (N - p)!}.
\end{equation}
Using the approximation for the short strands we obtain
\begin{equation}
\left< \bigotimes_{r = 1}^n \mathcal O^{- \dot -}_{(r)} \right> = \frac{A^n}{B^n} \frac{(N - |A|^2)!}{n! (N - |A|^2 - n)!}.
\end{equation}
Notice that for $n = 1$ we recover the result found in \cite{Giusto:2015dfa}.

\begin{figure}
\begin{subfigure}{0.5\textwidth}
\centering\includegraphics[height=4.5cm]{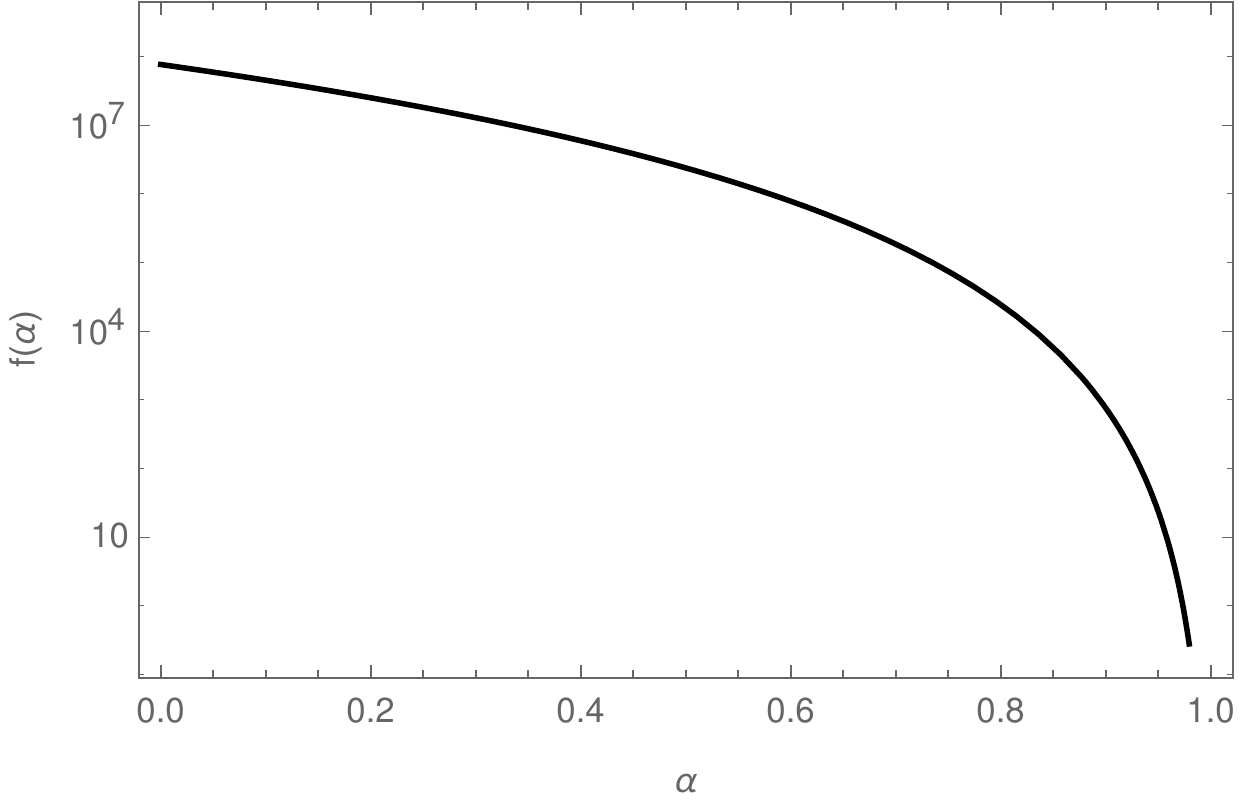}
\caption{$n$ of order one}
\label{1pffig:Oproduct1}
\end{subfigure}\hfill
\begin{subfigure}{0.5\textwidth}
\centering\includegraphics[height=4.5cm]{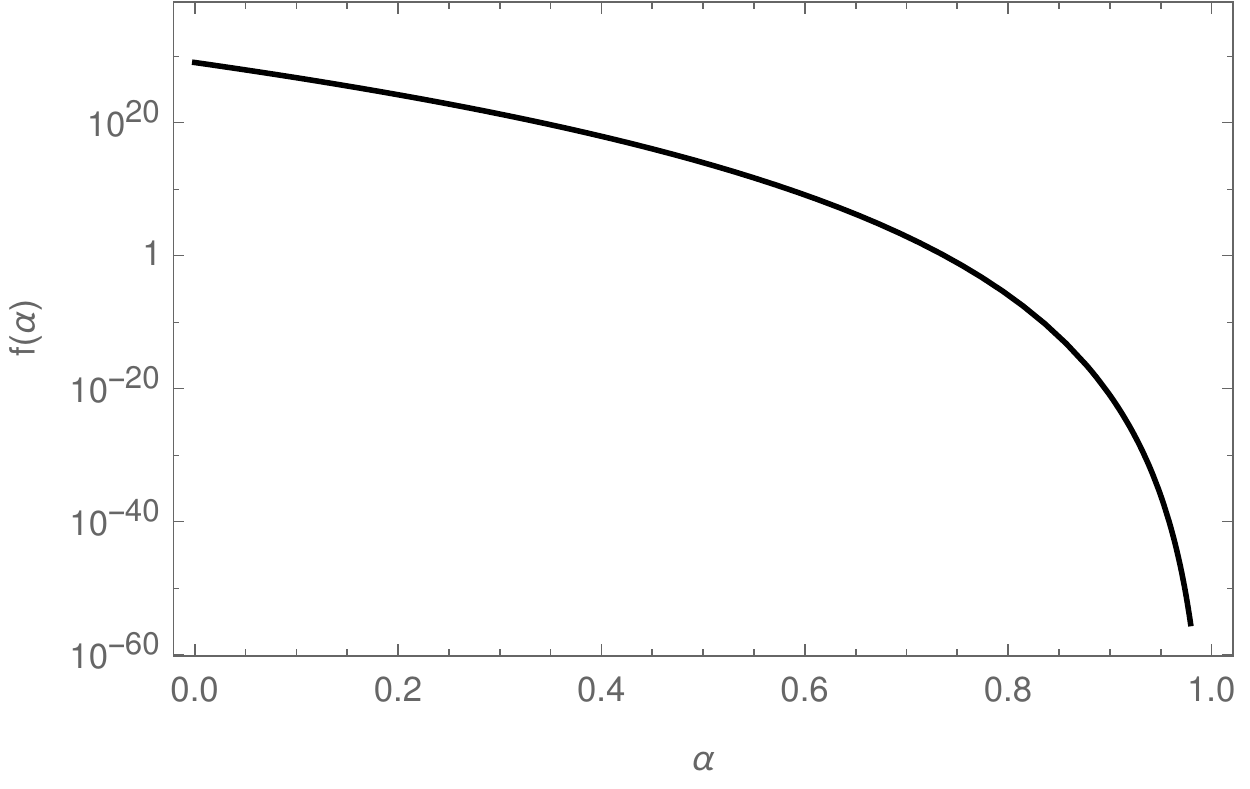}
\caption{$n$ comparable to $N$}
\label{1pffig:Oproduct2}
\end{subfigure}
\caption[Exact $n$-point function for $\otimes \mathcal O^{\al \dot \al}$]{Behaviour of the $n$-point function in equation \eqref{1pfeq:Oproductresult}. We see that it takes very big values. In figure \ref{1pffig:Oproduct1} we give the case where $n$ is of order one, by setting $n = 5$ and $N = 100$. In figure \ref{1pffig:Oproduct2} we give the case where $n$ is of order $N$, by setting $n = 50$ and $N = 100$. As we can see, in this second case the $n$-point function is bigger for small values of $\al$, but very close to zero for $\al$ close to one.}
\label{1pffig:Oproduct}
\end{figure}

We can also obtain the exact result using the same method as we will use in section \ref{1pfsection:long1pf}.
We refer to that section for the procedure, and we give the result directly in this case.
Setting $|A|^2 = \al N$, $|B|^2 = (1 - \al) N$, $0 <\al < 1$, where this $\al$ is unrelated to the coefficient matching the number of terms calculated above, we find that
\begin{equation}
\left< \bigotimes_{r = 1}^n \mathcal O^{- \dot -}_{(r)}\right> = A^n \bar B^n \frac{1}{n!} \frac{\sum_{p = 0}^{N - n} \al^p (1 - \al)^{-p} \frac{1}{p! (N - p - n)!}}{\sum_{p = 0}^N \left( \frac{\al}{1 - \al} \right)^p \frac{1}{p! (N - p)!}} = A^n \bar B^n \frac{N!}{n!} \frac{(1-\alpha )^n}{(N - n)!}.
\label{1pfeq:Oproductresult}
\end{equation}
We present the behaviour of the result in figure \ref{1pffig:Oproduct1}.
As we can see, the values it takes are very big for small $n$.
As we increase $n$, this answer becomes bigger close to $\al = 0$, and very close to zero as $\al$ approaches one.
We present this second case in figure \ref{1pffig:Oproduct2}.

So far, apart from this last example, we have illustrated with some examples how to obtain the CFT one point functions of chiral primary operators in the short strand case.
The tools that we have shown can be used to calculate more general one point functions as well, for example for the $\mathcal O^{\al \dot \al, *}_{\ka}$ operators defined in equation \eqref{1pfeq:twistedheavyO}.
We can also obtain one point functions for heavy and/or twisted $J$ operators.
Similarly, we can calculate one point functions for products of twist operators.
The method is exactly the same, but the combinatorics will be more involved.
In section \ref{section:joinings} we are interested in $n$-point functions for products of twist operators.
There we explain the combinatorics involved, reviewing some integer partition theory, and use the results to give an upper bound and a lower bound for the $\braket{(\Si^{- \dot -}_2)^n}$ $n$-point function.

In the next section we focus on the same one point functions as the ones we have explicitly calculated in this section, but in the opposite limit; in the long strand case.
Notice that some of the one point functions that we have calculated in this section, like the one above, are explicitly only possible in the short strand case.
Generalisations of these can be considered also in the long strand case.
We will not calculate all of them, as the examples provided should be enough to obtain these other cases.

\section{One point functions: long strand case}\label{1pfsection:long1pf}
In section \ref{1pfsection:1pf} we have calculated one point functions for chiral primaries in the short strand length case.
That is, in the case where we have a large amount of copies of each strand, with the lengths being of order one.
In this section we focus on the opposite limit, that is, in the limit where the strand length is large and the number of strands is small.
As in the short strand section first we will explain the approximation that we use, and then we will go case by case calculating the results.
As we will see, in this limit the one point functions give much smaller results than in the opposite one in general.

We consider now the case where the strand lengths are of order $N$, and the number of strands is of order one.
In this section we separate the results for the two and three-charge cases, as the method used differs slightly in both.
In the two-charge case we will be able to obtain exact results, whereas in the three-charge case we will strongly use the large $N$ limit.
We start with the 1/4-BPS states.

\subsection{Two-charge states}
As we just said, in this case we will be able to obtain exact results for the one point functions.
As we introduced in section \ref{1pfsubsection:statesreview} and used in section \ref{1pfsection:1pf}, we work with the state given in equation \eqref{gmr14dual},
\begin{equation}
\psi_{\{A_\ka^{(gs)}\}} = \sum_{\{N_\ka^{(gs)}\}} \left(\prod_{gs, \ka} A_\ka^{(gs)} \right)^{N_\ka^{(gs)}} \psi_{\{N_\ka^{(gs)}\}} = \sum_{\{N_\ka^{(gs)}\}} \prod_{gs, \ka} (A_\ka^{(gs)} \ket{S}_\ka)^{N_\ka^{(gs)}},
\label{eq:generalstate}
\end{equation}
which has norm
\begin{equation}
|\psi|^2 = \sum_{\{N_{\ka}^{(gs)}\}} \frac{N!}{\prod_{gs, \ka} N_{\ka}^{(gs)} \ka^{N_{\ka}^{(gs)}}} \prod_{gs, \ka} |A_{\ka}^{(gs)}|^{2 N_{\ka}^{(gs)}}.
\end{equation}
Recall also that the sum is constrained by
\begin{equation}
\sum_{gs, k} k N_k^{(gs)} = N,
\end{equation}
and the coefficients $A_k^{(gs)}$ satisfy
\begin{equation}
\sum_{k, gs} \left| A_k^{(gs)} \right|^2 = N.
\label{1pfeq:Acondition}
\end{equation}
In what follows, we first give an idea of how we do the calculation in this limit, and then we work case by case the answers for the one point functions.

\subsubsection{Method}
The final answer of each one point function is determined by the $A_i^{(gs)}$ coefficients, and so it is natural to set, given the condition that we have written in equation \eqref{1pfeq:Acondition},
\begin{equation}
|A^{(gs)}_i| = N \al_i^{(gs)}, \qquad \mathrm{with} \qquad \al_1^{(gs)} \in (0,1), \,\,\, \al_2^{(gs)} \in \left( 0, 1 - \al_1^{(gs)} \right), \,\,\, ...
\label{eq:Alongcondition}
\end{equation}
where the last $\al_i^{(gs)}$ coefficient is given in terms of the previous ones.
Since we are in the long strand case we consider the lengths $\ka$ of all the strands to be $N/m_{\ka}$, where $m_{\ka}$ is small compared to $N$.
Likewise, the number of strands $N_{\ka}^{(gs)}$ will also be a number of order one.
Now, the important thing to notice is that the only $N$-dependence of the norm of \eqref{eq:generalstate} is an $N!$ factor which comes out of the sum.
To see this, we substitute in the norm the parametrisation  \eqref{eq:Alongcondition}, which gives
\begin{equation}
|\psi|^2 =  N! \sum_{\{N_{\ka}^{(gs)}\}} \prod_{gs, \ka} \frac{m_{\ka}^{N_{\ka}^{(gs)}}}{N_{\ka}^{(gs)}!} \al_{\ka}^{N_{\ka}^{(gs)}}.
\label{eq:denominator}
\end{equation}
As we have seen from the previous section, this will be the denominator of all the one point functions.
If the numerator has at most an $(N - 1)!$ dependence in $N$ then this will mean that, up to numbers of order one, the long strand one point functions will be suppressed by a factor of $N$ with respect to the short strand length ones.
The expression of this numerator depends on the chiral primary we use, and so we will go one by one in the following sections.
As we will see the $N$ dependence will also factor out in all the numerators, and so we will be able to obtain the exact result up to a polynomial which depends only on the $\al_i^{(gs)}$.
We will give the polynomials for some cases as well, to see what is their behaviour for all values of the $\al_i^{(gs)}$.
Let us start by calculating the numerator for the $\Si^{- \dot -}_n$ operator with the most general 1/4-BPS state.

\subsubsection{\texorpdfstring{$\Si_n^{- \dot -}$}{Sin--} operator}\label{1pfsubsec:exactsin--}
\begin{figure}
\centering
\includegraphics[width=\textwidth/2]{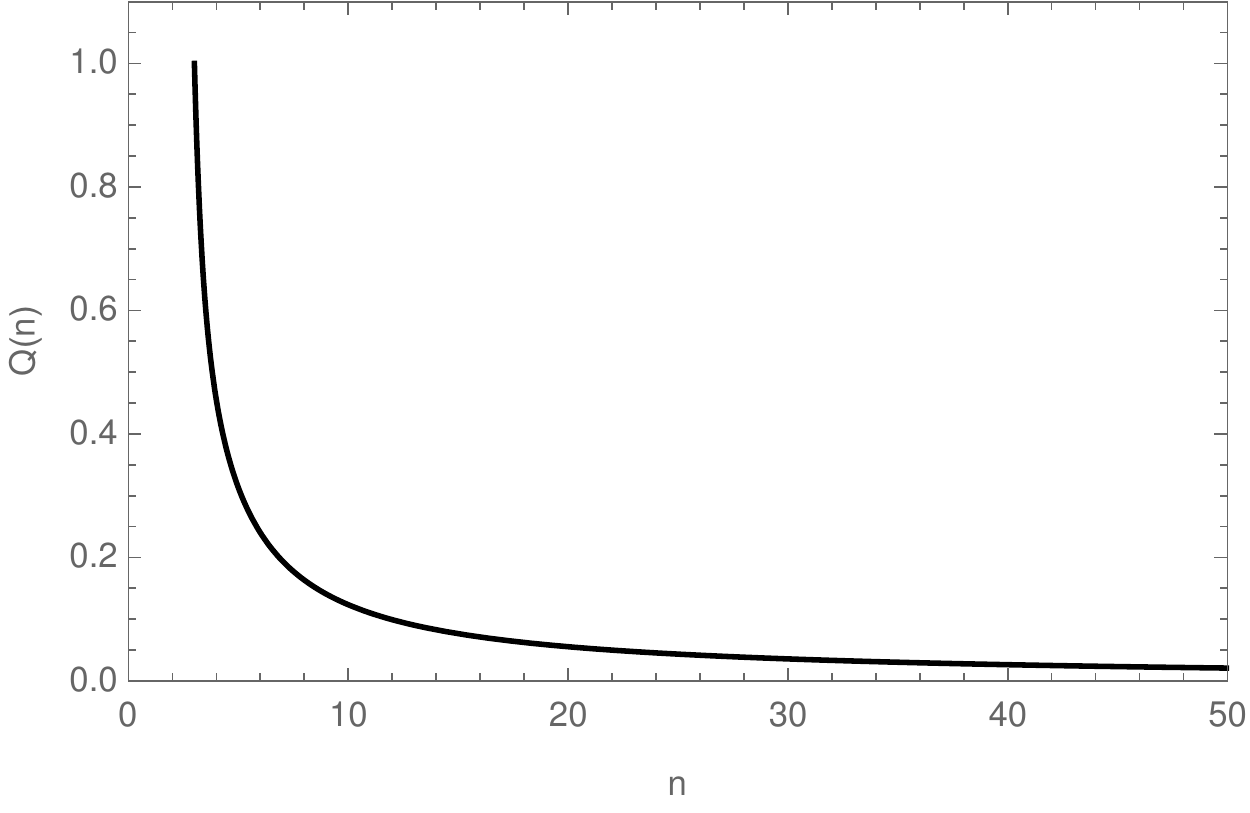}
\caption[$n$ behaviour of the $c_n$ coefficient]{Behaviour of $n/Q$, where $Q$ is defined in equation \eqref{1pfeq:Qdef}. As we can see, this function decreases exponentially in $n$. The $c_n$ coefficient will thus vanish exponentially fast for $n \geq 4$, as this function appears to the power of $N$.}
\label{1pffig:Qcoef}
\end{figure}

We can do the first part of this calculation in the most general case.
Also, we drop the $(gs)$ superscript, as it is redundant in this case and will only complicate the notation.
We start with the strands
\begin{equation}
\bigotimes_{i = 1}^n \left( A_i \ket{++}_{\frac{N}{m_1}} \right)^{p_i} \otimes \left( A \ket{++}_{\frac N m} \right)^p, \qquad \sum_{i = 1}^n \frac{1}{m_i} = \frac 1 m.
\end{equation}
After the action of the $\Si_n^{- \dot -}$ operator we have
\begin{equation}
\left( \prod_{i = 1}^n A_i^{p_i} \right) A^p  \bigotimes_{i = 1}^n \left( \ket{++}_{\frac{N}{m_1}} \right)^{p_i - 1} \otimes \left( \ket{++}_{\frac N m} \right)^{p + 1}.
\end{equation}
As we said, we are in the limit where $m_i, p_i, m, p$ are small numbers.
These strands contribute with the corresponding term in the in state $\bra{\psi}$ which has the same strands.
Thus, the contraction gives
\begin{equation}
\left( \prod_{i = 1}^n A_i \right) \overline A \left( \prod_{i = 1}^n |A_i|^{2 (p_i - 1)} \right) |A|^{2p} \norm{\bigotimes_{i = 1}^n \left( \ket{++}_{\frac{N}{m_1}} \right)^{p_i - 1} \otimes \left( \ket{++}_{\frac N m} \right)^{p + 1}}^2.
\label{eq:numerator}
\end{equation}
The norm is given by
\begin{equation}
\frac{N!}{\prod_{i = 1}^n \left[ \left( \frac{N}{m_i} \right)^{p_i - 1} (p_i - 1)! \right] \left( \frac N m \right)^{p + 1} (p + 1)!} \sim \frac{N!}{N^{p + \sum_{i = 1}^n p_i - n + 1}}.
\label{eq:numeratornorm}
\end{equation}
The product of the moduli of the $A_i$ yields
\begin{equation}
\al_1 \al_2 \cdot ... \cdot \al_{n - 1} (1 - \al_1 - \al_2 - ... - \al_{n - 1}) N^{p + \sum_{i = 1}^n p_i - n},
\label{eq:numeratoralpha}
\end{equation}
and thus the $N$-dependence of the product of the two is $(N - 1)!$.
Now, this is not the only part of the one point function that can give $N$ dependence.
We also have the $c_n$ and $\al$ coefficients, so we need to see what their product is.
Let us start by calculating $\al$, as it is straightforward to obtain.
Analogous to the previous section, we need to match the number of terms before and after, taking into account in how many ways can the gluing operator act.
Therefore,
\begin{equation}
\prod_{i = 1}^n p_i \frac{N}{m_i} \mathcal N_{\mathrm{initial}} = \al \mathcal N_{\mathrm{final}}.
\end{equation}
Certainly if we consider a process where we take more than one strand of the same kind the product of $p_i$ factor will be different, and we will have instead some combinatorial number.
However, as we said these are small numbers which do not affect the $N$ behaviour, and so we give the result only for this case.
The combinatorials needed for the exact result are explained in the previous section.
Solving for $\al$ we obtain
\begin{equation}
\al = (p + 1) \frac N m \sim N,
\end{equation}
{\it i.e.} it scales like $N$.
Let us now consider the $c_n$ coefficient.
We have written its expression in equation \eqref{1pfeq:cn} for strands of the same length, but let us recall its expression here in the most general case.
It is
\begin{equation}
c = \frac{\left( \frac N m \right)^{\frac{1}{2} ( \frac N m + 2 - n)}}{n} \prod_{i = 1}^n | 1- \bar{a}_i |^{\frac{1}{2} (\frac{N}{m_i} - 1)(n-1)} \prod_{j = 1}^n \left(\frac{N}{m_j}\right)^{-\frac{1}{2} \left(\frac{N}{m_j} + 1\right)} \prod_{j \neq k} | \bar{a}_j - \bar{a}_k |^{\frac{1}{2} \left( 1 - \frac{N}{m_k} \right)},
\label{eq:longc}
\end{equation}
where
\begin{equation}
\bar{a}_i = \frac{1 + \bar a e^{i\phi_i}}{1 - \frac{m}{m_i}}, \qquad \bar a = \left(\frac{m_1}{m} \prod_{i = 2}^n \left( 1 - \frac{m}{m_i} \right) - 1 \right)^{\frac{1}{n - 1}}
\end{equation}
and the phases are
\bea
(n-1) \in 2 Z : \quad \phi_i &=& \frac{(i-1) \pi}{n-1}, \quad  \phi_{i+1}  = - \frac{ (i-1) \pi}{n-1}, \quad i \in 2 Z, \quad i \ge 2 \\
n \in 2 Z : \quad \phi_2 &=& 0, \quad  \phi_i = \frac{(i-1) \pi}{n-1}, \quad  \phi_{i+1}  = - \frac{(i-1) \pi}{n-1}, \quad (i - 1) \in 2 Z, \quad  i \ge 3. \nonumber
\eea
Notice that the coefficients $\bar a$, $\bar a_i$ are independent of $N$.
Keeping only the factors with $N$ we see that
\begin{equation}
c \sim N^{1 - n} \left( \frac 1 m \right)^{\frac 1 2 \left( \frac N m + 2 - n \right)} |1 - \bar a_i|^{n \frac 1 2 \frac N m - \frac 1 2 \frac N m - \frac n 2 + \frac 1 2} \frac{1}{\prod_j m_j^{-\frac 1 2 \left( \frac{N}{m_j} + 1 \right)}} \prod_{j \neq k} |\bar a_j - \bar a_k|^{\frac 1 2 \left( 1 - \frac{N}{m_k} \right)}.
\label{eq:aproxlongc}
\end{equation}
We need to see that the product of all the factors that have an exponent dependent of $N$ go to zero in the large $N$ limit.
Let us check some simple cases first.
As noted in \cite{Tormo:2018fnt} for $n = 2,3$ we have simple expressions for the coefficient $c_n$, and we also have a compact formula when all the $m_i$ are equal.
For $n = 2$ the coefficient reads
\begin{equation}
c_2 = \frac 1 N \frac{m_1 + m_2}{2}.
\end{equation}
For $n = 3$ we have
\begin{equation}
c_3 = \frac{1}{N^2} \frac{m_1 m_2 + m_2 m_3 + m_1 m_3}{3}.
\end{equation}
The case of equal $m_i$ requires some more work.
Let $m_i = m$ for all $i \leq n$, so that we join $n$ strands of length $N/m$ into a single one of length $Nn/m$.
Then the general expression \eqref{eq:longc} reduces in this case to equation \eqref{1pfeq:cn}, which with the current notation is
\begin{equation}
c_n = \left(\frac{nN}{m}\right)^{1-n} n^{\frac{Nn}{2m} + \frac{n}{2} - \frac{N}{m}} \left (  (n-1)^2 (n \bar{a})^{(n-2)} \Lambda^{-1} \right )^{\frac{1}{2} (n-1) \left (1 - \frac{N}{m} \right )}
\end{equation}
where
\begin{equation}
\Lambda = \left ( 1 + n (n-1)^{(n-1)} - n^{n-1} \right )
\end{equation}
and
\begin{equation}
\bar a = \left( n \left( 1 - \frac 1 n \right)^{n - 1} - 1 \right)^{\frac{1}{n - 1}}.
\end{equation}
As we can see the $N$ factor can be read straight away and agrees with \eqref{eq:aproxlongc}.
We need to see what happens with the other factors.
Let
\begin{equation}
Q := (n-1)^2 (n \bar{a})^{(n-2)} \Lambda^{-1}. 
\label{1pfeq:Qdef}
\end{equation}
Using the large $N$ approximation we can drop some terms in the exponents and rewrite the coefficient as
\begin{equation}
c_n \approx N^{1 - n} \left( \left( \frac n Q \right)^{\frac N 2 (n - 1)} \left( \frac 1 n \right)^{\frac N 2} \right)^{\frac 1 m}
\end{equation}
Now, the function $n/Q$ is a monotonically decreasing function, which satisfies
\begin{equation}
\left[\frac{n}{Q}\right]_{n = 4} = \frac{5}{11} < 1, \qquad \lim_{n \to +\infty} \frac{n}{Q} = 0.
\end{equation}
More precisely, it is a exponentially decreasing function in $n$.
We show its behaviour in figure \ref{1pffig:Qcoef}.
Therefore,
\begin{equation}
\lim_{N \to \infty} c_n = 0, \qquad n \geq 4.
\end{equation}
If we add the $\al$ coefficient we obtain
\begin{equation}
c\al_{\mathrm{between}\; n \; \mathrm{strands\; of\; equal\; length}} \sim N^{2 - n} \left( \left( \frac n Q \right)^{\frac N 2 (n - 1)} \left( \frac 1 n \right)^{\frac N 2} \right)^{\frac 1 m},
\end{equation}
and so the answer for the one point function is, adding the normalisation of the twist,
\begin{align}
&\braket{\Si^{- \dot -}_n}_{\mathrm{long}} = \nonumber \\
& = \left( \prod_{i = 1}^n A_i \right) \bar A N^{1 - n} \left( \left( \frac n Q \right)^{\frac N 2 (n - 1)} \left( \frac 1 n \right)^{\frac N 2} \right)^{\frac 1 m} \left( \frac{n}{N (N - 1) ... (N - n + 1)} \right)^{\frac 1 2} f(\al_1, ..., \al_n),
\label{1pfeq:sin14gen}
\end{align}
where $f(\al_1, ..., \al_n)$ is an $N$-independent polynomial which numerator is given by equations \eqref{eq:numerator}, \eqref{eq:numeratornorm} and \eqref{eq:numeratoralpha} and with the denominator given by \eqref{eq:denominator}.
Notice that the one point function decreases exponentially with $N$ for $n \geq 4$, as for $n = 3$ we have $n/Q = 1$.
If we compare this result to the short strand case, equation \eqref{sigmanresult}, we see that the only $N$ dependence in the short strand case comes from the normalisation of the twist.
Therefore, we have the following relations for the one point functions,
\begin{align}
\braket{\Si_2^{- \dot -}}_{\mathrm{long}} & \sim \frac 1 N \braket{\Si_2^{- \dot -}}_{\mathrm{short}} \nonumber \\
\braket{\Si_3^{- \dot -}}_{\mathrm{long}} & \sim \frac{1}{N^2} \braket{\Si_3^{- \dot -}}_{\mathrm{short}} \nonumber \\
\braket{\Si_n^{- \dot -}}_{\mathrm{long}} & \sim  N^{1 - n} e^{-N \log n} \braket{\Si_n^{- \dot -}}_{\mathrm{short}} \qquad \mathrm{(same\; strand\; length)}.
\label{1pfeq:sis14gen}
\end{align}
We will now do this calculation for the single copy $\mathcal O^{- \dot -}_{(r)}$ operator, and afterwards we will show examples of the $f(\al_1, ..., \al_n)$ polynomials in both cases.

\subsubsection{\texorpdfstring{$\mathcal O^{- \dot -}_{(r)}$}{O--} operator}\label{1pf:O--14}
In this case we start with the strands
\begin{equation}
\left( A_1 \ket{++}_{\frac N m} \right)^{p_1} \left( A_2 \ket{00}_{\frac N m} \right)^{p_2}.
\end{equation}
After the action of the $\mathcal O^{- \dot -}_{(r)}$ operator we have
\begin{equation}
A_1^{p_1} A_2^{p_2} \left( \ket{++}_{\frac N m} \right)^{p_1 - 1} \left( \ket{00}_{\frac N m} \right)^{p_2 + 1}.
\end{equation}
The contraction gives
\begin{equation}
A_1 \bar A_2 |A_1|^{2p_1} |A_2|^{2(p_2 - 1)} \frac{N!}{(p_1 - 1)!(p_2 + 1)! \left( \frac N m \right)^{p_1 - 1} \left( \frac N m \right)^{p_2 + 1}}.
\end{equation}
The $\al$ coefficient in this case is
\begin{equation}
\al = p_2 + 1,
\end{equation}
and so the one point function is
\begin{equation}
\braket{\mathcal O^{- \dot -}_{(r)}}_{\mathrm{long}} = A_1 \bar A_2 \frac 1 N f(\al),
\label{1pfeq:gen14Ores}
\end{equation}
where, again, $f(\al)$ is a polynomial which is independent of $N$.
Comparing to the result of the short strand case we see that
\begin{equation}
\braket{\mathcal O^{- \dot -}_{(r)}}_{\mathrm{long}} \sim \frac 1 N \braket{\mathcal O^{- \dot -}_{(r)}}_{\mathrm{short}}.
\label{1pfeq:longshortOrel}
\end{equation}
Let us now write and plot some polynomials for both operators, to see how they behave.

\subsubsection{Exact answers for the one point functions (examples)}\label{1pfsubsec:14longcorrex}
We start with a couple of easy examples, and build up to more general ones.
Let us start with the $\Si_2^{- \dot -}$ operator in the easiest case possible.

\paragraph{Simplest non-trivial case}
Consider the state
\begin{equation}
\psi = \left(A_1 \ket{++}_{\frac N 4}\right)^4 + \left(A_1 \ket{++}_{\frac N 4}\right)^2 A_2 \ket{++}_{\frac N 2} + \left(A_2 \ket{++}_{\frac N 2}\right)^2.
\label{simplestlongstate}
\end{equation}
We have a non-zero one point function, which corresponds to
\begin{equation}
\langle \Si^{- \dot -}_2 \rangle = \left| \psi \right|^{-2} |\Si^{- \dot -}_2|^{-1} \bra{\psi} \Si^{- \dot -}_2 \ket{\psi}.
\end{equation}
Since we have just a few terms, we write them all out explicitly to see exactly how the calculation works.
Using the formulas of the previous sections, we see that the norm of this state is
\begin{equation}
\left| \psi \right|^2 = \frac{32 (N - 1)!}{3 N^3} |A_1|^8 + \frac{16 (N - 1)!}{N^2} \left| A_1 \right|^4 \left| A_2 \right|^2 + \frac{2 (N - 1)!}{N} \left| A_2 \right|^4.
\end{equation}
It is useful to write the generic expression for each term, $\mathcal N(p)$, to make some combinatorials easier afterwards.
They read
\begin{align}
\mathcal N(p) &= \frac{N!}{\left(\frac{N - p \frac N 2}{\frac N 4}\right)! \left(\frac N 4\right)^{\left(\frac{N - p \frac N 2}{\frac N 4}\right)} p! \left(\frac N 2 \right)^p} = \nonumber\\
& = \frac{N!}{(4 - 2p)! \left(\frac N 4\right)^{(4 - 2p)} p! \left(\frac N 2\right)^p}, \qquad p = 0,1,2.
\end{align}
It is also more convenient to write the state \eqref{simplestlongstate} as a sum.
We have
\begin{equation}
\psi = \sum_{p = 0}^2 \left(A_1 \ket{++}_{\frac N 4}\right)^{4 - 2p} \left(A_2\ket{++}_{\frac N 2}\right)^p.
\end{equation}
Now, the action of the gluing operator is
\begin{equation}
\Si^{-\dot -}_2 \left[\left(\ket{++}_{\frac N 4}\right)^{4 - 2p} \left(\ket{++}_{\frac N 2}\right)^p\right] = c_{\frac N 4, \frac N 4} \al \left[\left(\ket{++}_{\frac N 4}\right)^{2 - 2p} \left(\ket{++}_{\frac N 2}\right)^{p + 1}\right].
\end{equation}
The $c_2$ coefficient is
\begin{equation}
c_{\frac N 4, \frac N 4} = \frac 4 N.
\end{equation}
Now, for the combinatorial factor $\al$, we need to choose two strands of length $N/4$, and they each have $N/4$ positions on which we can insert the gluing operator.
Thus,
\begin{equation}
\binom{4 - 2p}{2} \left(\frac N 4\right)^2 \mathcal N(p) = \al \mathcal N(p + 1).
\end{equation}
Solving for $\al$ we obtain
\begin{equation}
\al = (p + 1) \frac N 4.
\end{equation}
The one point function then gives
\begin{align}
\langle \Si^{- \dot -}_2 \rangle &= |\psi|^{-2} |\Si_2^{- \dot -}|^{-1} \bra{\psi} \Si^{- \dot -}_2 \ket{\psi} = \nonumber \\
&= |\psi|^{-2} \Big[ A_1^4 (\bar A_1)^2 \bar A_2 \prescript{}{\frac N 2}{\bra{++}} \left( \prescript{}{\frac N 4}{\bra{++}} \right)^2 \left(\ket{++}_{\frac N 4}\right)^2 \ket{++}_{\frac N 2} + \nonumber \\
&\hspace{44pt} + A_1^2 A_2 (\bar A_2)^2 2 \left( \prescript{}{\frac N 2}{\bra{++}} \right)^2 \left(\ket{++}_{\frac N 2}\right)^2 \Big] = \nonumber \\
&= \frac{A_1^2 \bar A_2 \left(|A_1|^4 \frac{16 N!}{N^3} + 2 |A_2|^2 \frac{2N!}{N^2}\right)}{\frac{32}{3} \frac{N!}{N^4} |A_1|^8 + \frac{16N!}{N^3} |A_1|^4 |A_2|^2 + \frac{2N!}{N^2} |A_2|^4} \left( \frac{2}{N (N - 1)} \right)^{\frac 1 2}.
\label{resultsigma2long}
\end{align}

\begin{figure}
\centering
\includegraphics[width=\textwidth/2]{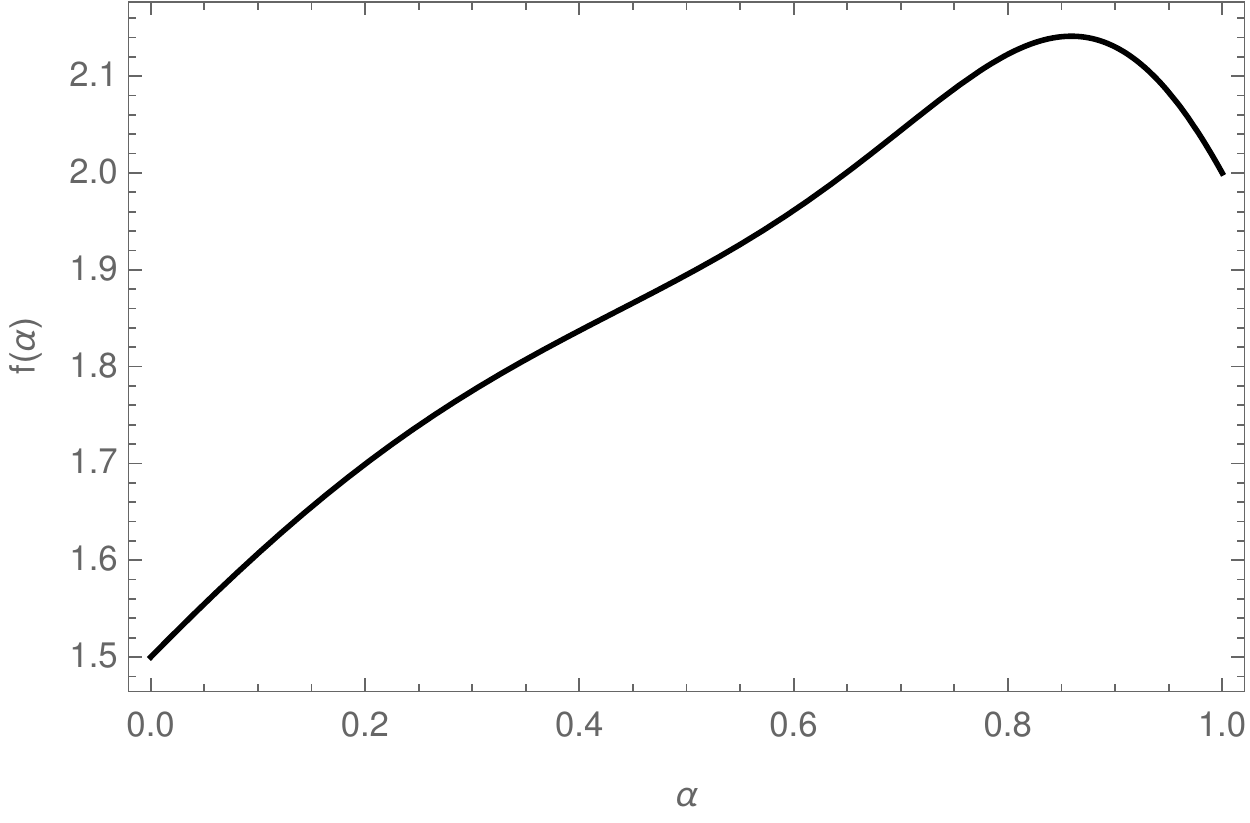}
\caption[Long strand one point function for $\Si_2$]{Behaviour of the coefficient $f(\alpha)$ which controls the relation between the one point functions for short and long strings.}
\label{fig:fsigma2long}
\end{figure}

We want to relate this result to its analogous in the short strand length, large number of copies calculated in \cite{Giusto:2015dfa}.
Let us recall that the expression we use for the one point function differs from the one in the paper, as we are including the normalisation for the twist operator.
In this section this normalisation does not play any crucial role, and so for the rest of the section we will leave it as $N_{\Si_{\ka}}$.
The result for the analogous short strand one point function is
\begin{equation}
\langle\Si^{- \dot -}_2 \rangle_{\mathrm{short}} = \frac{A_1^2 \bar A_2}{2} N_{\Si_2}.
\end{equation}
Using the substitutions $|A_1|^2 = N - |A_1|^2$ in \eqref{resultsigma2long}, and letting $|A_2|^2 = \alpha N$, where $\alpha \in \mathbb{R}$, $0 \leq \alpha \leq 1$, as in equation \eqref{eq:Alongcondition}, then \eqref{resultsigma2long} simplifies to
\begin{equation}
\langle \Si^{- \dot -}_2 \rangle_{\mathrm{long}} = N_{\Si_2} A_1^2 \bar A_2 \frac{6}{N} \frac{4 \alpha^2 - 7 \alpha + 4}{16\alpha^4 - 40\alpha^3 + 51\alpha^2 - 40\alpha + 16}.
\label{fsigma2first}
\end{equation}
If we define $f(\alpha)$ to be the division of the two polynomials in $\alpha$, we see that $f$ is bounded by 2.15 for $0 \leq \al \leq 1$, as we can see in figure \ref{fig:fsigma2long}.
Therefore, since $f(0) = 3/2$, we find that
\begin{equation}
A_1^2 \bar A_2 \frac{3}{2 N} < \frac{\langle \Si^{- \dot -}_2 \rangle_{\mathrm{long}}}{N_{\Si_2}} < A_1^2 \bar A_2 \frac{3}{N}
\end{equation}
for all possible values of $|A_1|^2$ and $|A_2|^2$.
Hence,
\begin{equation}
\langle \Si_2^{- \dot -}\rangle_{\mathrm{long}} \sim \frac{\langle \Si_2^{- \dot -}\rangle_{\mathrm{short}}}{N}.
\label{sigma2longshortrel}
\end{equation}

\begin{figure}
\begin{subfigure}{0.5\textwidth}
\centering\includegraphics[height=4.5cm]{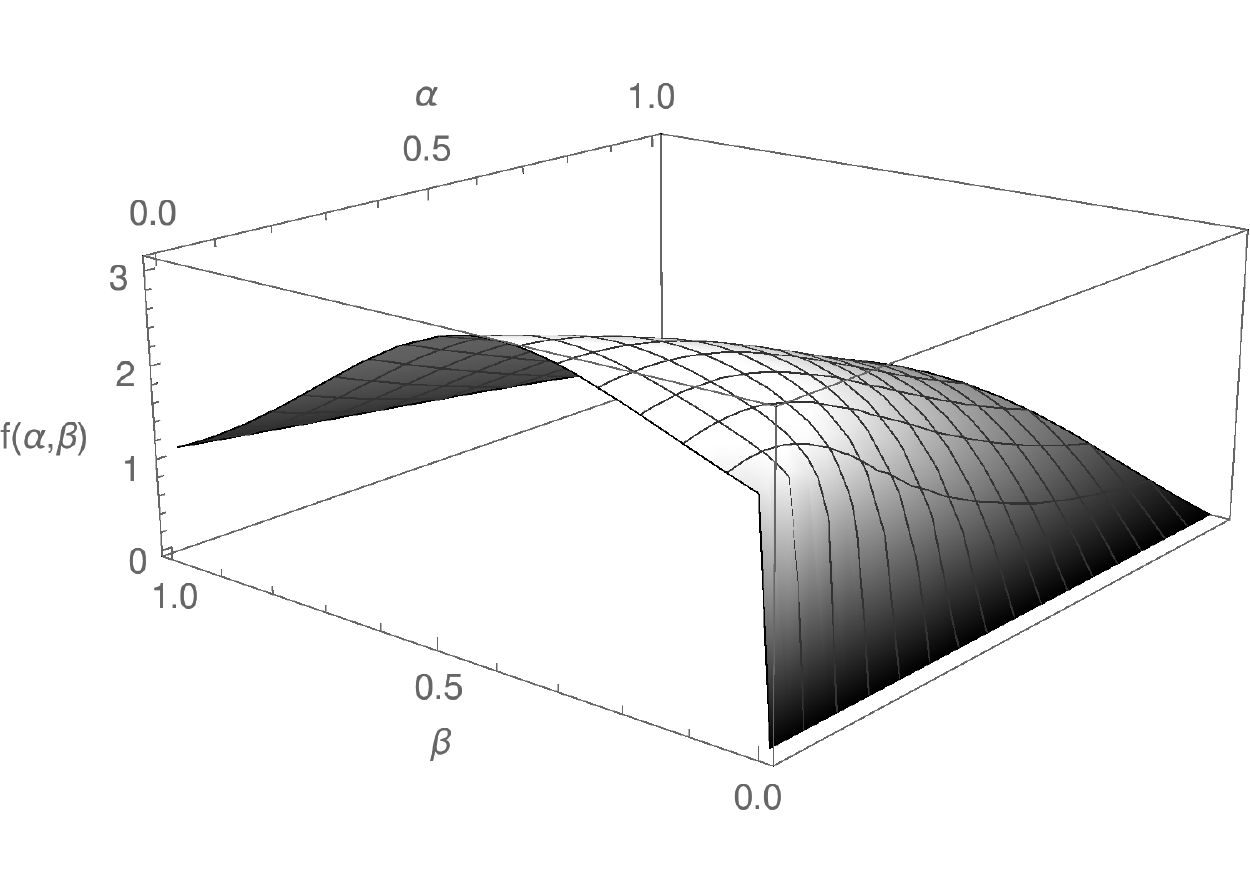}
\caption{First term}
\label{fig:sigma2dif1}
\end{subfigure}\hfill
\begin{subfigure}{0.5\textwidth}
\centering\includegraphics[height=4.5cm]{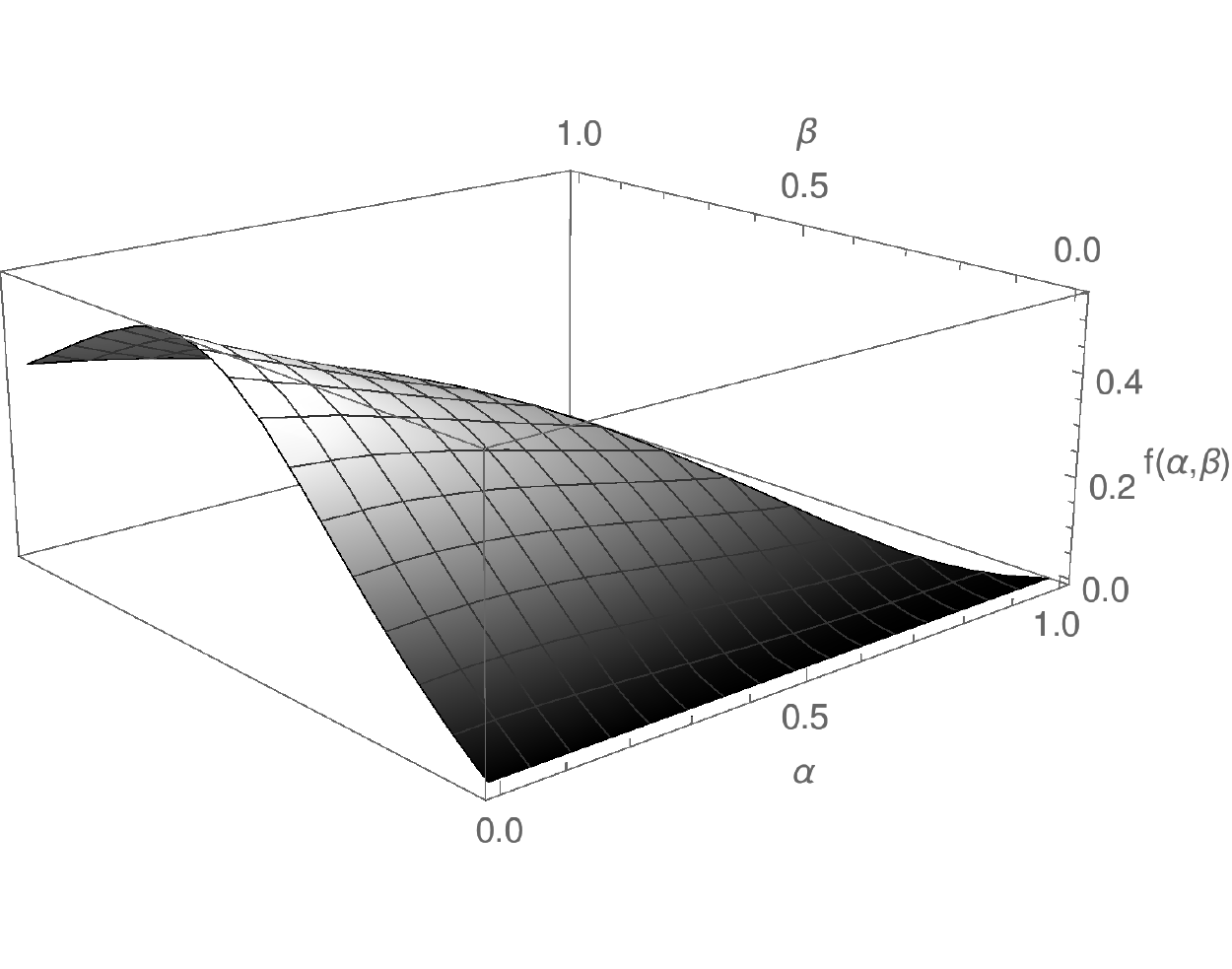}
\caption{Second term}
\label{fig:sigma2dif2}
\end{subfigure}
\caption[1-pf for $\Si_2^{- \dot -}$ for the easiest long case joining strands of different length]{$f$ coefficient comparing the long and short strand length one point functions for the $\Si_2^{- \dot -}$ operator, in the case where it joins two strands of lengths $N/4$ and $N/2$, as shown in equation \eqref{1pfeq:si2ed}. The state that needs to be considered for such one point function to not vanish also enables the action of the twist on two strands of length $N/4$. Subfigure \ref{fig:sigma2dif1} corresponds to the first term that contributes, the joining of two strands of different length, which is the process we are interested in. Subfigure \ref{fig:sigma2dif2} is the second term, where two strands of the same length are joined. As we can see, the second contribution does not change the behaviour of the result.}
\label{fig:simplesigma2dif}
\end{figure}

\paragraph{Joining strands of different length}
Let us now do a similar calculation, but joining two strands of different lengths.
The new feature in this example is that in this case we have two variables in the polynomial $f$.
Consider the state
\begin{equation}
\psi = \left( A_1 \ket{++}_{\frac N 2} \right)^2 + A_1 \ket{++}_{\frac N 2} \left( A_2 \ket{++}_{\frac N 4} \right)^2 + \left( A_2 \ket{++}_{\frac N 4} \right)^4 + A_2 \ket{++}_{\frac N 4} A_3 \ket{++}_{\frac{3N}{4}}.
\end{equation}
We do this calculation explicitly term by term as well, to see exactly all terms and how it works.
The norm of this state is
\begin{equation}
|\psi|^2 = \frac{N!}{N^2} \left( 2|A_1|^4 + |A_1|^2 |A_2|^4 \frac{16}{N} + |A_2|^8 \frac{32}{3 N^2} + |A_2|^2 |A_3|^2 \frac{16}{3} \right).
\end{equation}
Notice that we have two different terms now that are created with the $\Si_2^{- \dot -}$ operator,
\begin{equation}
\Si_2^{- \dot -} \psi \to A_1 A_2^2 \ket{++}_{\frac N 4} \ket{++}_{\frac{3N}{4}} + A_2^4 \ket{++}_{\frac N 2} \left( \ket{++}_{\frac N 4} \right)^2.
\label{1pfeq:si2ed}
\end{equation}
The first term is the one we are actually interested in, but as we will see the result is almost the same for both.
For the first term we have
\begin{equation}
c_{\frac N 4, \frac N 2} = \frac 3 N, \qquad \al_1 = \frac{3N}{4},
\end{equation}
and for the second one we have
\begin{equation}
c_{\frac N 4, \frac N 4} = \frac 1 N, \qquad \al_2 = \frac{N}{4}.
\end{equation}
Thus, the one point function is given by, analogous to the previous case,
\begin{align}
\braket{\Si_2^{- \dot -}} &= |\psi|^{-2} |\Si_2^{- \dot -}|^{-1} \bra{\psi} \left( A_1 A_2^2 \frac 9 4 \ket{++}_{\frac N 4} \ket{++}_{\frac{3N}{4}} + A_2^4 \frac 1 4 \ket{++}_{\frac N 2} \left( \ket{++}_{\frac N 4} \right)^2 \right) = \nonumber \\
& = N_{\Si_2} \frac{A_1 A_2 \overline A_3 12 |A_2|^2 + \overline A_1 A_2^2 \frac 4 N |A_2|^4}{2 |A_1|^4 + \frac{16}{N} |A_1|^2 |A_2|^4 + \frac{32}{3N^2} |A_2|^8 + \frac{16}{3} |A_2|^2 |A_3|^2}.
\end{align}
Now, we set the modulus of the coefficients $A_i$ to be
\begin{equation}
|A_1|^2 = N\al, \qquad |A_2|^2 = N\beta, \qquad |A_3|^2 = N(1 - \al - \beta),
\end{equation}
with
\begin{equation}
0 < \al < 1 \qquad \mathrm{and} \qquad 0 < \beta < 1 - \al.
\end{equation}
Plugging these expressions in the one point function yields
\begin{equation}
\frac 1 N \frac{A_1 A_2 \overline A_3 12 \beta + \overline A_1 A_2^2 4 \beta ^2}{2 \al^2 + 16\al \beta^2 + \frac{32}{3} \beta^4 + \frac{16}{3} \beta (1 - \al - \beta)}.
\end{equation}
Let $f(\al, \beta)$ be the coefficient without the $A_i$'s and the $N$ for both terms, just as in the previous case.
As we can see, the polynomials that we obtain agree with the general equations \eqref{eq:numerator} and \eqref{eq:denominator}.
Then, as we show in figure \ref{fig:simplesigma2dif} both terms are small in the range of interest.
The point $(\al, \beta) = (0,0)$ is singular, but let us recall that point is not a physical state and thus is not under consideration.
Let us now study the polynomial that results from the action of the $\Si_2^{- \dot -}$ operator in the case where it joins any two strands of equal length.

\begin{figure}
\centering
\includegraphics[width=\textwidth/2]{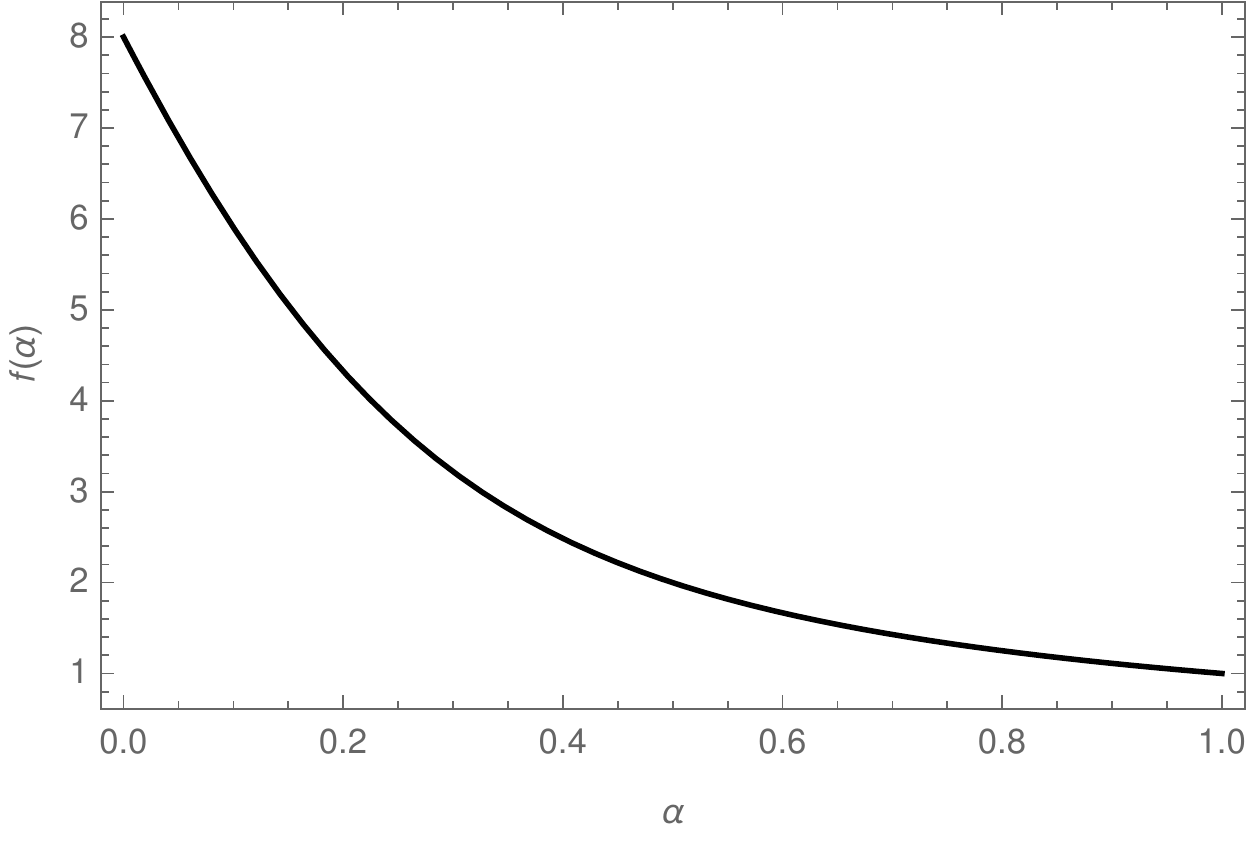}
\caption[1-pf for $\Si_2$ for general strand length.]{Behaviour of $N$ times the coefficient \eqref{fsigma2equal} relating the one point function for $\Si_2$ in the long and short strand case. At zero the function is ill-defined. In the plot $m$ = 17.}
\label{fig:sigma2samegeneral}
\end{figure}

\paragraph{Joining two strands of the same length}
Consider the state
\begin{equation}
\psi = \sum_{p = 0}^{\frac m 2} \left(A_1 \ket{++}_{\frac N m}\right)^{m - 2p} \left(A_2 \ket{++}_{\frac{2N}{m}}\right)^p.
\end{equation}
The polynomial is, in this case,
\begin{equation}
f(\al) = \frac{1}{\alpha} \frac{\sum_{p = 1}^{\frac m 2} \left(\alpha^p (1 - \alpha)^{-2p} \left(\frac{1}{2m}\right)^p \frac{1}{p! (m - 2p)!}\right)}{\sum_{p = 0}^{\frac m 2} \left(\alpha^p (1 - \alpha)^{-2p} \left(\frac{1}{2m}\right)^p \frac{1}{p! (m - 2p)!}\right)}.
\end{equation}
If we plug this expression in Mathematica we get
\begin{equation}
f(\alpha) = \frac{1-\frac{2^{-\frac{m}{2}} \left(-\frac{(\alpha-1)^2 m}{\alpha}\right)^{m/2}}{U\left(-\frac{m}{2},\frac{1}{2},-\frac{(\alpha-1)^2 m}{2 \alpha}\right)}}{\alpha},
\label{fsigma2equal}
\end{equation}
where $U$ is a hypergeometric function.
This function is of order one for most values of $\alpha$.
We present in figure \ref{fig:sigma2samegeneral} the plot of $f(\al)$ for $m = 17$.
We see that we have an asymptote at $\alpha = 0$, as it happens in the other cases.
Also, setting $m = 4$ in \eqref{fsigma2equal} we recover \eqref{fsigma2first}, as expected.
Let us recall, from equation \eqref{1pfeq:sin14gen}, that this one point functions also has a $1/N$ suppression with respect to the analogous short strand one.

\begin{figure}
\centering
\includegraphics[width=\textwidth/2]{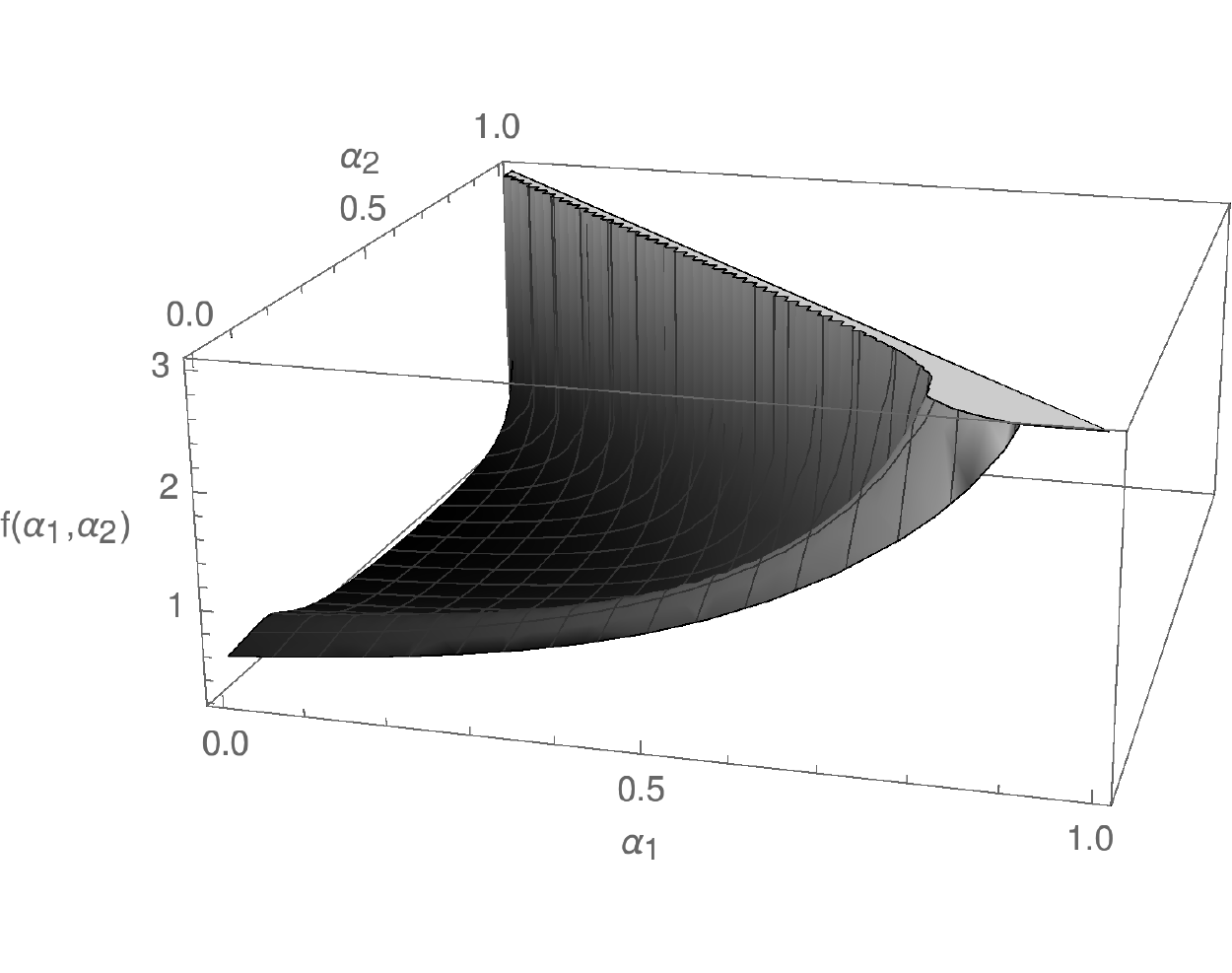}
\caption[1-pf for $\Si_2^{- \dot -}$ for any strand length.]{Behaviour of the coefficient $f(\al_1, \al_2)$ function for the $\braket{\Si_2^{- \dot -}}$ in the long strand case. We can see that, as in the previous cases, we get asymptotes in the boundary values. In the plot, $m_1 = 6$ and $m_2 = 12$.}
\label{fig:sigma2gen}
\end{figure}

\paragraph{Joining any two strands}
Consider the state
\begin{equation}
\psi = \sum_{p_1 = 0}^{m_1} \sum_{p_2 = 0}^{m_2\left(1 - \frac{p_1}{m_1}\right)} \left( A_1 \ket{++}_{\frac{N}{m_1}} \right)^{p_1} \left( A_2 \ket{++}_{\frac{N}{m_2}} \right)^{p_2} \left( A_3 \ket{++}_{\frac{N}{m_3}} \right)^{p_3},
\end{equation}
where
\begin{equation}
p_3 = \frac{N - \frac{N}{m_1} p_1 - \frac{N}{m_2} p_2}{\frac{N}{m_3}} = m_3 \left(1 - \frac{p_1}{m_1} - \frac{p_2}{m_2}\right)
\end{equation}
and
\begin{equation}
\frac{N}{m_3} = \frac{N}{m_1} + \frac{N}{m_2}.
\end{equation}
The norm of the state is
\begin{equation}
|\psi|^2 = \sum_{p_1 = 0}^{m_1} \sum_{p_2 = 0}^{m_2(1 - \frac{p_1}{m_1})} |A_1|^{2p_1} |A_2|^{2p_2} |A_3|^{2p_3} \mathcal N(p_1, p_2),
\end{equation}
where
\begin{equation}
\mathcal N(p_1, p_2) = \frac{N!}{p_1! p_2! p_3! \left(\frac{N}{m_1}\right)^{p_1} \left(\frac{N}{m_2}\right)^{p_2} \left(\frac{N}{m_3}\right)^{p_3}}.
\end{equation}
The $c_2$ coefficient in this case is
\begin{equation}
c_{\frac{N}{m_1},\frac{N}{m_2}} = \frac{m_1 + m_2}{2N}.
\end{equation}
The action of the gluing operator is
\begin{align}
& \Si_2^{- \dot -} \left[ \left( \ket{++}_{\frac{N}{m_1}} \right)^{p_1} \left( \ket{++}_{\frac{N}{m_2}} \right)^{p_2} \left( \ket{++}_{\frac{N}{m_3}} \right)^{p_3}\right] = \nonumber \\
=& \, c_{\frac{N}{m_1}, \frac{N}{m_2}} \al \left[ \left( \ket{++}_{\frac{N}{m_1}} \right)^{p_1 - 1} \left( \ket{++}_{\frac{N}{m_2}} \right)^{p_2 - 1} \left( \ket{++}_{\frac{N}{m_3}} \right)^{p_3 + 1}\right].
\end{align}
In this case the $\al$ coefficient is
\begin{equation}
\al = \frac{N}{p_3 m_3}.
\end{equation}
Mathematica is not able to perform the exact sums with all the coefficients for $f(\al_1, \al_2)$ in this case, and so we need to take approximations.
We are in the case where the lengths of the strands are big, that is, in the limit where $p_i$, $m_i$ are small integers.
Thus, since
\begin{equation}
p_3 = m_3 \left(1 - \frac{p_1}{m_1} - \frac{p_2}{m_2} \right),
\end{equation}
and recalling as well that $0 \leq p_i \leq m_i$, we see that the factor inside the parenthesis will be of order 1, and so
\begin{equation}
p_3 \lesssim m_3 = \frac{m_1 m_2}{m_1 + m_2}.
\end{equation}
Then, the total coefficient is approximated by
\begin{equation}
\frac{m_1 + m_2}{2 p_3 m_3} \gtrsim \frac{(m_1 + m_2)^3}{2 (m_1 m_2)^2}.
\end{equation}
The polynomials are too long to write out explicitly in this case, and the formulas do not give any deeper insights.
Also, we need to get rid of the factorials in the denominator so that Mathematica can perform the sums.
Notice that in this case it is not straightforward to recover the results from the previous sections, as now the combinatorial factors are not easily reduced to the case of two strands of the same length.
We present in figure \ref{fig:sigma2gen} a plot of the function for $m_1 = 6$ and $m_2 = 12$.
We observe again the asymptotes at the boundary values of $\al_1$ and $\al_2$.
Let us recall once again that $\al_1$ and $\al_2$ are defined only in the open interval (0,1), as otherwise we would not have the strands necessary to have the process studied.
Therefore, the polynomial is of order one in the range of interest, and so we confirm the behaviour described in the first line of equation \eqref{1pfeq:sis14gen}.
To finish this section let us study the case of the $\mathcal O^{- \dot -}_{(r)}$ operator.

\begin{figure}
\centering
\includegraphics[width=\textwidth/2]{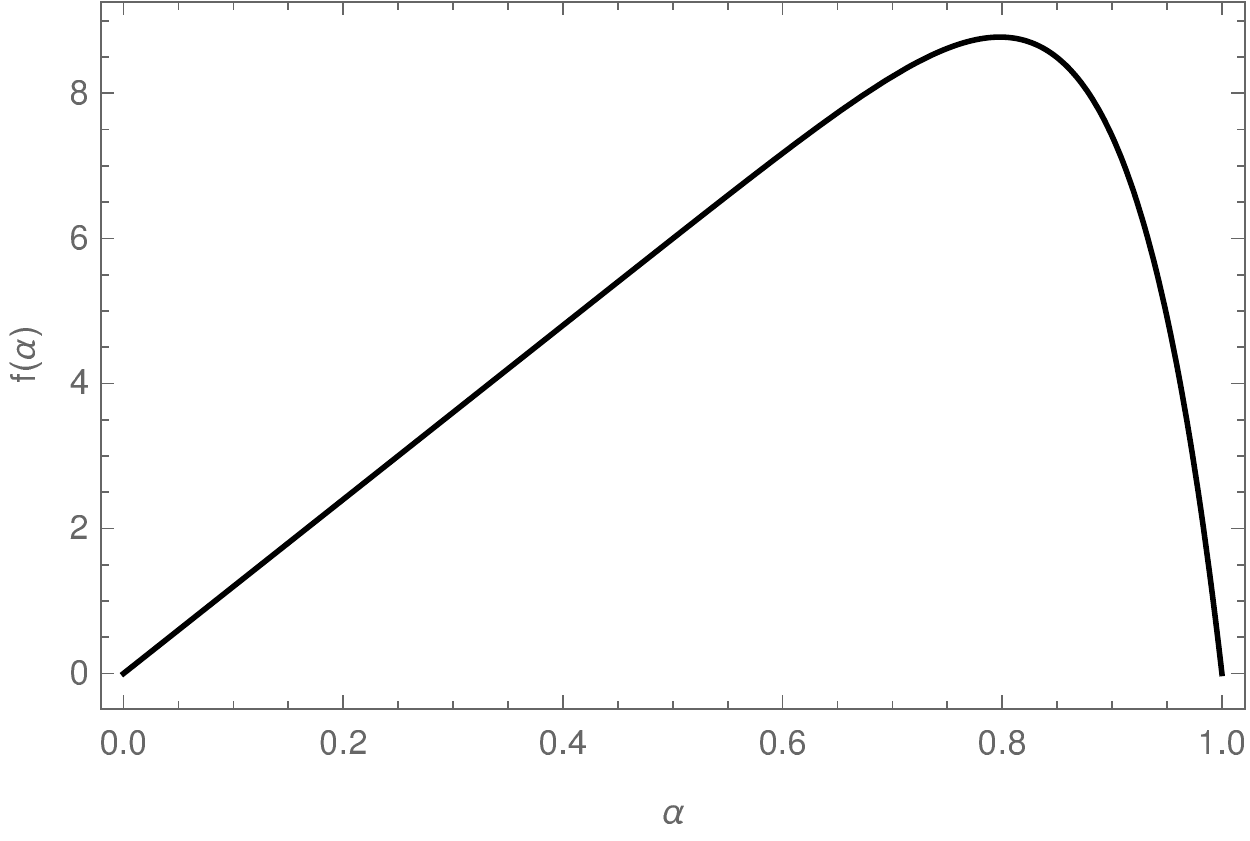}
\caption[$f(\al)$ for $\mathcal O^{- \dot -}$ in the 1/4-BPS case.]{Behaviour of the function $f(\al)$ for the untwisted $\braket{\mathcal O^{- \dot -}_{(r)}}$ one point function in the long strand case. As we can see we have a maximum of order one. Also, the function diverges for $\al = 1$ very rapidly, and so the divergence is not seen in this figure. In the plot $m = 12$.}
\label{fig:O14gen}
\end{figure}

\paragraph{$\mathcal O^{- \dot -}_{(r)}$ operator}
For the $\sum_{r = 1}^n \mathcal O^{- \dot -}_{(r)}$ operator consider the state
\begin{equation}
\psi = \sum_{p = 0}^m \left( A \ket{++}_{\frac N m} \right)^p \left( B \ket{00}_{\frac N m} \right)^{m - p}.
\end{equation}
Using the result \eqref{eq:denominator} and section \ref{1pf:O--14} we find that in this case the $f(\al)$ function is
\begin{equation}
f(\al) = \frac{\alpha  \left(-(m+1) \left(\frac{1}{1-\alpha }\right)^m (1-\alpha )^m+\alpha  m+1\right)-(\alpha -1) m \alpha ^m}{\alpha -1}.
\end{equation}
This function has an asymptote at $\al = 1$, but as always this is outside our range of interest.
Otherwise it is finite, being zero for $\al = 0$ and with a finite maximum (of order $m$) close to one.
We can see it in figure \ref{fig:O14gen} for $m = 12$.

\subsection{Three-charge states}
Consider now a three-charge state,
\begin{equation}
\psi_{\{N_{\ka ,m_{\ka}}^{(S)}\}} = \prod_{s = 1}^4 \prod_{\ka} (\ket{s}_{\ka})^{N_{\ka}^{(s)}} \prod_{\ka, m_{\ka}} \left(\frac{1}{m_{\ka}!} (J^+_{-1})^{m_{\ka}} \ket{00}_{\ka}\right)^{N_{\ka,m_{\ka}}^{(00)}}. \label{typ1/8}
\end{equation}
Its norm is
\begin{equation}
|\psi(\{A_\ka^{(s)}, B_{\ka, m_\ka}\})|^2 = \sum_{\{N_{\ka, m_\ka}^{(S)}\}} \mathcal N(\{N_{\ka, m_\ka}^{(S)}\}) \left(\prod_{s, \ka} |A_\ka^{(s)}|^{2N_\ka^{(s)}}\right) \left(\prod_{\ka, m_\ka} |B_{\ka, m_\ka}|^{2N_{\ka, m_\ka}^{(00)}} \right),
\end{equation}
where
\begin{equation}
\mathcal N(\{N_{\ka, m_\ka}^{(S)}\}) = \left(\frac{N!}{\prod_{s, \ka} N_\ka^{(s)}! \ka^{N_\ka^{(s)}}}\right)\left(\frac{1}{\prod_{\ka, m_\ka} N_{\ka, m_\ka}^{(00)}! \ka^{N_{\ka, m_\ka}^{(00)}}}\right) \prod_{\ka, m_\ka} \binom{\ka}{m_\ka}^{N_{\ka, m_\ka}^{(00)}}.
\end{equation}
Notice that due to the insertions of the $J^+_{-1}$ mode the $N$ dependence of the norm is now more complicated.
As in the two-charge case, we will first explain the approximation, and then we will go case by case giving the answers to the one point functions in this limit.

\subsubsection{Method}
The condition for the coefficients now reads \cite{Bena:2017xbt}
\begin{equation}
\sum_{gs, \ka} |A^{(gs)}_{\ka}|^2 + \sum_{\ka, m_{\ka}} \binom{\ka}{m_{\ka}} |B_{\ka, m_{\ka}}|^2 = N,
\end{equation}
and so, in addition to
\begin{equation}
|A^{(gs)}_i| = N \al_i^{(gs)}, \qquad \mathrm{with} \qquad \al_1^{(gs)} \in (0,1), \,\,\, \al_2^{(gs)} \in \left( 0, 1 - \al_1^{(gs)} \right), \,\,\, ...
\end{equation}
we need to set a parametrisation for the $|B_{\ka, m_{\ka}}|$ coefficients.
This parametrisation is more subtle, as we need to consider things like
\begin{equation}
|B_{\ka, m_{\ka}}|^2 = \frac{N}{\binom{\ka}{m_\ka}} \be_{\ka, m_{\ka}},
\end{equation}
with the $\be_{\ka, m_{\ka}}$ parameter running from zero to possibly a number bigger than one.
Notice that, since we are in the long strand case, $\ka$ will be of order $N$, and $m_\ka$ can take any value between one and $\ka$.
We work this out explicitly for each case in the following sections.

So, if we absorb these extra factors in the $\be_{\ka, m_{\ka}}$, plug all formulas in the norm of the state and write the strands length as $N/n_{\ka}$ instead of $\ka$ we obtain
\begin{equation}
|\psi|^2 = N! \sum_{\{N_{\ka, m_{\ka}}^{(S)}\}} \frac{\prod_{\ka, S} n_{\ka}^{N_{\ka}^{(S)}} \left( \al_{\ka}^{(S)} \right)^{N_{\ka}^{(S)}} \prod_{\ka, m_{\ka}} n_{\ka}^{N^{(00)}_{n_{\ka}, m_{\ka}}} \be_{n_{\ka}, m_{\ka}}^{N_{\ka, m_{\ka}}^{(00)}}}{\prod_{\ka, S} N_{n_{\ka}}^{(S)}! \prod_{\ka, m_{\ka}} N_{n_{\ka}, m_{\ka}}^{(00)}!} \prod_{\ka, m_{\ka}} \binom{\frac{N}{n_{\ka}}}{m_{\ka}}^{N_{\ka, m_{\ka}}^{(00)}}.
\end{equation}
As we said above, we have $N$-dependence inside of the sum in this case.
In some cases we will still be able to obtain an exact result, as we will be able to perform the sums.
However, in other cases we will not be able to do so.
Them, since the resulting polynomial will have now powers of $N$ in the factors, what we will do is keep only the dominant term in the polynomial and compare the numerator and the denominator one.
As shown in the previous cases, the denominator will always have a higher power of $N$ than the numerator, and so the $f$ function will be small.
We start with a simple example to see explicitly how it works, and then we will give more general results for the rest of the operators.

\subsubsection{Elementary example}\label{subsubsec:simplestOlong}
We start with a simple example, just as in the previous section, and then we build up to more general ones.
For this initial example we consider the untwisted $\mathcal O^{+ \dot -}_{(r)}$ operator.

\paragraph{\texorpdfstring{$\mathcal O^{+ \dot -}_{(r)}$}{O+-} operator}
Consider the state
\begin{equation}
\psi = \sum_{p = 0}^2 \left( A \ket{++}_{\frac N 2} \right)^p \left( B J^+_{-1} \ket{00}_{\frac N 2} \right)^{2 - p}
\end{equation}
which has norm
\begin{equation}
|\psi|^2 = \sum_{p = 0}^2 |A|^{2p} |B|^{2(2 - p)} \mathcal N(p), \qquad \mathrm{with} \qquad \mathcal N(p) = \frac{N!}{p! \left(\frac N 2\right)^p (2 - p)!}.
\end{equation}
We want to find the one point function of $\sum_{r} \mathcal O^{+ \dot -}_{(r)}$.
This operator transforms a strand $\ket{++}_{\frac N 2}$ into a strand $J^+_{-1} \ket{00}_{\frac N 2}$, and so we have
\begin{equation}
_{\frac N 2}\hspace*{-3pt}\bra{00} J^+_{-1} \sum_{r = 1}^{\frac N 2} \mathcal O^{+ \dot -}_{(r)} \ket{++}_{\frac N 2} \sim _{\frac N 2}\hspace*{-3pt}\bra{00} \sum_{r = 1}^{\frac N 2} \mathcal O^{- \dot -}_{(r)} \ket{++}_{\frac N 2} = e^{i\frac{\sqrt 2 v}{R}} {}_{\frac N 2}\braket{00|00}_{\frac N 2}.
\end{equation}
The combinatorial factor $\al$ in this case is
\begin{equation}
\al = \frac{(3 - p)!}{(2 - p)!}\frac 2 N,
\end{equation}
and thus we find
\begin{equation}
\braket{\mathcal O^{+ \dot -}_{(r)}}_{\mathrm{long}} = |\psi|^{-2} A \bar B \left[ |A|^2 \frac 2 N e^{i \frac{\sqrt 2 v}{R}} |\psi '|^2 + |B|^2 \frac 4 N e^{i\frac{\sqrt 2 v}{R}} |\psi ''|^2 \right],
\end{equation}
where
\begin{equation}
\psi ' = \left( \sum_r J^+_{-1 (r)} \ket{00}_{\frac N 2}\right) \ket{++}_{\frac N 2}, \qquad \psi '' = \left(\left(\sum_r J^+_{-1 (r)}\right) \ket{00}_{\frac N 2}\right)^2
\end{equation}
and so
\begin{equation}
|\psi '|^2 = \frac{N!}{\frac N 2}, \qquad |\psi ''|^2 = \frac{N!}{2}.
\end{equation}
Plugging these expressions in yields
\begin{equation}
\braket{\mathcal O^{+ \dot -}_{(r)}}_{\mathrm{long}} = \braket{\mathcal O^{+ \dot -}_{(r)}}_{\mathrm{short}} \frac{\frac 4 N \left( \frac{|A|^2}{N} + \frac{|B|^2}{2} \right)}{\frac{|B|^4}{2} + \frac 2 N |A|^2 |B|^2 + \frac{2}{N^2} |A|^4},
\label{1pfeq:O+-longcorr}
\end{equation}
where the short one point function is the one we computed before and is given by equation \eqref{eq:oshortgeneral}
\begin{equation}
\braket{\mathcal O^{+ \dot -}_{(r)}}_{\mathrm{short}} = A \bar B e^{i\frac{\sqrt 2 v}{R}}.
\end{equation}
Now, as mentioned above we use the relation
\begin{equation}
|A|^2 + \frac N 2 |B|^2 = N
\end{equation}
and we set
\begin{equation}
|B|^2 = \al, \qquad \al \in (0,2).
\end{equation}
Then we obtain
\begin{equation}
\braket{\mathcal O^{+ \dot -}}_{\mathrm{long}} = \frac 2 N \braket{\mathcal O^{+ \dot -}}_{\mathrm{short}}.
\end{equation}
Therefore, we see that in this initial case we obtain exactly the expected relation between both one point functions.
Next we start calculating one point functions for generic cases.

\subsubsection{Untwisted \texorpdfstring{$\mathcal O^{- \dot -}_{(r)}$}{O} operator}
\begin{figure}
\centering
\includegraphics[width=6cm]{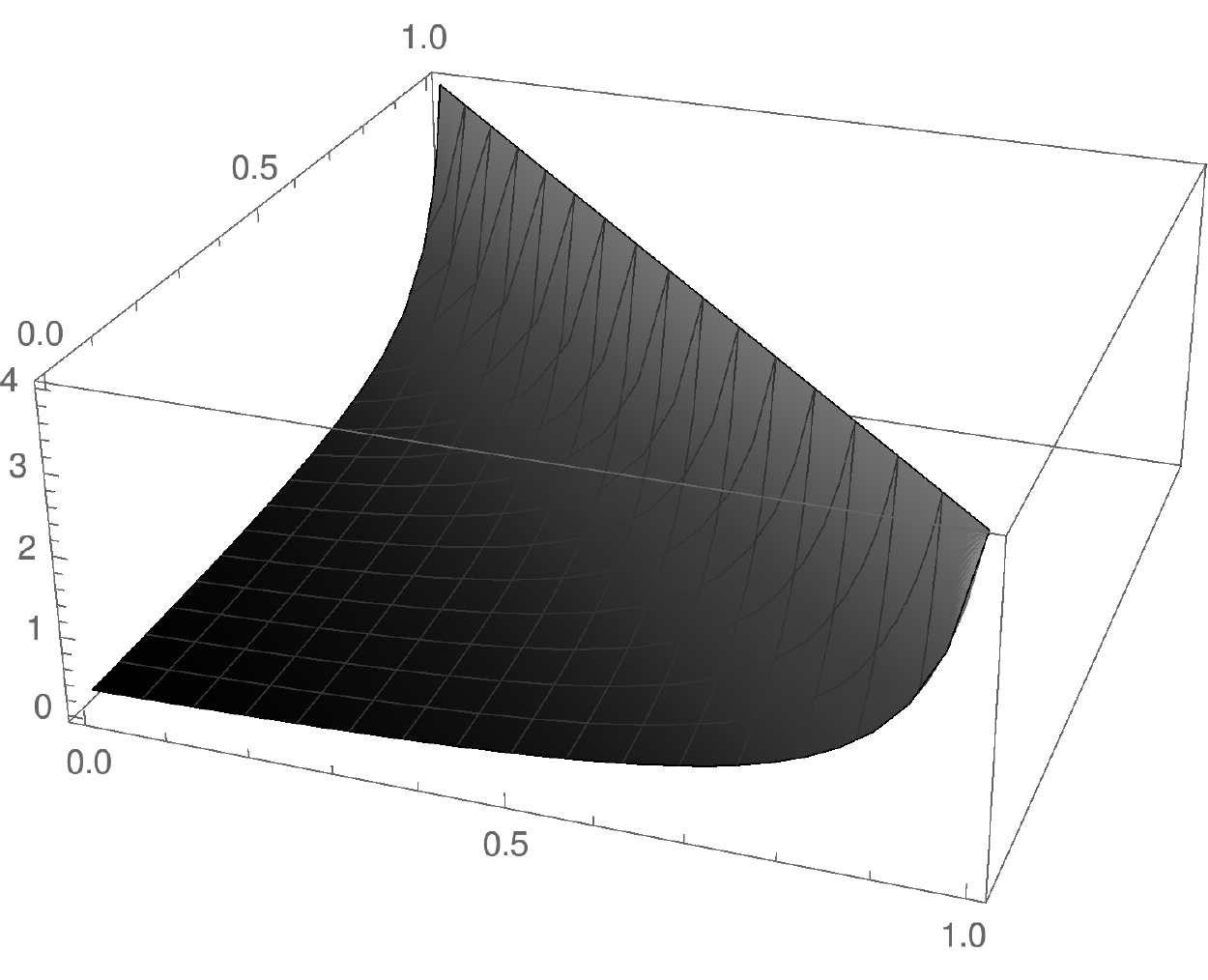}
\caption[1-pf for $\mathcal O^{- \dot -}$ in the long strand case]{Graphical representation of the numerical coefficient for the $\mathcal O^{- \dot -}$ one point function in the long strand case. In the plot, $n = 4$.}
\label{fig:Oolonggeneral}
\end{figure}

The aim of this section is to check that we recover the result from the two-charge case, by considering the same process as in the 1/4-BPS section but with a 1/8-BPS state now.
Consider the state
\begin{equation}
\psi = \sum_{N^{(++)} = 0}^{n} \sum_{N^{0(00)} = 0}^{n - N^{(++)}} \left( A^{(++)} \ket{++}_{\frac N n} \right)^{N^{(++)}} \left( B^{0(00)} \ket{00}_{\frac N n} \right)^{N^{0(00)}} \left( B^{1(00)} J^+_{-1} \ket{00}_{\frac N n} \right)^{N^{1(00)}}
\end{equation}
We want to calculate the one point function in the long strand case for the $\mathcal O^{- \dot -}$ operator.
The only on-trivial action of this operator on the state is when it performs the following transformation of strands,
\begin{equation}
\left( \sum_{r = 1}^{N/n} \mathcal O^{- \dot -}_{(r)} \right) \ket{++}_{\frac N n} \to \ket{00}_{\frac N n}.
\end{equation}
The $\al$ coefficient in this case is
\begin{equation}
\al = N^{0(00)} + 1,
\end{equation}
and after setting, as usual,
\begin{equation}
|A^{(++)}|^2 = \al N, \qquad |A^{0(00)}|^2 = \beta N, \qquad |A^{1(00)}|^2 = n(1 - \al - \beta)
\end{equation}
with
\begin{equation}
0 < \al < 1, \qquad 0 < \beta < 1 - \al
\end{equation}
we obtain
\begin{align}
&\braket{O^{- \dot -}}_{\mathrm{long}} = \frac{A^{(++)} \overline{A^{0(00)}}}{N}\cdot\nonumber \\
\cdot & \frac{\sum_{N^{(++)} = 0}^{n - 1} \sum_{N^{0(00)} = 1}^{n - N^{(++)}} N^{0(00)} \al^{N^{(++)}} \beta^{N^{0(00)} - 1} (1 - \al - \beta)^{N^{1(00)}} \frac{n^{N^{1(00)} - N^{(++)} - N^{0(00)}}}{N^{(++)}! N^{0(00)}! N^{1(00)}!}}{\sum_{N^{(++)} = 0}^{n} \sum_{N^{0(00)} = 0}^{n - N^{(++)}} \al^{N^{(++)}} \beta^{N^{0(00)}} (1 - \al - \beta)^{N^{1(00)}} \frac{n^{N^{1(00)} - N^{(++)} - N^{0(00)}}}{N^{(++)}! N^{0(00)}! N^{1(00)}!}}
\label{1pfeq:O--longres}
\end{align}
Mathematica can perform the sum ratio, giving
\begin{equation}
\braket{O^{- \dot -}}_{\text{long}} = \frac{A^{(++)} \overline{A^{0(00)}}}{N} \frac{n}{\alpha + \beta - n^2 (\alpha +\beta -1)}
\end{equation}
We present in figure \ref{fig:Oolonggeneral} this result for the second factor plotted.
We recover the expected $1/N$ behaviour except in the boundary values, as in the previous cases.

\subsubsection{\texorpdfstring{$\Si_2^{+ \dot -}$}{Si2+-} operator}\label{1pfsubsec:si+-18long}
This is the only twist operator one point function left to calculate in the long strand case, as we obtained results for the $\Si^{- \dot -}_n$ in the section \ref{1pfsubsec:exactsin--}.
Also, let us recall from section \ref{1pfsubsec:sinresult} that for the twist $n$, 1/8-BPS one point function we need an extra commutator to obtain it, which we are not giving in this paper.
Thus, we calculate the twist two case, as we did in section \ref{1pfsubsec:shortreview} for the short strand case.
Consider the state
\begin{equation}
\psi = \sum_{N^1 = 0}^m \sum_{N^0 = 0}^{m_2 \left( 1 - \frac{N^1}{m} \right)} \left( A \ket{++}_{\frac{N}{m_1}} \right)^{N^+} \left( B \ket{00}_{\frac{N}{m_2}} \right)^{N^0} \left( B^1 J^+_{-1} \ket{00}_{\frac N m} \right)^{N^1},
\end{equation}
where, as in the short strand case,
\begin{equation}
\frac{1}{m_1} + \frac{1}{m_2} = \frac 1 m \qquad \Longrightarrow \qquad m = \frac{m_1 m_2}{m_1 + m_2}
\end{equation}
and
\begin{equation}
\mathcal N = \frac{N!}{N^+! N^0! N^1! \left( \frac{N}{m_1} \right)^{N^+} \left( \frac{N}{m_2} \right)^{N^0}}.
\end{equation}
Analogous to section \ref{1pfsubsec:shortreview}, we have
\begin{equation}
\al = \frac{N^1 + 1}{2}
\end{equation}
and
\begin{equation}
c_2 = \frac{m_1 + m_2}{2N},
\end{equation}
which lead to the result
\begin{align}
& \braket{\Si_2^{+ \dot -}}_{\text{long}} = N_{\Si_2} \frac{e^{i\frac{\sqrt 2 v}{R}}}{4} \frac{A B \bar{B^1}}{N} \cdot \nonumber \\
& \cdot \frac{\sum_{N^1 = 1}^m \sum_{N^0 = 0}^{m_2 \left( 1 - \frac{N^1}{m} \right) - 1} (m_1 + m_2) m^{N^1} \al^{N^+} \be^{N^0} (1 - \al - \be)^{N^1} \frac{m_1^{N^+} m_2^{N^0}}{N^+! N^0! N^1!}}{\sum_{N^1 = 0}^m \sum_{N^0 = 0}^{m_2 \left( 1 - \frac{N^1}{m} \right)} m^{N^1} \al^{N^+} \be^{N^0} (1 - \al - \be)^{N^1} \frac{m_1^{N^+} m_2^{N^0}}{N^+! N^0! N^1!}}
\end{align}
These sums cannot be obtained with Mathematica, and so we need to approximate them.
However, we do not need the exact result; the leading behaviour with $N$ is enough.
Since the sums only have factors of order one within them, that is, since they do not have any $N$'s, their ratio will be a number much smaller than $N$, away from the boundaries of $\al$ and $\be$.
Therefore, we can approximate this one point function by
\begin{equation}
\braket{\Si^{+ \dot -}_2}_{\text{long}} \approx N_{\Si_2} \frac{e^{i\frac{\sqrt 2 v}{R}}}{4} \frac{A B \bar{B^1}}{N},
\label{1pfeq:si+-218approx}
\end{equation}
{\it i.e.} once again we learn that
\begin{equation}
\braket{\Si^{+ \dot -}_2}_{\text{long}} \approx \frac 1 N \braket{\Si_2^{+ \dot -}}_{\text{short}}.
\label{1pfeq:si2longshortrel}
\end{equation}

\subsubsection{Untwisted \texorpdfstring{$\mathcal O^{+ \dot -}_{(r)}$}{O+-} operator in full generality}\label{1pfsubsec:1/8statefull}
In this section we calculate the one point function of $\sum_{r = 1}^n \mathcal O^{+ \dot -}_{(r)}$ with the most general 1/8-BPS state.
That is, we consider the state
\begin{equation}
\psi = \sum_{\{\mathcal N\}} \left[ \left(A \ket{++}_{\frac{N}{\ka}} \right)^{N^+} \left( \prod_{m_{\ka}} B_{m_{\ka}} \frac{\left( J^+_{-1}\right)^{m_{\ka}}}{m_\ka!} \ket{00}_{\frac{N}{\ka}} \right)^{N^0_{\ka, m_{\ka}}} \right].
\label{1pfeq:longOgenstate}
\end{equation}
As we will see in a moment, the equations in this case cannot be written out in a closed form.
However, we will still be able to give a result.
We will work out the example $\ka = 4$, and the general result follows easily from induction using the same process.
So, let us consider $\ka = 4$.
The state $\psi$ in this case reads
\begin{align}
\psi =& \left( A \ket{++}_{\frac N 4} \right)^4 + \sum_{p = 0}^{3} \left( A \ket{++}_{\frac N 4} \right)^p \otimes \nonumber \\
& \otimes \left[ \sum_{i_1 = 1}^{\frac N 4} ... \sum_{i_{4 - p} = i_{3 - p}}^{\frac N 4} \frac{B_{i_1}\cdot ... \cdot B_{i_{4 - p}}}{i_1!\cdot ... \cdot i_{4 - p}!} \left( J^+_{-1} \right)^{i_1} \ket{00}_{\frac N 4} \otimes ... \otimes \left( J^+_{-1} \right)^{i_{4 - p}} \ket{00}_{\frac N 4} \right].
\label{1pfeq:18bps4state}
\end{align}
We have not found any easy way to write the norm of this state in a compact form that is understandable, so we write it all out explicitly.
Once that is done, how terms are constructed in general will be clear from this example.
That is why we leave all the factorials explicit in what follows.
The norm for this state is
\begin{align}
\mathcal N = \frac{N!}{\left(\frac N 4 \right)^4} \Biggl\{& \frac{1}{4!} + \frac{1}{3!} \sum_{i = 1}^{\frac N 4} \binom{\frac N 4}{i} + \frac{1}{2!} \left( \frac{1}{2!} \sum_{i = 1}^{\frac N 4} \binom{\frac N 4}{i}^2 + \sum_{i = 1}^{\frac N 4} \sum_{j = i + 1}^{\frac N 4} \binom{\frac N 4}{i} \binom{\frac N 4}{j} \right) + \nonumber \\
& + \left( \frac{1}{3!} \sum_{i = 1}^{\frac N 4} \binom{\frac N 4}{i}^3 + \frac{1}{2!} \sum_{i = 1}^{\frac N 4} \sum_{j = i + 1}^{\frac N 4} \binom{\frac N 4}{i}^2 \binom{\frac N 4}{j} +\right. \nonumber \\
& \;\;\;\;\;\;\; \left. + \sum_{i = 1}^{\frac N 4} \sum_{j = i + 1}^{\frac N 4} \sum_{l = j + 1}^{\frac N 4} \binom{\frac N 4}{i} \binom{\frac N 4}{j} \binom{\frac N 4}{l} \right) + \nonumber \\
& + \left( \frac{1}{4!} \sum_{i = 1}^{\frac N 4} \binom{\frac N 4}{i}^4 + \frac{1}{3!} \sum_{i = 1}^{\frac N 4} \sum_{j = i + 1}^{\frac N 4} \binom{\frac N 4}{i}^3 \binom{\frac N 4}{j} + \frac{1}{2!} \frac{1}{2!} \sum_{i = 1}^{\frac N 4} \sum_{j = i + 1}^{\frac N 4} \binom{\frac N 4}{i}^2 \binom{\frac N 4}{j}^2 \right. + \nonumber \\
& \;\;\;\;\;\;\; \left. + \frac{1}{2!} \sum_{i = 1}^{\frac N 4} \sum_{j = i + 1}^{\frac N 4} \sum_{l = j + 1}^{\frac N 4} \binom{\frac N 4}{i}^2 \binom{\frac N 4}{j} \binom{\frac N 4}{l} + \right. \nonumber \\
& \;\;\;\;\;\;\; \left. + \sum_{i = 1}^{\frac N 4} \sum_{j = i + 1}^{\frac N 4} \sum_{l = j + 1}^{\frac N 4} \sum_{t = l + 1}^{\frac N 4} \binom{\frac N 4}{i} \binom{\frac N 4}{j} \binom{\frac N 4}{l} \binom{\frac N 4}{t} \right) \Biggr\}.
\label{1pfeq:18bps4norm}
\end{align}
From this equation we can deduce how the terms look like for general $\ka$.
Looking at the state \eqref{1pfeq:18bps4state} it is clear that, in the three-charge strands terms, for each term with $i_n$ sums we will have $p(n)$ terms, that is, the number of integer partitions of $n$ terms.
Each term will have as many sums as elements has every partition of $n$, and the binomial coefficient's powers correspond to the numbers in each partition (to its parts).
Take the last parenthesis in \eqref{1pfeq:18bps4norm} for instance.
We have four sums in the state, and so we have $p(4) = 5$ terms in the norm.
And the exponents of the binomials coefficients are given by the five possible partitions of four (see equation \eqref{1pfeq:intpartex} or figure \ref{fig:refiningpartitions4} for the partitions).
Let us also recall that the coefficients in the state satisfy
\begin{equation}
|A|^2 + \sum_{i = 1}^{\frac N 4} \binom{\frac{N}{\ka}}{m_{\ka}} |B_{m_\ka}|^2 = N.
\end{equation}
In this case we have all the possibilities for the different number of insertions of the $J^+_{-1}$ mode, and so we need to be more careful with the parametrisation of the equation above.
Expanding the sum and dividing all the equation by $N$ yields
\begin{equation}
1 = \frac{|A|^2 + |B_{\frac N 4}|^2}{N} + |B_1|^2 + |B_{\frac N 4 - 1}|^2 + N \left( |B_2|^2 + |B_{\frac N 4 - 2}|^2\right) + O(N^2).
\end{equation}
Recalling that we are in the large $N$ limit, all coefficients except the first ones must be very small.
Therefore, we can work with the parametrisation
\begin{equation}
|A|^2 = \al N, \qquad |B_i|^2 = \be_{\frac N 4} N, \qquad |B_1|^2 = \be_{1}, \qquad |B_{\frac N 4 - 1}|^2 = \be_{\frac N 4 - 1},
\label{1pfeq:18fourier4}
\end{equation}
and analogous for the rest coefficients.
To simplify the expressions later, we will absorb all the powers of $N$ of these parameters in the $\be_i$.
The running parameters labelled by the Greek letters taking values between zero and one, as usual.
As in previous cases we use the commutator \eqref{1pfeq:JOcommutator} to see that
\begin{equation}
_{\frac N 4} \bra{00} J^+_{-1} \sum_{r = 1}^{\frac N 4} \mathcal O^{+ \dot -}_{(r)} \ket{++}_{\frac N 4} = {}_{\frac N 4} \braket{00|00}_{\frac N 4} e^{i \frac{\sqrt 2 v}{R}}.
\end{equation}
We are just missing the $\al$ coefficient to be able to get the answer for the one point function.
Again, the expressions cannot be written easily in a compact way, so we go term by term to see the general result.
Let $\al_i$ be the coefficient for the $p = i$ term, where $p$ refers to the sum index in equation \eqref{1pfeq:18bps4state}.
Clearly
\begin{equation}
\al_4 = \frac 4 N.
\end{equation}
For the rest of the terms we will need to use the large $N$ limit.
We need to keep in mind that the operator $\mathcal O^{+ \dot -}_{(r)}$ only generates states with one insertion of the $J^+$ mode.
For $p = 3$ we have
\begin{equation}
3 \frac{N!}{\left( \frac N 4 \right)^4} \frac{1}{3!} \sum_{i = 1}^{\frac N 4} \binom{\frac N 4}{i} = \al_3 \frac 1 2 \frac{N!}{\left( \frac N 4 \right)^4} \left( \frac N 4 \sum_{i = 2}^{\frac N 4} \binom{\frac N 4}{i} + \frac 1 2 \left( \frac N 4 \right)^2 \right).
\end{equation}
Solving for $\al_3$ we obtain
\begin{equation}
\frac 4 N \left( 1 + \frac{\frac N 8}{\sum_{i = 1}^{\frac N 4} \binom{\frac N 4}{i} - \frac N 8 } \right) \approx \frac 4 N,
\end{equation}
where in the last step we used the large $N$ limit to approximate the second term in the parenthesis by zero.
Similarly, for $p = 2$ we find
\begin{equation}
\al_2 = \frac 4 N \left( 1 + \frac{\frac 2 3 \left( \frac N 4 \right)^2 + \frac 1 2 \frac N 4 \sum_{i = 2}^{\frac N 4} \binom{\frac N 4}{i}}{\frac{1}{2!} \sum_{i = 1}^{\frac N 4} \binom{\frac N 4}{i}^2 + \sum_{i = 1}^{\frac N 4} \sum_{j = i + 1}^{\frac N 4} \binom{\frac N 4}{i} \binom{\frac N 4}{j}} \right) \approx \frac 4 N,
\end{equation}
and an analogous expression for $p = 1$.
Therefore the value of $\al$ is the same for all terms, 
\begin{equation}
\al \approx \frac 4 N
\end{equation}
Since in the long strand limit $\ka$ (in equation \eqref{1pfeq:longOgenstate}) is a small number, we thus learn that in general
\begin{equation}
\al \approx \frac{\ka}{N}.
\end{equation}
We are now ready to calculate the one point function.
Again, the expressions are complicated, so we will be very explicit.
As usual, we have
\begin{equation}
\braket{\mathcal O^{+ \dot -}_{(r)}} = |\psi|^{-2} \bra{\psi} \sum_{r = 1}^{\frac N 4} \mathcal O^{+ \dot -}_{(r)} \ket{\psi}.
\end{equation}
As we mentioned above, we need to keep in mind that the operator only generates strands with one insertion of the R-symmetry current, and so we need to be careful with the contractions.
This also means that we do not have to worry about the factorials dividing when taking the modulus of the out state, as all factors that are not repeated in the bra and the ket and survive will be a one.
Keeping all this in mind and using the substitutions \eqref{1pfeq:18fourier4} we find
\begin{align}
\braket{\mathcal O^{+ \dot -}} \approx & |\psi|^{-2} A \bar B_1 e^{i \frac{\sqrt 2 v}{R}} \frac{4^5 N!}{N} \left\{ \frac{\al^3}{3!} + \al^2 \sum_{i = 1}^{\frac N 4} \be_i \binom{\frac N 4}{i} + \right. \nonumber \\
& + \al \left[ \frac 1 2 \sum_{i = 1}^{\frac N 4} \be_i^2 \binom{\frac N 4}{i}^2 + \sum_{i = 1}^{\frac N 4} \sum_{j = i + 1}^{\frac N 4} \be_i \be_j \binom{\frac N 4}{i} \binom{\frac N 4}{j} \right] + \nonumber \\
& + \frac{1}{3!} \sum_{i = 1}^{\frac N 4} \be_i^3 \binom{\frac N 4}{i}^3 + \frac 1 2 \sum_{i = 1}^{\frac N 4} \sum_{j = i + 1}^{\frac N 4} \be_i^2 \be_j \binom{\frac N 4}{i}^2 \binom{\frac N 4}{j} + \nonumber \\
& \left. \;\;\;\; + \sum_{i = 1}^{\frac N 4} \sum_{j = i + 1}^{\frac N 4} \sum_{k = j + 1}^{\frac N 4} \be_i \be_j \be_k \binom{\frac N 4}{i} \binom{\frac N 4}{j} \binom{\frac N 4}{k} \right\} =: |\psi|^{-2} A \bar B_1 e^{i \frac{\sqrt 2 v}{R}} \frac{4}{N} |\varphi|^2.
\end{align}
If we now let $|\psi_{\al}|^2$ be the polynomial resulting from the norm of $\psi$ after substituting with \eqref{1pfeq:18fourier4} without the common factors in front, we obtain
\begin{equation}
\braket{\mathcal O^{+ \dot -}} \approx \frac{4}{N} A \bar B_1 e^{i \frac{\sqrt 2 v}{R}} \frac{|\varphi|^2}{|\psi|^2}.
\end{equation}
And so, with the general observations done throughout this section we learn that, for general $\ka$,
\begin{equation}
\left< \sum_{r = 1}^{\frac{N}{\ka}} \mathcal O^{+ \dot -}_{(r)} \right> \approx \frac{\ka}{N} A \bar B_1 e^{i \frac{\sqrt 2 v}{R}} \frac{|\varphi|^2}{|\psi|^2}.
\label{1pfeq:olonggeneral}
\end{equation}
Comparing this result to its short strand case counterpart, equation \eqref{eq:oshortgeneral} we see that, up to numbers of order one,
\begin{equation}
\frac{\braket{\mathcal O^{+ \dot -}}_{\mathrm{long}}}{\braket{\mathcal O^{+ \dot -}}_{\mathrm{short}}} \approx \frac{1}{N} f(N, \al, \be_i),
\end{equation}
where $f$ is the polynomial described above.
Notice that, maybe except for boundary values of the $\al$, $\be_i$ variables, the polynomial $f$ is a very small number, as the denominator has terms with higher powers of $N$ than the numerator.
Therefore, we see once again that the long strand one point function is parametrically smaller in $N$ than the short strand one, \emph{i.e.} at most
\begin{equation}
\braket{\mathcal O^{+ \dot -}_{(r)}}_{\text{long}} \approx \frac{1}{N} \braket{\mathcal O^{+ \dot -}_{(r)}}_{\text{short}}.
\label{1pfeq:O+-longshortrel}
\end{equation}
Let us finish this section with the calculation of the one point function for the modes of the R-symmetry current.

\subsubsection{Untwisted \texorpdfstring{$\left( J^+_{-1}\right)^m$}{J+-1m} operator}
\begin{figure}
\centering
\includegraphics[width=\textwidth/2]{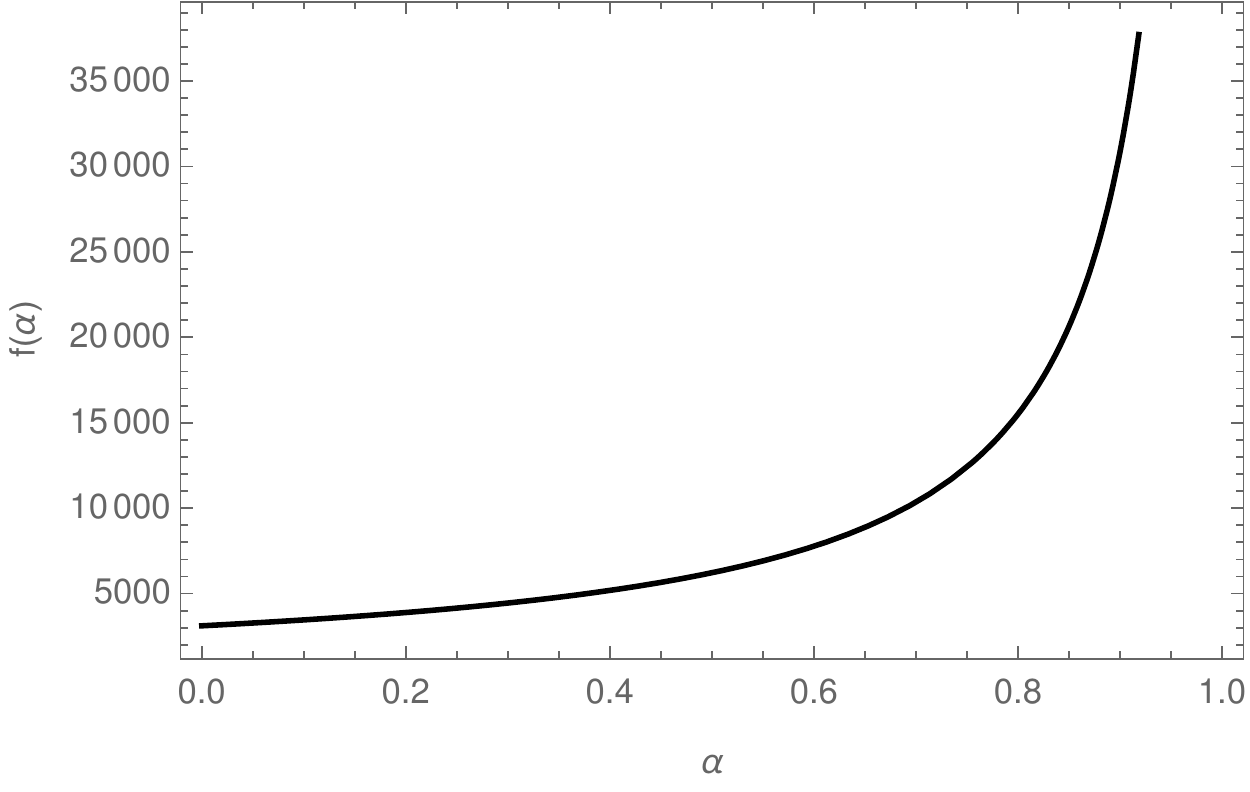}
\caption[Answer for the $\left( J^+_{-1} \right)^m$ one point function in the long strand case.]{Behaviour of the function $f(\al)$ for the $\left(J^+_{-1}\right)^m$ one point function in the long strand case. There is no $N$ suppression in this case. In the plot, $N = 100$, $n = 9$, $m = 4$. The function has the same behaviour for all the values of the parameters.}
\label{1pffig:Jlong}
\end{figure}

We will now calculate the same one point function as in section \ref{1pfsubsec:Jshort}.
Just as in the short strand case we also calculate all $m-$point functions for $m \leq n$, as the calculation is analogous in all cases.
Thus, consider the state
\begin{equation}
\psi = \sum_{p = 0}^n \left( B \ket{00}_{\frac N n} \right)^p \left( B_m \frac{\left( J^+_{-1} \right)^m}{m!} \ket{00}_{\frac N n} \right)^{n - p}.
\end{equation}
As before, we have
\begin{equation}
\mathcal N(p) = \frac{N!}{\left( \frac N n \right)^n} \frac{1}{p! (n - p)!} \binom{\frac N n}{m}^{n - p}
\end{equation}
and
\begin{equation}
\al = n - p + 1.
\end{equation}
Setting $|B|^2 = \al N$ as usual (where let us remind that this $\al$ is unrelated to the one above), and recalling that in this case we have
\begin{equation}
|B|^2 + \binom{\frac N n}{m} |B_m|^2 = N,
\end{equation}
we see that we can sum exactly the expressions for the numerator and denominator.
Notice that we can obtain the result for any $m$, even with the combinatorial factor of the equation above, as this factor will just increase the power of the corresponding factor in the norm of the state.
The exact result for this one point function is thus
\begin{equation}
\left< \left( J^+_{-1} \right)^m \right> = B \bar B_m \frac{n \binom{\frac{N}{n}}{m}^3}{\alpha  N-(\alpha -1) \binom{\frac{N}{n}}{m}^2} =: B \bar B_m f(\al).
\label{1pfeq:Jmlongexactres}
\end{equation}
As we can see in figure \ref{1pffig:Jlong} $f(0) = n \binom{\frac N n}{m}$, and then it rapidly increases until we get to $\al = 1$, where $f(1) = \frac n N \binom{\frac N n}{m}^3$.
Let us compare this result with the short strand case one, equation \eqref{1pfeq:Jshort}.
Since the result for that case was independent of $N$, we see that in this case the long strand result is much bigger than the short strand one, for any value of the parameters.

\section{Brief discussion of the results}\label{1pfsection:discussion}
Now that we have calculated some exemplary one point functions for short and long strands we can compare the results we obtained.
We give a summary of the results and of the relations between them in section \ref{section:resultreview}, but let us outline the main highlights here.

Let $O$ be a chiral primary of the theory.
The main result that we have obtained so far in this paper is that
\begin{equation}
\braket{O}_{\text{long}} \sim \frac{\braket{O}_{\text{short}}}{N^r},
\end{equation}
where $r \geq 1$.
This means that $\braket{O}_{\text{long}}$ is comparable to the supergravity corrections, that is, to higher derivative modes.
Indeed, one could think of altering the states with which we calculate the one point functions to get bigger results, but then we are only rephrasing the issue.
In that case we would need to act on the states with operators which are not in the supergravity theory.
Therefore, the conclusion of this first part is that one point functions of chiral primary operators in the long strand limit are suppressed by powers of $N$ with respect to the short strand ones.

\section{Different ways of joining strands}\label{section:joinings}
\begin{figure}[t]
\begin{subfigure}{0.3\textwidth}
\centering\includegraphics[height=3.5cm]{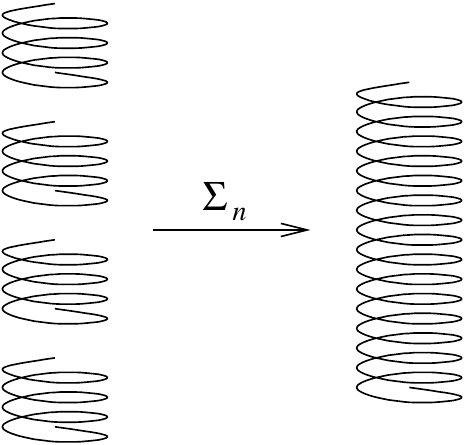}
\caption{All in one step}
\label{fig:sigman}
\end{subfigure}\hfill
\begin{subfigure}{0.7\textwidth}
\centering\includegraphics[height=3.5cm]{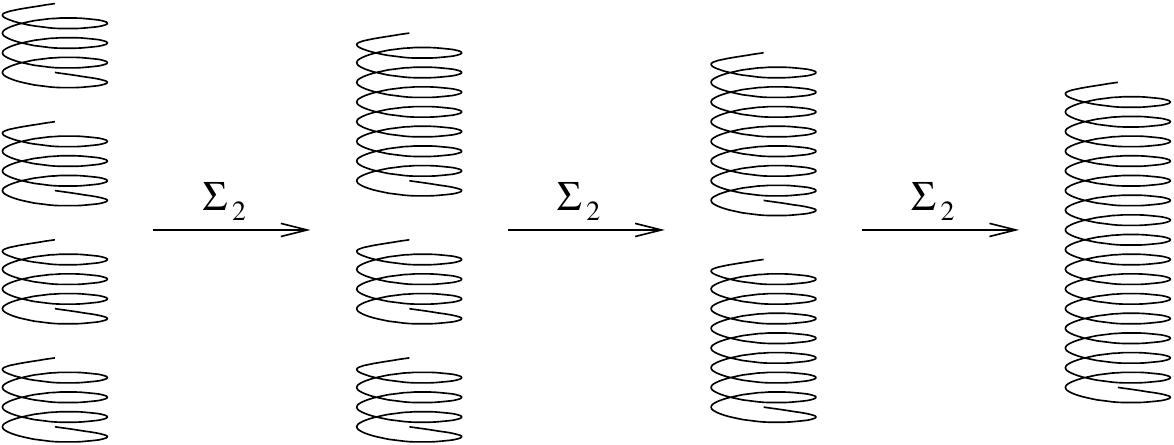}
\caption{Two by two}
\label{fig:sigma2}
\end{subfigure}
\caption[Joining of strands using different twists.]{Joining of strands using different operators. In (a) we join them using a single operator, $\Si_n$, which joins them all in one step. In (b) we join them two by two, in $n-1$ steps.}
\label{fig:sigmas}
\end{figure}

In the previous sections we have calculated correlation and one point functions for single trace chiral primary operators.
In this section we take a slightly different direction, and study all the possible ways to join strands.
To do so, we compare two values: the one point function $\langle\Si_n^{- \dot -}\rangle$ and the one point function of multi-particle states $\langle(\Si_2^{- \dot -})^{n-1}\rangle$.
The first one translates in joining $n$ strands (of any length) together at the same time, with one step.
The second one consists in joining the $n$ strands in $n-1$ steps, namely, joining them two by two.
Both cases represented in figure \ref{fig:sigmas}.

As is obvious from the figure, in the first case, subfigure \ref{fig:sigman}, the number of ways in which the joining can be done is much smaller, as it only depends where within each strand the operator is inserted.
However, in the other case, subfigure \ref{fig:sigma2}, there are many ways to join, as the second application of the gluing operator could have joined the pair which was joined in the previous step to one of the two strands that was alone for instance.
Hence, when joining the strands two by two the combinatorics are more complicated and will give a non-trivial contribution.
We present work in this section which aims to compare both cases.
Normalisations are the key point for this calculation, and they are related to integer partitions.
The different countings are obtained in section \ref{subsection:sigma2corfun}.

\subsection{Setting up}\label{joiningsec:setup}
The procedure of this section is the same as in the previous one.
The state that we start with in both cases is
\begin{equation}
\psi = \sum_{p=0}^{\frac{nN}{M}} \left(A\ket{++}_{\frac M n}\right)^p \left(B\ket{++}_M\right)^{\frac N M -\frac p n}.
\end{equation}
The norm of this state is
\begin{equation}
\left|\psi\right|^2 = \sum_{p = 0}^{\frac{nN}{M}} |A|^{2p} |B|^{\frac N M - \frac p n} \mathcal N(p),
\end{equation}
where
\begin{equation}
\mathcal N(p) = \frac{N!}{p! \left(\frac N M - \frac p n\right)! \left(\frac M n\right)^p M^{\frac N M - \frac p n}}.
\end{equation}

We will be able to give an exact answer for the relation between the one and the $n$-point functions, as both describe the same process at the end.
Namely, all the calculations will be exactly the same, except for the $\al$ and $c_n$ coefficients.
Furthermore, the two results will only differ in factors proportional to $n$ and $M$, and so all other contributions will cancel.
We give more details and the results at the end of the section, after we have done all the combinatorics.
In what follow we assume $n \gg 1$ but not necessarily comparable to $N$.
The final result will be valid for any $n \gg 1$.

\begin{figure}
\centering
\includegraphics[height=3.5cm]{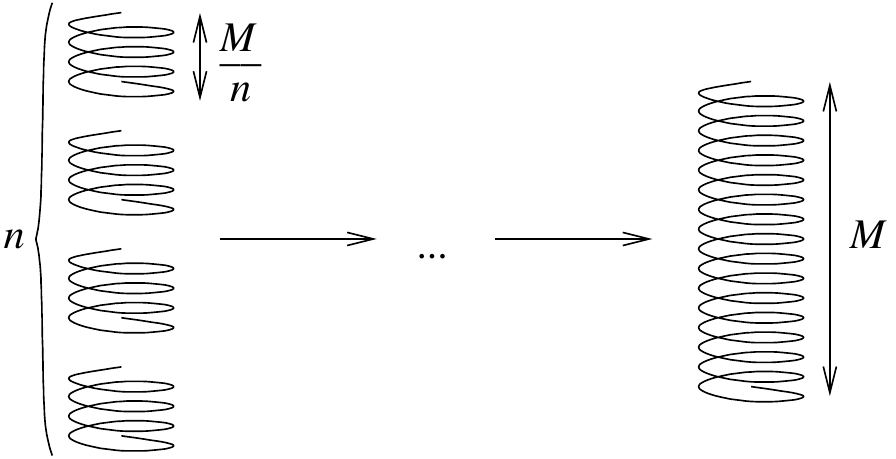}
\caption[Initial and final states of the gluing process.]{Initial and final states of the gluing process. The process starts with $n$ strands of length $M/n$ and finishes with all of them together, {\it i.e.} with a single strand of length $M$.}
\label{fig:strandsstructure}
\end{figure}

In figure \ref{fig:strandsstructure} we present a cartoon to visualise the state and the gluing process.
Drawing only the strands involved in the gluing, we start with $n$ strands of length $M/n$ and end up with a single strand of length $M$.
In both cases we need the $\Si^{- \dot -}$ operator, so that the final state has the correct charges.
In this section we write out the normalisations of the twist operators explicitly, as they play an important role.
Let us now compute the $\al$ coefficient and the $c$ coefficient for both cases, to be able to compare both results.

\subsection{\texorpdfstring{$\braket{ \Si_n^{- \dot -}}$}{sigman} coefficients}\label{subsection:sigmancorfun}
We have done this calculation in section \ref{1pfsubsec:sigman}; let us just rewrite the result here to have it at hand.
The $\al$ coefficient is
\begin{equation}
\al = \frac{M}{n!} \left( \frac N M - \frac p n + 1 \right),
\end{equation}
and the $c_n$ coefficient is
\begin{equation}
c_n = M^{1-n} n^{\frac{M}{2} + \frac{n}{2} - \frac{M}{n}} \left (  (n-1)^2 (n \bar{a})^{(n-2)} \Lambda^{-1} \right )^{\frac{1}{2} (n-1) \left (1 - \frac{M}{n} \right )}
\label{1pfeq:cn2}
\end{equation}
with
\begin{equation}
\Lambda = \left ( 1 + n (n-1)^{(n-1)} - n^{n-1} \right )
\end{equation}
and
\begin{equation}
\bar{a}^{n-1} = n \left ( 1 - \frac{1}{n} \right )^{n-1} - 1.
\end{equation}
Let us now calculate these factors for the $(n-1)$-point function, which will have more complicated combinatorics.

\subsection{Calculation of \texorpdfstring{$\langle(\Si_2^{- \dot -})^{n-1}\rangle$}{sigma2}}\label{subsection:sigma2corfun}
\begin{figure}
\centering
\includegraphics[height=3.5cm]{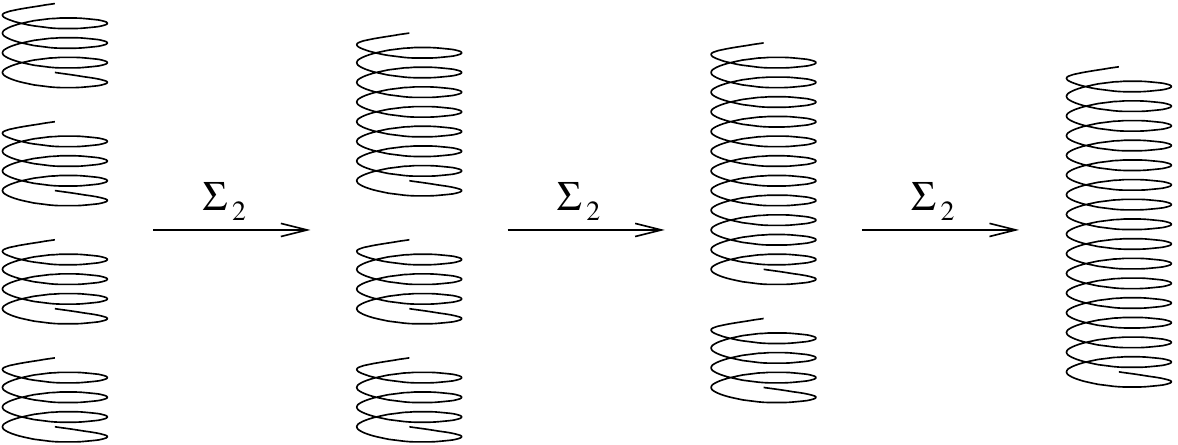}
\caption[Joining the strands two by two.]{Joining the strands two by two. The order in which we do it here is the opposite from figure \ref{fig:sigma2}. Here we keep gluing all the strands to one same strand, which keeps growing while the rest remain untouched.}
\label{fig:sigma2t}
\end{figure}

The initial and final states are, of course, the same ones as before, and thus we have
\begin{equation}
\left(\Si^{- \dot -}_2\right)^{n - 1} \left[\left(\ket{++}_{\frac M n}\right)^p \left(\ket{++}_M\right)^{p_2}\right] = c_{n2b2} \al \left[\left(\ket{++}_{\frac M n}\right)^{p-n} \left(\ket{++}_M\right)^{p_2 + 1}\right].
\end{equation}
We need the coefficients $c_{n2b2}$ and $\al$.
However, looking at the picture \ref{fig:sigma2} again we notice that this case is more subtle.
In the picture we joined strands two by two, however we chose a particular ordering.
That is, we joined all strands by pairs, to have $n/2$ strands of length $2M/n$, then we joined these in pairs again to have $n/4$ strands of length $4M/n$, then to have $n/8$ strands of length $8M/n$, and so on until the end, where we have only one strand of length $M$.
Clearly, this is only one way of doing the gluing process.
Instead, we could have joined all new strands to the same one, creating an increasingly long strand, and leaving the rest untouched.
We depict this process in figure \ref{fig:sigma2t}, again for $n = 4$.

The final result for the coefficients may depend on the way in which we join the strings.
We calculate the two limit cases, which are the ones depicted in figures \ref{fig:sigma2} and \ref{fig:sigma2t}.
In what follows we find which one gives the leading contribution, we check in how many ways the gluing can be done for general $n$ and then we approximate the result for the $(n - 1)$-point function, obtaining a lower bound and an estimate for the upper bound.

\subsubsection{\texorpdfstring{$c_{n2b2}$}{cn2b2} coefficient}
We start the calculation by obtaining the $c$ coefficient.
First let us recall what the coefficient for two strands is.
As we said in the previous sections, it was first computed in \cite{Carson:2014ena}, and in our case at hand is
\begin{equation}
c_2 = \frac{\frac M n + \frac M n}{2 \frac M n \frac M n} = \frac n M.
\end{equation}
The $c$ coefficient depends on the length of both strands that are being joined together, and so it might different for each case.
Let us do first the case depicted in figure \ref{fig:sigma2t} which we denote by $c_{2b2t}$.
In this case we simply need to apply that coefficient repeatedly, with one of the strands increasing in length one by one at each step.
The product gives
\begin{equation}
c_{n2b2t} = \frac{\frac M n + \frac M n}{2 \frac M n \frac M n} \cdot \frac{\frac{2M}{n} + \frac M n}{2 \frac{2M}{n} \frac M n} \cdot\frac{\frac{3M}{n} + \frac M n}{2 \frac{3M}{n} \frac M n} \cdot ... \cdot \frac{\frac{(n - 1)M}{n} + \frac M n}{2 \frac{(n - 1)M}{n}\frac M n} = \frac{M}{2^{n - 1}\left(\frac M n\right)^n}.
\end{equation}
Let us now compute it for the other case, the one shown in figure \ref{fig:sigma2}.
In this case, we join all the strands in pairs, then we join all the pairs in pairs again, and so on until we reach the final state.
Therefore, we have the coefficient $c_2$ $n/2$ times, that same coefficient for strands of double length $n/4$ times and so on.
More concretely, we have
\begin{align}
c_{n2b2p} & = \left(\frac{1}{\frac M n}\right)^{\frac n 2} \cdot \left(\frac{1}{2\frac M n}\right)^{\frac n 4} \cdot \left(\frac{1}{4\frac M n}\right)^{\frac n 8} \cdot \left(\frac{1}{8\frac M n}\right)^{\frac{n}{16}} \cdot ... \cdot \frac{1}{\frac n 2 \frac M n} = \nonumber \\
& = \prod_{j = 1}^{\log_2 n} \left(\frac{1}{2^{j - 1}\frac M n}\right)^{\frac{n}{2^j}}.
\end{align}
We can easily calculate this product taking its logarithm and summing the resulting equation.
After doing so we obtain
\begin{equation}
c_{n2b2} = n \left( \frac{n}{2M} \right)^{n - 1}.
\end{equation}
As we can see, this result is exactly equal to $c_{n2b2t}$, {\it i.e.} the $c$ coefficient is the same for both joining processes.
So, to simplify notation from now on let us define
\begin{equation}
c_{n2b2} := c_{n2b2p} ( = c_{n2b2t}).
\end{equation}
Since the $c$ coefficients are the same all the difference in the result will come from the comparison of the number of terms; from the $\al$ coefficient.
Let us calculate it.

\subsubsection{Counting the number of terms}
Counting the number of terms is the most involved part of the computation of this $n$-point function.
We need to take into account all the possibilities for the application of each $\Si_2$ operator, and also the different ways in which we can join the strings.
As before, we calculate the contribution for the two limit cases, figures \ref{fig:sigma2t} and \ref{fig:sigma2}, and see which one is dominant.
Then we calculate the total number of ways in which we can join the strands two by two in order to get the final strand of length $M$, and then we approximate the result.
As is being hinted, all this also involves some integer partition theory.

\begin{figure}
\centering
\includegraphics[height=5cm]{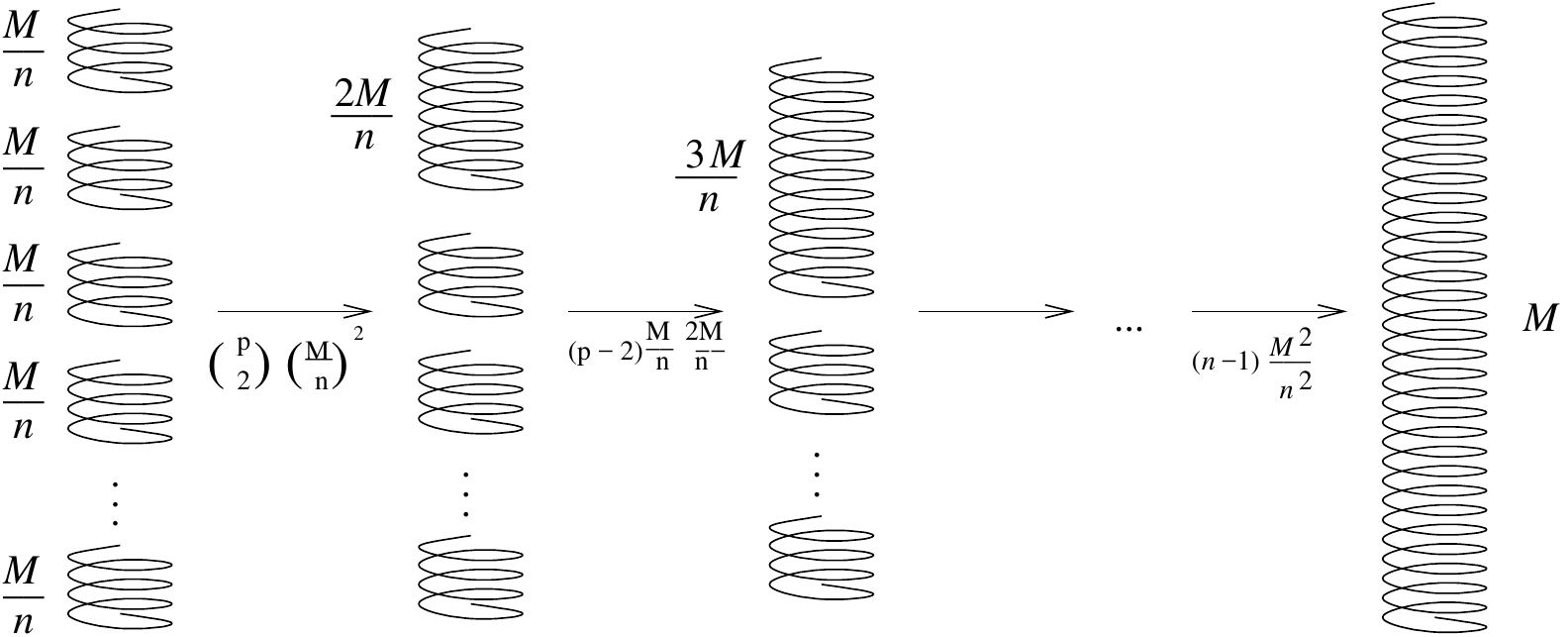}
\caption[Joining strands accumulating them all in the top one.]{Joining strands accumulating them all in the top one. We include some of the combinatorial factors here. The gluing operator can be inserted, for each strand, at any position within it.}
\label{fig:nsigma2t}
\end{figure}

Let us first count in how many ways the $\Si_2$ operators can act in the case where we accumulate all strands in one, that is, in the case of figure \ref{fig:sigma2t}.
To make things more visual we include a picture again with some of the factors involved, figure \ref{fig:nsigma2t}.
We start with $p$ strands of length $M/n$.
In the first step we join two of these, so we need to pick two of them, $\binom{p}{2}$.
Then, each of the two strands has $M/n$ insertion points for the operator.
Thus, the combinatorial factor for the first step is
\begin{equation}
\binom{p}{2} \left(\frac M n\right)^2.
\end{equation}
Similarly, for the second step we need to pick one of the $p - 2$ strands of length $M/n$ and join it to the $2M/n$ strand, giving
\begin{equation}
(p - 2) \frac M n \frac{2M}{n},
\end{equation}
and so on.
We notice that all factors after the first one will have $(M/n)^2$ and a product of the length of the long strand times how many single strands we have left.
Therefore, the total number of ways in which we can do this kind of joining is
\begin{equation}
\binom{p}{2} \left( \frac{M}{n} \right)^2 \prod_{i = 2}^{n - 1} \left(\frac M n\right)^2 i (p - i) = \frac 1 2 \left(\frac M n\right)^{2n} (n - 1)! \frac{p!}{(p - n)!}.
\label{nsigma2tresult}
\end{equation}
Notice that this result is valid for any $n$.
We will now calculate the opposite case, the one shown in figure \ref{fig:sigma2}.
In this case we will need to assume $n$ to be a power of 2, to simplify the expressions.
As we will comment later in section \ref{subsubsection:nnopower2}, nothing qualitatively new happens when $n$ is not a power of 2.

\begin{figure}
\centering
\includegraphics[height=5cm]{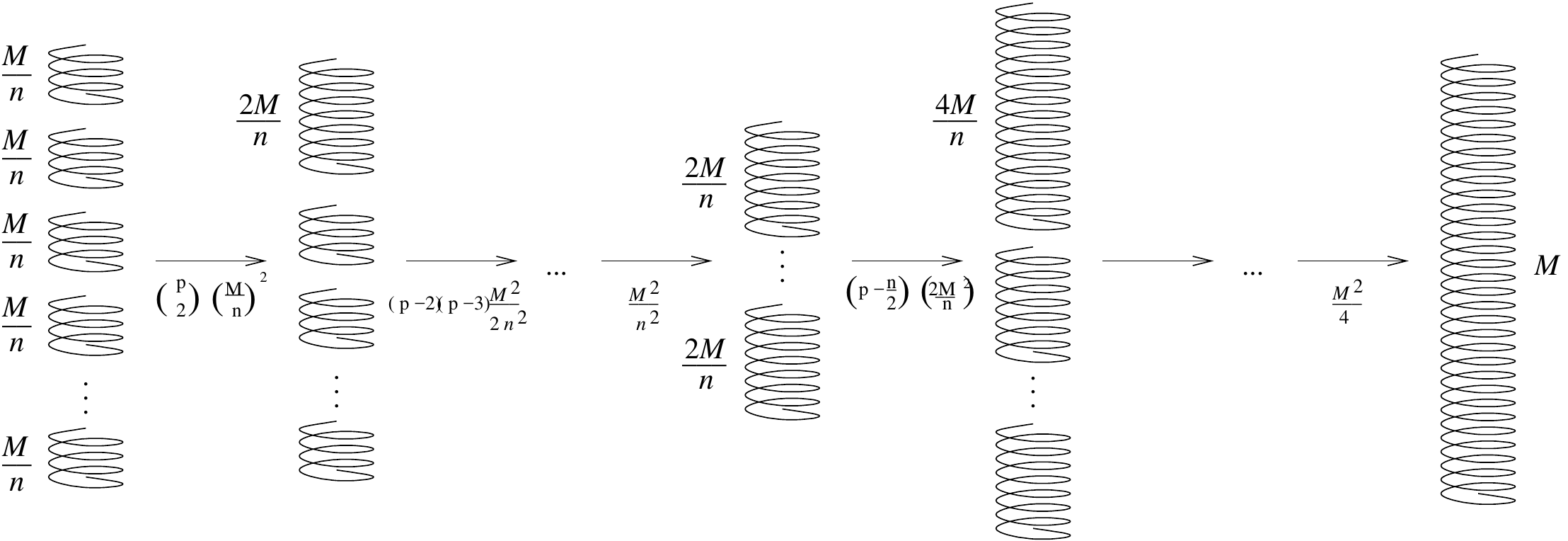}
\caption[Joining strands creating pairs of equal size.]{Joining strands creating pairs of equal size. We sketch the start of the procedure only. We include some of the combinatorial factors here. The gluing operator can be inserted, for each strand, at any position within it.}
\label{fig:nsigma2}
\end{figure}

The counting is completely analogous to the one above, but the product will be harder to compute this time.
Again, to make things visual we include a picture of the gluing process with some numbers, figure \ref{fig:nsigma2}.
The first step is the same one as before.
The second step this time will be to take two of the remaining $p - 2$ strands of length $M/n$ and join them in another pair of length $2M/n$.
We can do this in
\begin{equation}
\binom{p - 2}{2} \left(\frac M n\right)^2
\end{equation}
ways, as we can see in figure \ref{fig:nsigma2}.
Next, we repeat the process but now we will have two strands of length $M/n$ less, as we joined two in the previous step.
Repeating this process, we see that the factor that we obtain from creating $n/2$ strands of length $2M/n$ is
\begin{equation}
\left(\prod_{i = 0}^{\frac n 2 - 1} \binom{p - 2i}{2} \left(\frac M n\right)^2\right) = \left( \frac M n \right)^n \frac{1}{2^{\frac n 2}} \frac{p!}{(p - n)!}.
\end{equation}
Now we need to repeat the same process, but we start with $n/2$ strands of length $2M/n$ and join them in pairs to create strands of length $4M/n$.
We can look for all the coefficients again and multiply them, or just use the equation above and change it accordingly.
In any case, the result for moving our state from $n/2$ $2M/n$ strands to $n/4$ $4M/n$ strands is
\begin{equation}
\left(\prod_{i = 0}^{\frac n 4 - 1} \binom{p - \frac n 2 - 2i}{2} \left(\frac{2M}{n}\right)^2\right) = \left( \frac{2M}{n} \right)^{\frac n 2} \frac{1}{2^{\frac n 4}} \frac{\left( p - \frac n 2 \right)!}{(p - n)!}.
\end{equation}
We need to keep repeating this process until we have just two strands of length $M/2$, and then the final step is just to join them.
We have also included this last coefficient in figure \ref{fig:nsigma2}.
Let us recall that we are assuming $n$ to be a power of two, and so we end up using all the initial strands following this process.
If we write all the contributions together we obtain
\begin{align}
&\left[  \left( \frac M n \right)^n \frac{1}{2^{\frac n 2}} \frac{p!}{(p - n)!} \right] \cdot \left[ \left( \frac{2M}{n} \right)^{\frac n 2} \frac{1}{2^{\frac n 4}} \frac{\left( p - \frac n 2 \right)!}{(p - n)!} \right] \cdot \nonumber \\
&\cdot \left[ \left( \frac{4M}{n} \right)^{\frac n 4} \frac{1}{2^{\frac n 8}} \frac{\left( p - \frac n 2 - \frac n 4 \right)!}{(p - n)!} \right]  \cdot ... \cdot \left[ \binom{ p - n + 2}{2} \left( \frac{(n - 2)M}{n} \right)^2 \right] = \nonumber \\
= & \prod_{j = 1}^{\log_2 n} \left[ \left(\prod_{i = 0}^{\frac{n}{2^j} - 1} \binom{p - \sum_{k = 1}^{j - 1} \frac{n}{2^k} - 2i}{2} \left(\frac{2^{j - 1} M}{n}\right)^2\right)\right].
\end{align}
The product in $i$ is straightforward to compute, but the product in $j$ is more involved.
After doing the $i$ multiplication, we separate the product above in three products, which we will calculate separately.
They are
\begin{equation}
\left(\prod_{j = 1}^{\log_2 n} 2^{(2j - 3) \frac{n}{2^j}}\right) \cdot \left(\prod_{j = 1}^{\log_2 n} \left(\frac M n\right)^{\frac{n}{2^{j-1}}}\right) \cdot \left(\prod_{j = 1}^{\log_2 n} \frac{\left( p - \frac{n}{2^j} (2^j - 2) \right)!}{(p - n)!} \right).
\label{productnsigma2}
\end{equation}
The first two factors are easy to obtain.
The first one is
\begin{equation}
\left(\prod_{j = 1}^{\log_2 n} 2^{(2j - 3) \frac{n}{2^j}}\right) = 2^{\sum_{j = 1}^{\log_2 n} \frac{n j}{2^{j - 1}}} 2^{-\sum_{j = 1}^{\log_2 n} \frac{3 n}{2^j}} = \frac{2^{n - 1}}{n^2},
\end{equation}
and the second one is
\begin{equation}
\left(\prod_{j = 1}^{\log_2 n} \left(\frac M n\right)^{\frac{n}{2^{j-1}}}\right) = \left(\frac M n\right)^{2(n - 1)}.
\end{equation}
We need some more work to obtain the result for the third product.
The product as it is cannot be calculated for generic $p$ as far as we are aware, as there is no formula that gives its result.
We will consider the case $p = n$, as that one can be summed.
Notice that in this case the product reduces to the product of factorials of all the powers of 2 until $n$,
\begin{equation}
\left(\prod_{j = 1}^{\log_2 n} \left(\frac{n}{2^{j - 1}}\right)! \right).
\end{equation}
As far as we are aware, there is no exact expression for this product either.
However it grows very fast, and so Stirling's formula, which we gave in equation \eqref{stirlingweak}, will be a good approximation\footnote{Using Stirling's approximation with the exact product (not setting $p = n$) also results in a sum which cannot be performed by any methods we are aware of.}.
Let us rewrite, for convenience, the product above as
\begin{equation}
F(k) := \prod_{j = 0}^{k - 1} \left(2^{k - j}\right)!,
\end{equation}
where we are using $n = 2^k$, for some $k \in \mathbb N$. 
Then,
\begin{equation}
\ln F(k) = \sum_{j = 0}^{k - 1} \ln \left[\left(2^{k - j}\right)!\right].
\end{equation}
Using Stirling's approximation, this can be rewritten for large $n$ as
\begin{equation}
\ln F(k) \approx \sum_{j = 0}^{k - 1} \left( 2^{k - j} \ln \left(2^{k - j}\right) - 2^{k - j} \right) = 2n \ln \left(\frac n 2\right) + 2\ln 2 + 2 - 2n,
\end{equation}
where in the second equality we computed the sums and substituted back to $n$.
Then, the third factor gives
\begin{equation}
\prod_{j = 1}^{\log_2 n} \left(\frac{n}{2^{j - 1}}\right)! \approx n^{2n} 4^{1 - n} e^{2 - 2n}.
\end{equation}
We have now calculated the three factors that we needed in equation \eqref{productnsigma2}.
Multiplying them, we see that the total number of ways in which we can do this kind of joining is, for large $n$ and $n$ being a power of 2,
\begin{equation}
\left(\frac{M^2}{2 e^2}\right)^{n - 1}.
\label{nsigma2result}
\end{equation}
We now need to compare both results, \eqref{nsigma2tresult} and \eqref{nsigma2result}.
Clearly \eqref{nsigma2result} is bigger, as we are in the large $n$ limit.
This means that there are many more ways to join the strands by creating pairs of equal size than accumulating them all together in one big strand, which was to be expected.
However, we are not finished with the counting.
We just saw that the counting is bigger when we join them in pairs, and in the previous section we saw that the $c$ coefficient is the same for both cases.
However, these are the two limit cases.
There are many ways in which we can join all the strands.
For instance, we could create three pairs of length $2M/n$, and then join everything together in one big strand by accumulating them, just as we did above.
That is, we can have a process which is a combination of both limit processes that we just calculated.

It is important to notice though that we calculated the two limit cases, and any other case will be a mixture of the two, and so will give a smaller result than \eqref{nsigma2result} and bigger than \eqref{nsigma2tresult}.
However, there are a lot of intermediate cases; there are a lot of ways in which the $n$ strands can be joined.
The number of ways may in fact change the scaling with $n$ of the result, and so we need to calculate it.

\begin{figure}
\centering
\includegraphics[height=3cm]{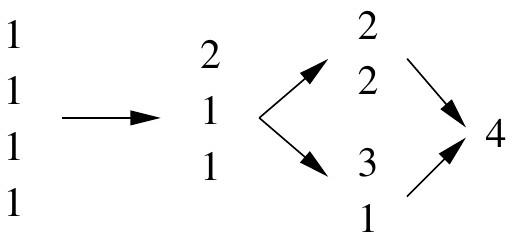}
\caption[Refinements of the partition 4.]{There are two different ways in which we can go from the partition 1,1,1,1 to the partition 4.}
\label{fig:refiningpartitions4}
\end{figure}

\begin{figure}
\centering
\includegraphics[height=7cm]{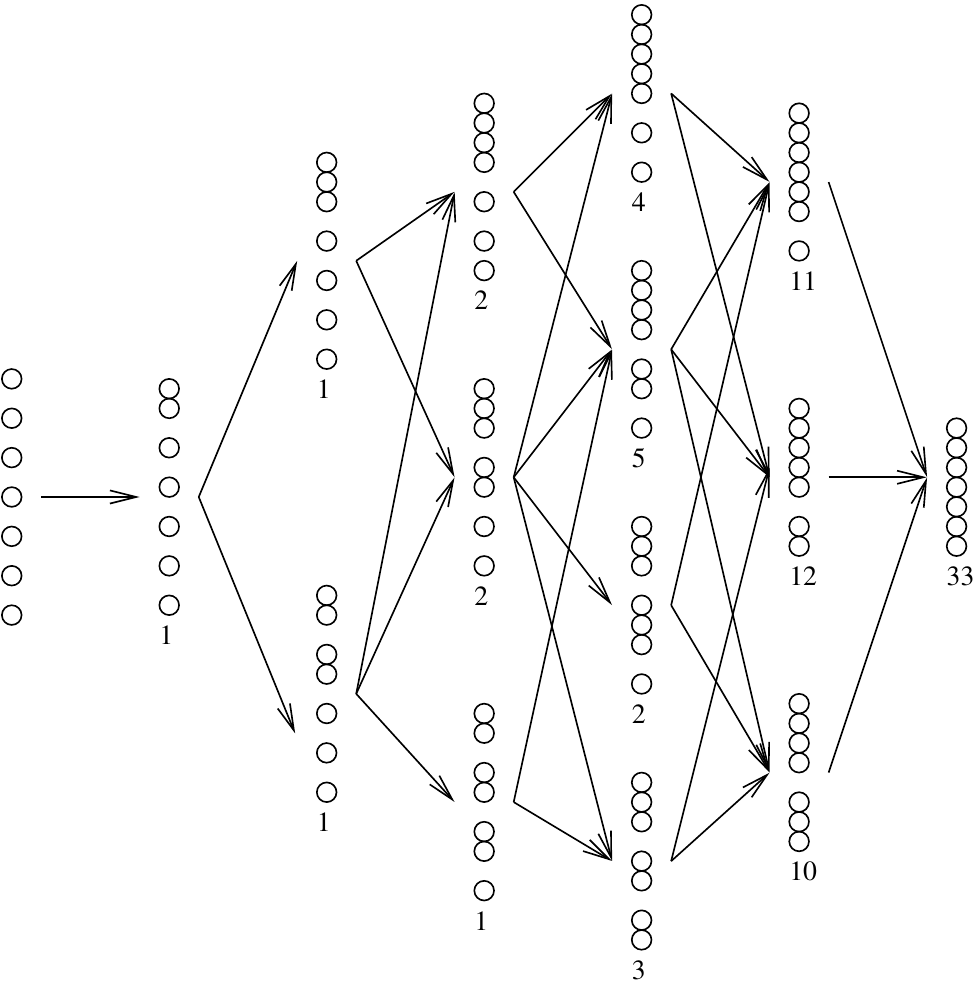}
\caption[Refinements of the partition 7.]{Number of ways of going from the partition 1,...,1 to 7. The numbers next to the partitions are the number of distinct paths from the original set. In this case we have $f(7) = 33$. Figure taken from \cite{erdos1975refining}.}
\label{fig:refiningpartitions}
\end{figure}

So, the problem at hand now is to find in how many ways we can join all the strands in one.
As it seems natural at this point, this problem is also within the integer partitions theory.
This counting is equivalent to finding in how many ways we can go from the partition of $n$ 1, ..., 1 to the partition $n$ by adding numbers in pairs.
For example, for four we have only two ways, as we show in figure \ref{fig:refiningpartitions4}.
Finding this for arbitrary $n$ is again a very difficult problem in number theory.
And again, as far as we are aware there is no exact formula for this counting.
However, the scaling with $n$ is known for the case of large $n$.
Paul Erdős and collaborators found that, if $f(n)$ is the number we want to find and $c_1$ and $c_2$ are constants, then \cite{erdos1975refining}
\begin{equation}
c_1^n n^{\frac n 2} < f(n) < c_2^n n^{\frac n 2}.
\label{erdosrefiningresult}
\end{equation}
They suggest $c_1$ to be 0.75 and find $c_2$ to be $8\sqrt 2$.
For details on the proof we refer to that paper.
We just want to mention that in the proof Stirling's formula is also used, and so our previous assumption of $n$ being large is consistent.
We do include a figure inspired by that paper to illustrate the problem better, and show that $f(n)$ grows very fast by studying the case $n = 7$; figure \ref{fig:refiningpartitions}.
For our case at hand we will ignore the constants, as we are only concerned about the scaling with $n$ and the main contribution comes from the other factor.
Therefore, we approximate it by saying that the number of ways in which we can join $n$ strands two by two to get a single strand is $n^{\frac n 2}$.

We now have all the coefficients that we need to compute the one point function, so let us put them all together.

\subsubsection{Result}
In the previous section we learned that the dominant way in which the strings are joined two by two is the one depicted in figure \ref{fig:nsigma2}.
However, we were able to calculate the $\al$ coefficient in that case, as we could not do the counting for generic $p$.
Therefore, we will use the coefficients that we obtained for the other case.
Since we know that is the smallest case, the result that we obtain is a lower bound.
First let us find $\al$, which corresponds to the matching of number of terms.
We calculated the number of ways in which we can apply the $n - 1$ $\Si_2$ operators in the case where we keep gluing the single length strands to the same one in \eqref{nsigma2tresult}.
Therefore, we have
\begin{equation}
\frac 1 2 \left( \frac M n \right)^{2n} (n - 1)! \frac{p!}{(p - n)!} \mathcal N(p) = \al \mathcal N(p - n).
\end{equation}
Solving for $\al$ we obtain
\begin{equation}
\al_{2b2t} = M \left( \frac N M - \frac p n + 1 \right) \frac 1 2 \left( \frac M n \right)^n (n - 1)!.
\end{equation}
Before we compare both results we give the result of this $(n - 1)$-point function in the short strand case, so that we can already compare its result to the one obtained in section \ref{subsection:sigmancorfun}.
It is
\begin{equation}
\left\langle \left( \Si_2^{- \dot -} \right)^{n - 1} \right\rangle_t = \left( N_{\Si_2} \right)^{n - 1} A^n \bar B c_{n_{2b2t}} \frac 1 2 \left( \frac M n \right)^n (n - 1)!.
\label{sigma2tresult}
\end{equation}
Now, this is the one point function if the only gluing process we consider is the one depicted in figure \ref{fig:nsigma2t}.
As we have seen, this process is the one that will give the smallest answer to the one point function.
Even if it is the only one that we have calculated exactly, as we have argued we have $n^{n/2}$ ways of doing this process, and the $\al$ factors for all the other processes will be bigger.
We can check this if we set $p = n$ in \eqref{nsigma2tresult} and compare it to \eqref{nsigma2result}.
If we want to obtain an answer for the full one point function we need to consider all joinings.
Clearly, if we multiply the result \eqref{sigma2tresult} by the number of ways we can join, we will obtain a lower bound for this $(n - 1)$-point function.
Hence, the answer is
\begin{equation}
\left\langle \left( \Si_2^{- \dot -} \right)^{n - 1} \right\rangle > A^n \bar B M (n - 1)! \left(\frac{\sqrt{n}}{2}\right)^n \left( \frac{n}{N (N - 1)} \right)^{\frac{n - 1}{2}}.
\label{sigma2lowerbound}
\end{equation}

\subsubsection{\texorpdfstring{$n$}{n} not being a power of two}\label{subsubsection:nnopower2}
We have now computed both $n$-point functions.
As we saw in section \ref{subsection:sigmancorfun}, the result of the one point function of $\Si_n^{- \dot -}$, equation \eqref{sigmanresult} is valid for any $n$.
However, as we have just seen, \eqref{sigma2lowerbound} is only valid when $n$ is a power of two, because we join the strands in pairs.
This does not mean that we can only calculate this $n - 1$-point function in that case, though.
If $n$ is not a power of two we simply need to look for the biggest power of two smaller than $n$, do the process for that subset, and do the same for the smaller subset which is not a power of two.
Clearly the result will be longer to write, but the calculation is exactly the same, and we will just end up with a product of results of the form \eqref{sigma2lowerbound}.
For tidiness we keep assuming that $n$ is a power of two.

\subsection{Comparison of results}\label{joinsec:compare}
We now have all the results we needed, so the last thing that we have to do is to compare the result of both calculations, to see which one is bigger.
Let us recall again that \eqref{sigma2lowerbound} is a lower bound, and so the scaling of $n$ of this one point function is higher than what we use in this section.
Also, notice that to compare both results we do not need to use the answers we obtained in the short strand limit, nor we need to take the long strand limit or worry about the $f(\al)$ polynomial.
The process described by both operators is the same, and so the calculations are exactly the same except for the $\al$ and $c$ coefficients.
Since these are independent of the sum index (except for a term in the $\al$ coefficient, which is the same in both cases), all factors which are different for both calculations come out of the sums, and thus the sums cancel.
Therefore, to compare both answers we only need to divide the $c$ and $\al$ factors and the normalisations of the twists.
By doing so we obtain
\begin{equation}
\frac{\left\langle \left( \Si_2^{- \dot -} \right)^{n - 1} \right\rangle}{\left\langle \Si_n^{- \dot -} \right\rangle} \gtrsim \frac{n! (n - 1)! M^n n^{\frac M n} \left( \frac{2}{N (N - 1)} \right)^{\frac{n-1}{2}}}{2^n n^{\frac M 2 + \frac n 2} \left( (n - 1)^2 (n \bar a)^{n - 2} \Lambda^{-1} \right)^{\frac 1 2 (n - 1) \left( 1 - \frac M n \right)} \left( \frac{n}{N (N - 1) ... (N - n + 1)} \right)^{\frac 1 2}},
\label{1pfeq:joincompare}
\end{equation}
where
\begin{equation}
\Lambda = \left ( 1 + n (n-1)^{(n-1)} - n^{n-1} \right )
\end{equation}
and
\begin{equation}
\bar{a}^{n-1} = n \left ( 1 - \frac{1}{n} \right )^{n-1} - 1.
\end{equation}
As we can see, depending on the values of $M$ and $n$ the fraction above will either go to zero or infinity in the large $N$ limit.
Namely, for large values of $M$ it will go to infinity.
To see this more clearly, let us simplify the result.
Let us recall that we are assuming $n \gg 1$, however we can have the case $1 \ll n \ll N$.
Let us assume this is the case.
Then, the equation above simplifies to
\begin{equation}
\frac{\left\langle \left( \Si_2^{- \dot -} \right)^{n - 1} \right\rangle}{\left\langle \Si_n^{- \dot -} \right\rangle} \gtrsim \frac{n! (n - 1)! M^n n^{\frac M n}}{2^{\frac{n+1}{2}} n^{\frac M 2 + \frac n 2 + \frac 1 2} \left( (n - 1)^2 (n \bar a)^{n - 2} \Lambda^{-1} \right)^{\frac 1 2 (n - 1) \left( 1 - \frac M n \right)}} N^{1 - \frac n 2},
\end{equation}
which leads to
\begin{equation}
\frac{\left\langle \left( \Si_2^{- \dot -} \right)^{n - 1} \right\rangle}{\left\langle \Si_n^{- \dot -} \right\rangle} \to +\infty \qquad \text{for} \qquad M \approx N \to \infty.
\end{equation}
So, if $M$ is of order $N$ then $\braket{(\Si_2^{- \dot -})^{n - 1}}$ will be bigger.
If $M$ is orders of magnitude smaller than $N$ then the one point function $\braket{\Si_n^{- \dot -}}$ has a bigger value.
Thus, the joining of strands with multiple twist operators should also need to be considered, as the expectation values of both calculations can be of the same order of magnitude depending on the case.

As we have seen there are many contributions that play a role here, but the main difference comes from combinatorics.
There are many ways in which all the $\Si_2$ operators can be inserted, and that gives a very big contribution, whereas for $\Si_n$ the process is much more restricted, the number of ways in which the gluing can be done is much smaller, and so the coefficients also are.
Let us recall that we have done the comparison with a lower bound of $\braket{(\Si_2^{- \dot -})^{n - 1}}$.
By looking at equations \eqref{productnsigma2} and \eqref{nsigma2result} we see that its upper bound will have higher powers of $n$ and $M$.
Therefore, the qualitative comparison done above holds in the same way for the upper bound as well.

There is an obvious extension to this result to obtain a stronger one, which we consider in the next section.
Up until now we have been concerned about two ways of joining strands: all at the same time, or by pairs.
However, there are many more ways to join strands.
So far we have only used the $\Si_2$ twist and the $\Si_n$, but there are also twist operators $\Si_i$, for $2 \leq i \leq n$.
Let us see how this is translated in terms of $n$-point functions.

\subsection{All possible ways of joining the strands}\label{1pfsubsec:allwaystojoin}
As we just said, we have computed the $n$-point function when we join strands two by two, and when we join them all at the same time.
However, we could have also joined them all three by three, if $n$ was a power of three.
Or with any combination of gluing operators, up to $\Si_n$.
That is, for every $n$ we have as many possibilities for joining them as possible combinations of twist operators are there that will join them all together.
Rephrasing this in terms of integer partitions, there are as many ways of joining the $n$ strands as there are partitions of $n$ that do not contain 1 as a part.
This is another hard problem in number theory related to integer partitions.
The sequence that results from this counting is recorded in the On-Line Encyclopedia of Integer Sequences (OEIS).
It is the sequence A002865 \cite{OEISA002865}.
Again, there is no exact formula for this counting, but it grows exponentially fast with $n$.
There is an approximate formula for this counting, which is
\begin{equation}
\frac{\pi e^{\sqrt{\frac{2n}{3}} \pi}}{12 \sqrt 2 n^{\frac 3 2}} \left[ 1 - \frac{3 \frac{\sqrt{\frac 3 2}}{\pi} + \frac{13 \pi}{24 \sqrt 6}}{\sqrt n} + \frac{\frac{217 \pi^2}{6912} + \frac{9}{2\pi^2} + \frac{13}{8}}{n} \right].
\label{partitionswithout1}
\end{equation}
So, there are many ways to do the joining, and in the previous sections we have calculated the two limit cases.
Let us recall that the final answer for the one point functions is highly dependant on the combinatorics of the joining of the strands, that is, in how many ways we can apply the gluing operators.
Clearly joining the strands two by two is the case where we have the most combinations, and joining them all together in one step is the case where we have the least.
We also need to take into account the normalisations of the twists, which will give different powers of $N$.
Thus, just as it happened in the previous case, equation \eqref{1pfeq:joincompare}, in all the intermediate cases the corresponding one point function will have a value comparable to $\langle \Si_n \rangle$ depending on how big $M$ (and $n$) are.

Calculating the $n$-point functions in the middle by the same methods we used would not be straightforward, as we would need to count all the possible ways of joining.
For $\Si_2$ we used the result from Erdős' paper \cite{erdos1975refining}, but that counting has not been studied for any other integer as far as we are aware.
To finish this section let us give an example to illustrate what the counting problem is.

Assume that $n$ is a power of $m$, where $m$ is a natural number bigger than two.
Thus, let $n = m^k$, for some positive integer $k$.
It is straightforward to see that if we want to join all the strands using only $\Si_m$ we will need to do $(n-1)/(m-1)$ steps, that is, that the $n$-point function that we would want to calculate is $\langle (\Si_m)^{\frac{n - 1}{m - 1}} \rangle$.
However, to calculate it we would need to know in how many ways we can join the $n$ strands by joining $m$ in each step.
As we said above, this has only been studied for $m = 2$ so far, which is the calculation we did in the previous section.
Just to get an idea, the number of possible ways grows very fast for $m > 2$ as well, but slower than for two, as would be expected.
For example, if $m = 3$, then we have one combination for $k = 1$, five combinations for $k = 2$ and 5,026,161 combinations for $k = 3$.
Also, we would need to consider all combinations of twist operators that add up to $n$, which again results in problems within the theory of integer partitions that have not been solved, as far as we are aware.

These countings need to be studied carefully in all their limits, as they might point towards other relevant subclasses of microstates.
The bounds given for the results might also help in the holographic calculations, as some of these multi-particle one point functions are also relevant in supergravity.

\section{Review of results}\label{section:resultreview}
In this section we give a summary of all the results presented in this paper.
Let us remind that this is not supposed to be a comprehensive list of all possible one point functions for chiral primaries.
Rather, we rewrite the results we have obtained for some exemplary cases, which can be easily extended to calculate many more one point functions.

\subsection{Short strand one point functions}
In the short strand case we have obtained the following results:
\begin{align}
\braket{\Si_2^{+ \dot -}}_{\text{short}} & = \frac{e^{i\frac{\sqrt 2 v}{R}}}{4} \frac{(n_l + p_l)}{n_l\; p_l} A^{(++)}_{n_l} B^{0(00)}_{p_l} \overline{B^{1(00)}_{n_l + p_l}} \left( \frac{2}{N (N - 1)} \right)^{\frac 1 2}, \nonumber \\
\left\langle \sum_{r = 1}^n \mathcal O^{- \dot -}_{(r)} \right\rangle_{\text{short}} & = \frac{A \bar B}{n}, \nonumber \\
\left\langle \sum_{r = 1}^n \mathcal O^{+ \dot -}_{(r)} \right\rangle_{\text{short}} & = \frac 1 n A \overline{B^1} e^{i \frac{\sqrt 2 v}{R}}, \nonumber \\
\braket{\Si_n^{- \dot -}}_{\text{short}} & = |\Si_n^{- \dot -}|^{-1} c_n \frac{N - \frac{\bar p M}{n}}{n!} A^n A_2^{-1} = \nonumber \\
& = A^n \bar A_2 \frac{c_n}{n!} \left( \frac{n}{N (N - 1) \cdot ... \cdot (N - n + 1)} \right)^{\frac 1 2}, \nonumber \\
\left< \left( J^+_{-1} \right)^m \right>_{\text{short}} & = B \bar B_m \frac 1 n \binom{n}{m}, \nonumber \\
\left< \bigotimes_{r = 1}^n \mathcal O^{- \dot -}_{(r)}\right> & = A^n \bar B^n \frac{N!}{n!} \frac{(1-\alpha )^n}{(N - n)!}.
\end{align}
These results are equations \eqref{simplestSigma}, \eqref{1pointfunO--}, \eqref{eq:oshortgeneral}, \eqref{sigmanresult}, \eqref{1pfeq:Jshort} and \eqref{1pfeq:Oproductresult} respectively.
We have also given some combinatorics to calculate the $\braket{\Si_n^{+ \dot -}}$ one point function for 1/8-BPS states.
However we have not given the final answer for that correlator, as to do that we need the commutator $[\Si_n^{+ \dot -}, \otimes_r J_{(r)}]$.

\subsection{Long strand one point functions}
In the two-charge case we have obtained general results for the $\Si_n^{- \dot -}$ and for the $\mathcal O^{- \dot -}_{(r)}$ operators.
They are
\begin{align}
&\braket{\Si^{- \dot -}_n}_{\mathrm{long}} = \nonumber \\
& = \left( \prod_{i = 1}^n A_i \right) \bar A N^{1 - n} \left( \left( \frac n Q \right)^{\frac N 2 (n - 1)} \left( \frac 1 n \right)^{\frac N 2} \right)^{\frac 1 m} \left( \frac{n}{N (N - 1) ... (N - n + 1)} \right)^{\frac 1 2} f(\al_1, ..., \al_n), \nonumber \\
&\braket{\mathcal O^{- \dot -}_{(r)}}_{\mathrm{long}} = A_1 \bar A_2 \frac 1 N f(\al),
\end{align}
which are equations \eqref{1pfeq:sin14gen} and \eqref{1pfeq:gen14Ores}.
$f(\al_1, ..., \al_n)$ are polynomials independent of $N$, with the $\al_i$ parameters taking values in the $(0,1)$ interval.
The polynomials are of order one for all values of the $\al_i$, and they diverge at the (excluded) boundary values.
We have given explicit examples of the polynomials in section \ref{1pfsubsec:14longcorrex}.

In the three-charge case we have obtained the following results,
\begin{align}
\braket{\mathcal O^{+ \dot -}_{(r)}}_{\mathrm{long}} & = A \bar B e^{i \frac{\sqrt 2 v}{R}} \frac{2}{N} \qquad \text{(initial example)}, \nonumber \\
\braket{\mathcal O^{- \dot -}_{(r)}}_{\mathrm{long}} & = \frac{A^{(++)} \overline{A^{0(00)}}}{N} \frac{n}{\alpha + \beta - n^2 (\alpha +\beta -1)}, \nonumber \\
\braket{\Si^{+ \dot -}_2}_{\text{long}} & \approx N_{\Si_2} \frac{e^{i\frac{\sqrt 2 v}{R}}}{4} \frac{A B \bar{B^1}}{N}, \nonumber \\
\left< \sum_{r = 1}^{\frac{N}{\ka}} \mathcal O^{+ \dot -}_{(r)} \right>_{\text{long}} & \approx \frac{\ka}{N} A \bar B_1 e^{i \frac{\sqrt 2 v}{R}} \frac{|\varphi|^2}{|\psi|^2} \qquad \text{(see subsec. \ref{1pfsubsec:1/8statefull})}, \nonumber \\
\left< \left( J^+_{-1} \right)^m \right>_{\text{long}} & = B \bar B_m \frac{n \binom{\frac{N}{n}}{m}^3}{\alpha  N-(\alpha -1) \binom{\frac{N}{n}}{m}^2},
\end{align}
which are equations \eqref{1pfeq:O+-longcorr}, \eqref{1pfeq:O--longres}, \eqref{1pfeq:si+-218approx}, \eqref{1pfeq:olonggeneral} and \eqref{1pfeq:Jmlongexactres}.

\subsection{Comparison between short and long strand results}
Comparing the results we have seen that
\begin{align}
\braket{\Si_2^{- \dot -}}_{\mathrm{long}} & \sim \frac 1 N \braket{\Si_2^{- \dot -}}_{\mathrm{short}}, \nonumber \\
\braket{\Si_3^{- \dot -}}_{\mathrm{long}} & \sim \frac{1}{N^2} \braket{\Si_3^{- \dot -}}_{\mathrm{short}}, \nonumber \\
\braket{\Si_n^{- \dot -}}_{\mathrm{long}} & \sim  N^{1 - n} \left( \frac 1 n \right)^N \braket{\Si_n^{- \dot -}}_{\mathrm{short}}, \qquad n \geq 4, \quad\mathrm{(same\; strand\; length)}, \nonumber \\
\braket{\mathcal O^{- \dot -}_{(r)}}_{\mathrm{long}} & \sim \frac 1 N \braket{\mathcal O^{- \dot -}_{(r)}}_{\mathrm{short}}, \nonumber \\
\braket{\mathcal O^{+ \dot -}_{(r)}}_{\mathrm{long}} & \sim \frac{1}{N} \braket{\mathcal O^{+ \dot -}_{(r)}}_{\mathrm{short}}, \nonumber \\
\braket{\Si^{+ \dot -}_2}_{\text{long}} & \sim \frac 1 N \braket{\Si_2^{+ \dot -}}_{\text{short}}, \nonumber \\
\braket{(J^+_{-1})^m}_{\text{long}} & \approx N \braket{(J^+_{-1})^m}_{\text{short}},
\end{align}
which are equations \eqref{1pfeq:sis14gen}, \eqref{1pfeq:longshortOrel}, \eqref{1pfeq:O+-longshortrel} and \eqref{1pfeq:si2longshortrel}.
As we can see, all the long strand ones are suppressed by at least $1/N$ with respect to the short ones, except for the R-symmetry current mode.
As we said, this indicates that one point functions for long strand states are comparable to supergravity corrections.

\subsection{Different ways of joining strands}
In section \ref{subsection:sigma2corfun} we have been concerned about the $n$-point function for the $\Si_2^{- \dot -}$ twist operator.
We have obtained a lower bound for it, equation \eqref{sigma2lowerbound}, which is
\begin{equation}
\left\langle \left( \Si_2^{- \dot -} \right)^{n - 1} \right\rangle > A^n \bar B M (n - 1)! \left(\frac{\sqrt{n}}{2}\right)^n \left( \frac{n}{N (N - 1)} \right)^{\frac{n - 1}{2}}.
\end{equation}
We have also estimated the upper bound, counted all other $n$-point functions of products of twist operators and commented the result.
In the last section we have found the exact result, using the lower bound for $\braket{(\Si_2^{- \dot -})^{n-1}}$,
\begin{equation}
\frac{\left\langle \left( \Si_2^{- \dot -} \right)^{n - 1} \right\rangle}{\left\langle \Si_n^{- \dot -} \right\rangle} \gtrsim \frac{n! (n - 1)! M^n n^{\frac M n} \left( \frac{2}{N (N - 1)} \right)^{\frac{n-1}{2}}}{2^n n^{\frac M 2 + \frac n 2} \left( (n - 1)^2 (n \bar a)^{n - 2} \Lambda^{-1} \right)^{\frac 1 2 (n - 1) \left( 1 - \frac M n \right)} \left( \frac{n}{N (N - 1) ... (N - n + 1)} \right)^{\frac 1 2}},
\end{equation}
and we have also seen that the number of combinations of twist operators that study the same process scales like
\begin{equation}
\frac{\pi e^{\sqrt{\frac{2n}{3}} \pi}}{12 \sqrt 2 n^{\frac 3 2}} \left[ 1 - \frac{3 \frac{\sqrt{\frac 3 2}}{\pi} + \frac{13 \pi}{24 \sqrt 6}}{\sqrt n} + \frac{\frac{217 \pi^2}{6912} + \frac{9}{2\pi^2} + \frac{13}{8}}{n} \right],
\end{equation}
and so we have a large number of $n$-point functions which, depending in the lengths of the strands we join, will have values comparable (or bigger) to $\braket{\Si^{- \dot -}_n}$.

\section{Conclusions and outlook} \label{section:conc}

In this paper we have calculated one point functions of chiral primary operators in the D1-D5 orbifold CFT, in classes of two and three charge black hole microstates. Three charge microstates are obtained by adding momentum excitations to the Ramond ground states, as discussed in section \ref{1pfsubsec:1/8bpsstatedef}. The typical structure of the three charge microstates that we consider in this paper is shown in \eqref{typ1/8}: these states involve adding excitations with integer momentum to Ramond ground states. As reviewed in section \ref{1pfsubsec:1/8bpsstatedef}, the majority of three charge microstates are obtained by acting with fractional modes on long string states and it would be interesting to extend the calculations of this paper to generic states involving fractional modes.  

Black hole microstates of the type \eqref{typ1/8} have been explored before, as dual descriptions for the classes of supergravity solutions analysed in \cite{Bena:2015bea, Giusto:2015dfa, Bena:2016agb,Bena:2016ypk,Bena:2017geu}. The results of this paper indicate that when such microstates involve excitations of long string Ramond ground states the one point functions of chiral primaries are suppressed by factors of $N$. Holographically this implies that the characteristic scales in the supergravity solutions are very small: the supergravity solutions differ from the naive black hole geometry at horizon scales, by contributions that are comparable to the scale of higher derivative corrections to supergravity. Thus, even when such microstates have a representation in supergravity, the supergravity description may just be an extrapolation of a string background. 

Another important issue is distinguishability of microstates: the only information encoded in supergravity geometries is the expectation values of single particle chiral primary operators in the CFT. Complete information about the microstates requires the expectation values of all operators: multi particle chiral primaries and (for three charge black hole microstates) 1/8 BPS operators. In this paper we have computed expectation values of multi particle chiral primaries and it would be interesting to explore how this information can be encoded into the holographic description of a black hole microstate. 

The results derived in this paper have applications beyond the black hole microstate programme. A recent paper \cite{Eberhardt:2018ouy} proposed a worldsheet dual for the symmetric product CFT itself and the correlation functions calculated here could be used to test this duality. Efficient methods to compute correlation functions holographically were developed in \cite{Rastelli:2017udc}; it would be interesting to use these methods to calculate the correlation functions discussed here from dual $AdS_3 \times S^3$ backgrounds.

\section*{Acknowledgements}

This work is funded by the STFC grant ST/P000711/1. This project has received funding and support from the European Union's Horizon 2020 research and innovation programme under the Marie Sklodowska-Curie grant agreement No 690575.

\bibliographystyle{JHEP}
\bibliography{refs}

\end{document}